\documentclass[12pt]{article}

\usepackage{amsthm}
\usepackage{amsmath}
\usepackage{natbib}
\usepackage[colorlinks,citecolor=blue,urlcolor=blue,filecolor=blue,backref=page]{hyperref}
\usepackage{graphicx}

\usepackage{enumerate}
\usepackage{amsfonts}
\usepackage{bm}
\usepackage[T1]{fontenc}

\usepackage{booktabs}
\usepackage{subcaption}
\usepackage{float}

\usepackage{authblk}
\usepackage{lmodern}

\begin{document}

\title{\Large \bf Joint Species Distribution Modeling with Competition for Space}

\author[1]{Juho Kettunen}
\author[2]{Lauri Meht\"atalo}
\author[3]{Eeva-Stiina Tuittila}
\author[4,5]{\authorcr Aino Korrensalo}
\author[6]{Jarno Vanhatalo}
\affil[1]{School of Computing, University of Eastern Finland, Joensuu, Finland}
\affil[2]{Bioeconomy and Environment Unit, Natural Resources Institute Finland (Luke), Joensuu, Finland}
\affil[3]{School of Forest Sciences, University of Eastern Finland, Joensuu, Finland}
\affil[4]{Department of Environmental and Biological Sciences, University of Eastern Finland, Joensuu, Finland}
\affil[5]{Natural resources unit, Natural Resources Institute Finland (Luke), Joensuu, Finland}
\affil[6]{Department of Mathematics and Statistics, Organismal and Evolutionary Biology Research Program, University of Helsinki, Helsinki, Finland}

\date{}
\maketitle

\begin{abstract}
Joint species distribution models (JSDM) are among the most important statistical tools in community ecology. However, existing JSDMs cannot model mutual exclusion between species. We tackle this deficiency in the context of modeling plant percentage cover data, where mutual exclusion arises from limited growing space and competition for light. We propose a hierarchical JSDM where latent Gaussian variable models describe species’ niche preferences and Dirichlet-Multinomial distribution models the observation process and competition between species. We also propose a decision theoretic model comparison and validation approach to assess the goodness of JSDMs in four different types of predictive tasks. We apply our models and methods to a case study on modeling vegetation cover in a boreal peatland. Our results show that ignoring the interspecific interactions and competition reduces models’ predictive performance and leads to biased estimates for total percentage cover. Models’ relative predictive performance also depends on the predictive task highlighting that model comparison and assessment should resemble the true predictive task. Our results also demonstrate that the proposed JSDM can be used to simultaneously infer interspecific correlations in niche preference as well as mutual competition for space and through that provide novel insight into ecological research.
\end{abstract}

\noindent%
\vfill

\section{Introduction}
\label{sec:intro}

Community ecology has seen fast and significant development of statistical methods for joint species distribution modeling in recent years \citep{Warton:2015,Ovaskainen+Abrego:2020,Nordberg2019,Vanhatalo2020}. 
Species distribution models (SDMs) are statistical models that describe and predict the variation in the occurrence and abundance of species in space and time \citep{Gelfand2006}. 
Joint species distribution models (JSDMs) are their extensions to modeling multivariate species communities \citep{Ovaskainen+Abrego:2020,Vanhatalo2020}. 
These models are used in a wide variety of applications, ranging from applied use, such as natural resources management and conservation planning \citep{Kallasvuo_etal:2017,guisan:2013}, to scientific inference, such as studying species' responses to environmental filtering \citep{clark:2016,Kottaetal:2019} and interspecific relationships \citep{Tikhonov_etal:2017}.
(J)SDMs are routinely used also for prediction. The most common type of prediction is species distribution mapping where predictions from (J)SDMs are turned into thematic maps showing either the occurrence probability or abundance of species over a region of interest \citep{Gelfand2006,Elith2009,Makinen2018}. Other common predictive applications of (J)SDMs are biomass estimation over spatiotemporal domains \citep{Kallasvuo_etal:2017} and predictions concerning biodiversity at unobserved locations \citep{clark:2016}.

The state-of-the-art JSDMs are built using hierarchical multivariate generalized linear models appended with Gaussian latent factors \citep{Pollock+etal:2014,Warton:2015,Nordberg2019,Ovaskainen+Abrego:2020}. 
The underlying assumption in these models is that species observations are linked to a latent Gaussian model, which includes a description for species' environmental niche through a linear model of environmental covariates, and a multivariate residual process modeled through latent factors. 
Modifications and extensions to this general structure include, for example, models where species are clustered together into archetypes for which the environmental responses are shared \citep{Dunstan2013,Hui2013,Johnson+Sinclar:2017,Sollmann+etal:2021}, models that cluster species' according to similar dependence pattern in their latent factors \citep{Taylor-Rodriguezetal:2017,Shirota2019}, and models where the covariate effects are described by non-parametric models \citep{Vanhatalo2020}. 
Even though this general approach allows for flexible modeling of species dependencies through interspecific correlations in both covariate effects and random factors, interpreting their results is challenging. 
The environmental covariate effects and latent factors are commonly claimed to separate the environmental filtering from biotic interactions.
However, as demonstrated by \citet{Poggiato+etal:2021}, latent factor part of an JSDM is confounded with the environmental covariate effects so that current JSDMs cannot really achieve this.
Similarly, another common claim that interspecific correlations in the latent factor part of the model can give hints on species interactions \citep[see, e.g.,][]{Pollock+etal:2014,Ovaskainen+etal:2016,Wilkinson+etal:2019} has been heavily criticized \citep{Clark+etal:2014,Blanchet+etal:2020}.
For example, positive interspecific correlations can arise from both competition and mutualism which arise from very different biological processes \citep{Poggiato+etal:2021}. 
 
A specific example of species-to-species interaction that cannot be accounted for by current JSDMs, and which we tackle in this work, is exclusive competition for space. 
This is a process that commonly arises especially among plant species who compete for growing space and light.
As a motivating example for our work, we consider inference and prediction for vegetation data that are measured through percentage cover. 
By definition, percentage cover is the proportion of area occupied by a species. 
Since a given space can be occupied only by one specimen at a time, increasing percentage cover of one species necessarily decreases the space available for other species that grow on the same vegetation layer and, hence, compete for the same common space. 
Percentage cover data, thus, reveal inherent negative dependence among species occupying a common vegetation layer whereas similar, direct, negative effects are not anticipated between species occupying different layers. 

We build a hierarchical JSDM with an explicit description for interspecific exclusion.
We use latent Gaussian variable models and multivariate Gaussian processes \citep{Gelfand+Schmidth+Banerjee+Sirmans:2004,Vanhatalo2020} to describe species specific niche preference and relative competitive performance over the study area. 
However, instead of modeling each species as conditionally independent given the Gaussian latent variable, as done in contemporary JSDMs \citep[see, e.g.,][]{Warton:2015,Ovaskainen+Abrego:2020}, we model the observed percentage covers of mutually exclusive species with Dirichlet-Multinomial model. 
Dirichlet distributions model interspecific competition for space -- the exclusion effect arising from the fact that only one specimen can be in one location at a time. 
Multinomial distributions model the uncertainty in the observation process related to measuring the percentage covers over inventory plots.
Dirichlet distributions have been used in the context of JSDMs earlier, e.g., by \citet{Taylor-Rodriguezetal:2017} and \citet{Shirota2019} to cluster species according to their responses to latent factors and by \citet{Johnson+Sinclar:2017} and \citet{Sollmann+etal:2021} to cluster species according to their responses to environmental covariates. 
Our model differs from these earlier works since we do not use Dirichlet process to cluster species but we use Dirichlet distributions to describe a conditional exclusion process given the latent factors and environmental effects. 
A Dirichlet-Multinomial model similar to our model has earlier been used to model vegetation cover data in ordination settings by \citet{Damgaard+Hansen+Hui:2020}. 
However, our model is the first one combining it with formal joint species distribution modeling framework.

We test and demonstrate the properties of our model with a simulation study after which we apply them to a case study on plant community data. In the case study, we make predictive percentage cover maps and estimate the total vegetation cover of moss and vascular plant species over a study site around an eddy covariance tower in a boreal peatland in southern Finland (Figure~\ref{fig:grid_design}). 
Eddy covariance tower is a measurement device to monitor vertical fluxes between biosphere and atmosphere \citep{Korrensalo2019}. 
They collect data on, e.g., GHG exchange rates between the biosphere and atmosphere over natural and artificial ecosystems.
Our estimates on plant percentage cover will subsequently be used to calibrate the carbon gas flux estimates made by the eddy covariance tower. 
Out of the modeled plant species, mosses grow in one layer and therefore compete for space against others alike whereas vascular plants can grow in multiple layers and, thus, do not express exclusive interspecific competition for space. 

We compare our models to the contemporariry JSDMs, and traditional stacked SDMs. The latter model each species independently and then combine their predictions into a community prediction. 
Since we are interested in model performance in several different types of predictive tasks, we propose to use rigorous decision theoretic approach to formulate model comparison and validation method for each of these tasks. 
We use the log score to measure model's predictive performance and uniform Q-Q plots of the probability integral transform (PIT) to assess their probabilistic calibration. 
We use then structured cross-validation (CV) to estimate models expected predictive performance in a given prediction scenario. That is, we divide the data into training and test data-blocks such that they structurally resemble the properties of the training data and predicted outcomes in the true predictive task.

The rest of the paper is organized as follows. In Section~\ref{sec:site}, we introduce the motivating case study and, in Section~\ref{sec:models_and_inference}, we introduce the models used in this work. We then propose the model comparison and validation methods in Section~\ref{sec:validation} and the simulation experiments and the case study analyses in Section~\ref{sec:experiments}. We present the results in Section~\ref{sec:results} and end by discussion in Section~\ref{sec:discussion}.

\section{Motivating case study: plant community modeling}\label{sec:site}

\begin{figure}
\centering
\includegraphics[]{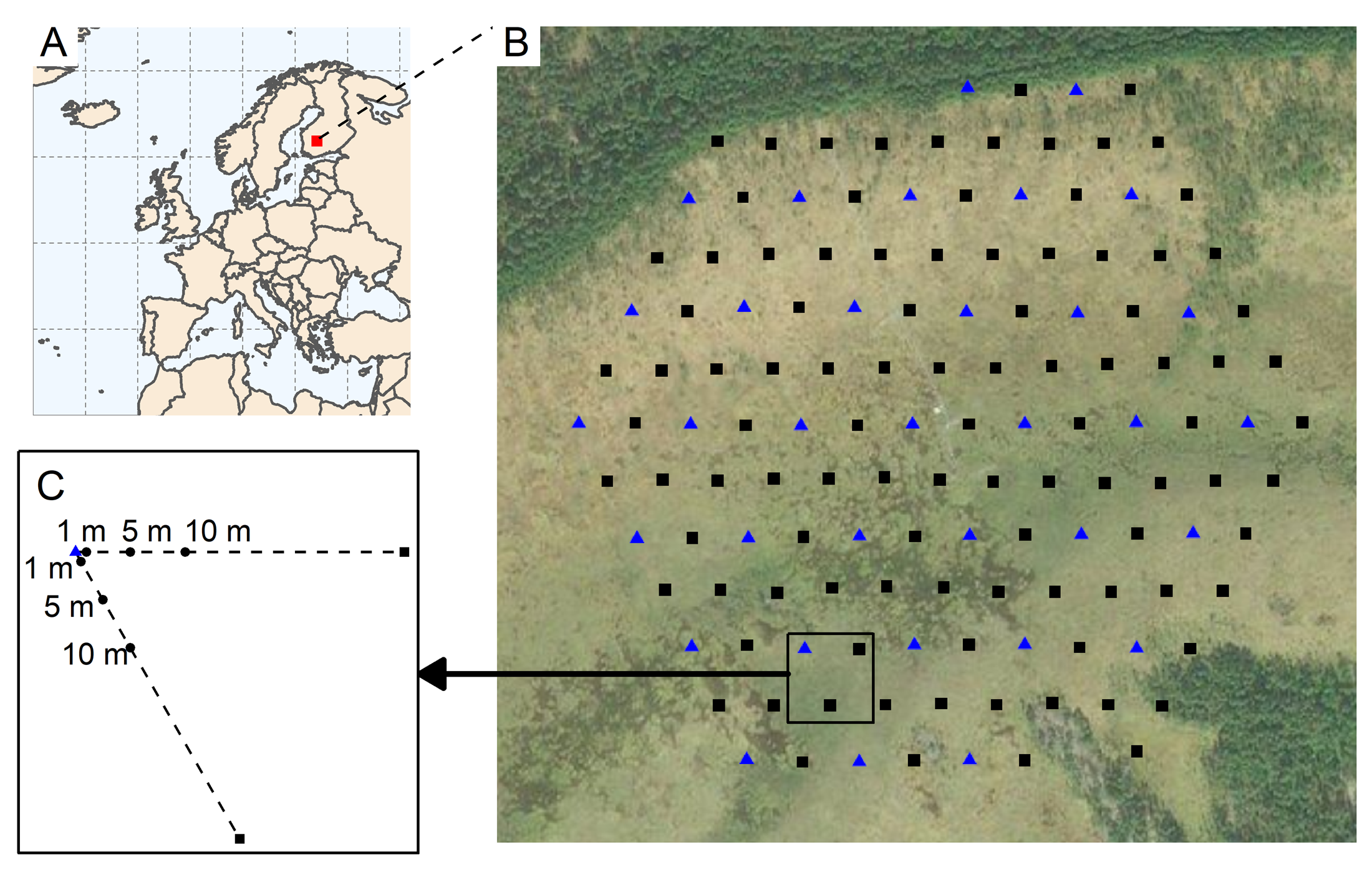}
\caption{The location of the Siikaneva peatland (A), locations of the main inventory plots with aerial photo of the peatland (B) and the design for the additional inventory plots (C). Triangles show the locations of the main plots having additional plots and squares present the locations of the main plots without additional plots.}
\label{fig:grid_design}
\end{figure}

Our case study site is located at a boreal poor fen, which is part of Siikaneva peatland complex in Southern Finland (Figure \ref{fig:grid_design}).
 The study site has an eddy covariance tower, which monitors the $\text{CO}_{2}$ and $\text{CH}_{4}$ fluxes within a 200-meter radius footprint around the tower that extends from the margin of the peatland towards the center. 
The carbon gas exchange between an ecosystem and atmosphere is primarily determined by the amount of photosynthetically active biomass and the area of green leaves – even though abiotic factors, such as the light and water availability and temperature, play a role as well \citep[e.g.][]{Peichl2018}.
Therefore, knowledge about the abundance and structure of the vegetation is an essential input for modeling the ecosystem climatic impact \citep[e.g.][]{Korrensalo2019} and for ecosystem models in general.
 In this context, peatland ecosystems are of special interest because they store approximately one third of global terrestrial carbon in their peat layer \citep{Gorham1991} and are the largest natural source of atmospheric methane \citep[e.g.][]{Heilig1994}. 
Different peatland species have distinct photosynthetic capacities and differ by the substrate quality they provide for decomposition processes. Further, the abundance of different vascular plant species is an important control for ecosystem-scale methane efflux, as certain species act as conduits of methane from the peat to the atmosphere \citep{Bhullar2013}, bypassing the microbial methane oxidation in the oxidation peat layers \citep{Larmola2010}. Plant species groups also differ by their emissions of biogenic organic compounds, that have a net cooling effect on climate \citep{Tiiva2009, Faubert2011}. 

Typical to aapa mires, the margin of the study site is poorer in nutrients and slightly drier than the wetter center. 
The bottom layer of the site is formed by mosses (mainly Sphagnum L. species) and the field layer consists of vascular plants adapted to the conditions where water table prevails close to the surface. Both of these vegetation layers have a significant role in the greenhouse gas cycling of the site. For this work, we selected the six most common Sphagnum (\textit{S.}) mosses: \textit{Sphagnum papillosum}, \textit{S. balticum}, \textit{S. fallax}, \textit{S. magellanicum}, \textit{S. majus} and \textit{S. angustifolium} and the following eight ecologically interesting vascular plants: \textit{Carex lasiocarpa}, \textit{Carex limosa}, \textit{Carex rostrata}, \textit{Empetrum nigrum}, \textit{Eriophorum vaginatum}, \textit{Pinus sylvestris}, \textit{Rubus chamaemorus} and \textit{Scheuchzeria palustris}. The Sphagnum species grow in the same vegetation layer and are mutually exclusive so that only one species can occupy a spatial location at any one time \citep{Gong2020}. The vascular plants, however, can grow in layers that overlap with each other and the Sphagnum layer so that they are not mutually exclusive to others. For species names we followed \citet{Plantlist2013}.

To quantify the spatial variation of the vegetation within the eddy covariance tower footprint, a vegetation inventory was done in the summer of 2017. The grid sampling design with 328 inventory plots was applied within 200 m distance from the eddy covariance tower (Figure \ref{fig:grid_design}). 
Locations that were in the drainage ditch and mineral soil in the border of the region were excluded from the grid and thus the formed grid is not symmetric. 

For every inventory plot, the percentage cover of each species was estimated within a circular frame of 0.071 m$^{2}$ (radius 15 cm). The percentage cover was estimated as a total horizontal projection on the ground and reported in 0.25 percentage unit accuracy at cover below 1\% and in 1 percentage unit accuracy at cover above it. 
As is typical, the percentage cover observations are expert assessments based on visual inspection of the inventory plots. To aid the visual assessment researchers divide the inventory plot (either mentally or using a mesh) into a homogeneous lattice. A reported percentage cover then corresponds to the proportion of lattice nodes occupied by a species. 
Hence, we denote by $\mathbf{y}_i=(y_{i1},\dots,y_{iJ})^{\top}$
the number of these lattice nodes occupied by all species at the $i$th inventory plot so that the estimated percentage covers are given as

$$\text{[percentage cover of species $j$]} = y_{ij} / N_{i,j},$$ 

where $N_{i,j}$ corresponds to the number of lattice nodes used to estimate that species. 
The lattice size is, thus, related to the accuracy of the estimation. 
For example, a one percentage unit accuracy corresponds to $N_{i,j}=100$ whereas 0.25 percentage unit accuracy corresponds to $N_{i,j}=400$.
This leads naturally to Binomial marginal distribution for $y_{ij}$ conditionally to the true percentage cover in an inventory plot. 
However, the estimation for percentage covers of sphagnum species were synchronized so that their total did not exceed 100\% leading to Multinomial joint distribution for the numbers of occupied lattice nodes by all sphagnum species conditionally to their true percentage covers (see Section~\ref{sec:model}).
To minimize human-related errors, such as misclassified or omitted species \citep{Kennedy1987}, each plot was inventoried by two persons.

\section{Statistical modeling and inference}\label{sec:models_and_inference}

\subsection{Observation model}
\label{sec:model}

\begin{figure} 
\begin{center}
\includegraphics[]{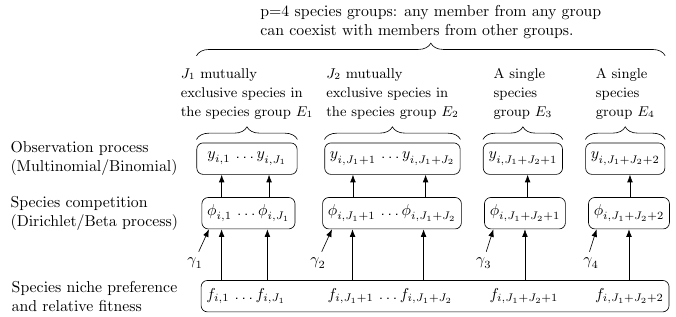}
\end{center}
\caption{The directed acyclic graph (DAG) representation of the joint species distribution model. The columns represent the species groups and rows the model layers. } \label{fig:DAG_LMC+DM}
\end{figure}

We model the species specific cover jointly using hierarchical Bayesian approach where each hierarchical layer represents a part of the modeled process (Figure~\ref{fig:DAG_LMC+DM}). 
We denote by $\mathcal{D}\subset \mathbb{R}^{2}$ the bounded study region of interest and by a set $\mathbf{S}=\{\mathbf{s}_{1},\mathbf{s}_{2},...,\mathbf{s}_{n}\}$ the locations (centroids) of the inventory plots such that $\mathbf{s}_{i}\in \mathcal{D} \ \forall i\in\{1,...,n\}$, where $n$ is the total number of plots (see Figure~\ref{fig:grid_design}).
The set of modeled species is denoted by $E=\{e_{1},...,e_{J}\}$, where $e_{j}$ is the identifier of the $j$'th species and $J$ is the total number of species. 
We then allocate species into species groups such that each species group is formed by mutually exclusive species; that is, by species that cannot grow over each other. These species groups form nonempty disjoint sets $E_{g}$ which are defined such that $E=\mathop{\cup}_{g=1}^{p} E_{g}$ where $p$ is the total number of species groups (Figure \ref{fig:DAG_LMC+DM}).

Motivated by the data collection process (Section~\ref{sec:site}), we assume that, at each inventory plot, a researcher has counted the number of mesh nodes occupied by species and collected them into vector $\mathbf{y}_{i}$. 
We denote by a vector $\bm{\phi}_{i}=(\phi_{i1},\phi_{i2},...,\phi_{iJ})^{\top}$ the true species specific percentage covers in the inventory plot; that is, $\phi_{ij}$ is the percentage cover of species $j$ at plot $i$. 
We further group the true percentage covers and node counts into vectors formed by mutually exclusive species groups: $\bm{\phi}_{i,g} = (\phi_{ij})_{j\in E_g}$ and $\mathbf{y}_{i,g}=(\mathbf{y}_{ij})_{j\in E_g}$, $g=1,\dots,p$.
Now, conditional on $\bm{\phi}_{i}$ the species specific node counts at location $\mathbf{s}_{i}$ can be modeled as 
\begin{equation}\label{eq:observation_model}
\pi\left(\mathbf{y}_{i}|\mathbf{N}_{i},\bm{\phi}_i \right)= \prod_{g=1}^p \text{Multinomial}([\mathbf{y}_{i,g},y_{i,g_0}]\vert [\bm{\phi}_{i,g}, \phi_{i,g_0}], N_{ig})
\end{equation}
where $y_{i,g_0}=N_{i,g}-\sum_{j\in E_g}y_{ij}$ and $\phi_{i,g_0}=1-\sum_{j\in E_g}\bm{\phi}_{ij}$ correspond to the number of mesh nodes and the proportion of the inventory plot where none of the species from species group $E_{g}$ is present, and $\mathbf{N}_{i}=[N_{i1},\dots,N_{ip}]$. 
Hence, our observation model differs importantly from the contemporary (J)SDMs where the observations $\bm{y}_i$ are conditionally independent given the underlying probability of presence parameter \citep[see e.g.,][]{Gelfand2006,Taylor-Rodriguezetal:2017,Ovaskainen+Abrego:2020}. 
For our data, this contemporary approach would mean that each species forms its own group with Binomial observation model so that $\pi\left(\mathbf{y}_{i}|\mathbf{N}_{i},\bm{\phi}_i \right)= \prod_{j=1}^J \text{Bin}(y_{i,j}|\phi_{ij},N_{i,j})$.

Our observation model \eqref{eq:observation_model} is similar to the observation model for pin-point vegetation cover data used by \citet{Damgaard+Hansen+Hui:2020} in their model-based ordination study. 
However, they modeled all species counts with single Multinomial model whereas, by species grouping, we explicitly account for the fact that some species can grow on top of others, which leads to conditional independence in species counts between different species groups given $\bm{\phi}_{i}$.
Parameters $N_{ig}, g=1,\dots,p$ are formally sample sizes of a Multinomial distribution but they also corresponds to the percentage cover measurement accuracy. 
The smaller $N_{ig}$ is, the more uncertainty is assumed in the expert assessments. 
In the limit as $N_{ig}=1$, the resulting observation model is a categorical distribution. In the other limit, as $N_{ig}\rightarrow \infty$ we would have $\text{E}[y_{ij}/N_{ig}]\rightarrow\phi_{ij}$ and $\text{Var}[y_{ij}/N_{ig}]\rightarrow0$ corresponding to exact percentage cover observations.

\subsection{Model for percentage covers}
The assumption that species compete for space only among species in the same species group leads to (conditional) independence between percentage covers in different groups, so that we can decompose the percentage cover model as
\begin{equation}
\pi(\bm{\phi}_{i} | \cdot) =
\prod_{g=1}^p \pi(\bm{\phi}_{i,g}|\cdot ).
\end{equation}
We then model the group-wise percentage covers with a Dirichlet distribution 
\begin{equation}
\pi([\bm{\phi}_{i,g}, \phi_{i,g_0}]|\mathbf{f}_{i,g}, \gamma_{g} ) = \text{Dir}([\bm{\phi}_{i,g},\phi_{i,g_0}]|\bm{\alpha}(\mathbf{f}_{i,g}) \times \gamma_{g} )
\end{equation}
where $\gamma_g$ is a scale parameter and $\bm{\alpha}(\mathbf{f}_{i,g})=\bm{\alpha}_{i,g} = \left[(\alpha_{ij})_{j\in E_g}, 1-\sum_{j\in E_g}\alpha_{ij}\right]$ is a vector of expected percentage covers defined through the soft-max function
\begin{equation}
\alpha_{ij} = \frac{\exp(f_{ij}) }{1+\sum_{j\in E_g} \exp(f_{ij})},
\end{equation}
where $\mathbf{f}_{i,g}=(f_{ij})_{j\in E_g}$ is a vector of species specific latent variables.
In single species groups, the Dirichlet distribution reduces to Beta distribution and the softmax function reduces to the inverse logit-link function.

The choice of Dirichlet distribution is justified since it is a natural prior for proportions. 
The stick breaking generative model of Dirichlet distribution has also an intuitive ecological interpretation as exclusive competition for space. 
The expected percentage cover, $\alpha_{ij}$, is a summary of a species' relative strength, among others in its group, to occupy a location.
The bigger $\alpha_{ij}$, the more likely it is that a species "breaks" a large proportion of an inventory plot for itself. 
The scale parameter $\gamma_g$ governs then the level of randomness in this breaking process so that the randomness decreases with increasing $\gamma_g$. 
Moreover, since the expected percentage cover depends on latent variables $\mathbf{f}_{i,g}$, we can use covariates to explain species relative competition strengths (see Section~\ref{sec:spatlatpros}).
Hence, a natural measure for exclusive competition for space between two species is the correlation between their percentage covers, which conditionally on latent factors is:
\begin{equation}\label{eq:competition_measure}
\mathrm{Corr}(\phi_{ij},\phi_{ij'}|\mathbf{f}_i) = -\sqrt{ \frac{\alpha_{ij}\alpha_{ij'}}{(1- \alpha_{ij})(1- \alpha_{ij'})} }.
\end{equation}
This correlation is the strongest when $\alpha_{ij}=\alpha_{ij'}=1/2$ and decreases towards zero when either or both of the $\alpha$ terms decrease to zero. Moreover, as $\alpha_{ij}$ varies in space the correlation varies in space as well and, hence, the model can capture spatial variation in the strength of space competition.
Note though, that we use the term competition for space in a rather abstract manner since it does not contain explicit mechanistic description for possible biological processes underlying it.

\subsection{Gaussian latent variable models} \label{sec:spatlatpros}

We model the species specific latent processes with latent Gaussian variable models 
\begin{equation}\label{eq:LGVM}
f_{j}(\bm{s}_i)=\bm{\beta}_{j}^{\top}\bm{x}_i+\epsilon_{ij},
\end{equation}
where $\bm{\beta}_{j}$ is a vector of covariate effects (including an intercept) for species $j$, $\bm{x}_i$ is a vector of (environmental) covariates and $\epsilon_{ij}$ is a Gaussian random effect for species $j$ at a location $\mathbf{s}_{i}$.
This is a typical Gaussian latent variable formulation for JSDMs where environmental covariates are used to explain the environmental niche of a species and random effects are used to describe the residual variation not explainable by the environmental covariates \citep{Warton:2015,Ovaskainen+Abrego:2020,Vanhatalo2020}. In single species SDMs, both the covariate effects, $\bm{\beta}_j$, and random effects, $\epsilon_{ij}$, are given mutually independent (Gaussian) priors across species reflecting an implicit assumption that the processes behind species' niches are mutually independent.
However, this is an unrealistic assumption in practice since species typically show positive and negative associations in their niche preferences \citep{Ovaskainen+Abrego:2020}.
For this reason, JSDMs model these associations with hierarchical priors that contain interspecific correlations between the covariate effects and random effects, which can significantly improve the predictive accuracy of species distribution models \citep{Nordberg2019}. 
In this work, we follow \citet{Vanhatalo2020} in building JSDMs by giving a joint multivariate Gaussian priors for covariate effects so that the prior for species specific effects of $d$'th covariate is $[\beta_{d,1},\dots,\beta_{d,J}]^{\top}\sim N(0,\Sigma_d)$ where $\Sigma_d$ is a dense covariance matrix. The priors for the random effects are introduced next.

\subsubsection{Stationary spatial Gaussian processes}

We model the spatial latent processes $\epsilon_{ij}$, either as mutually independent or jointly dependent among species. In the former approach, a species specific latent function is given a univariate Gaussian process prior
\begin{equation}
\epsilon_{ij}\sim GP(0,k^{(\text{e})}_{j}(\mathbf{s}_{i},\mathbf{s}_{i'})),
\label{eq:independentGP}
\end{equation}
where $k^{(\text{e})}_{j}(\mathbf{s}_{i},\mathbf{s}_{i'})$ is the species specific spatial covariance function. The superscript $(\text{e})$ stands for stationary exponential covariance function
\begin{equation}
k^{(\text{e})}_{j}(\mathbf{s}_{i},\mathbf{s}_{i'})=\sigma_{j}^{2}\exp\left(-\frac{||\mathbf{s}_{i}-\mathbf{s}_{i'}||}{l_{j}} \right)
\label{eq:exp-cov}
\end{equation}
with a species specific length scale parameter, $l_{j}$, governing how fast the spatial correlation decays, and a variance parameter $\sigma_{j}^{2}$ governing the magnitude of the variation. 

For JSDMs, we extend the baseline model by allowing for interspecific correlations between the spatial latent processes. 
We model dependencies through linear model of coregionalization \citep[LMC;][]{Gelfand+Schmidth+Banerjee+Sirmans:2004,Banerjee2015} such that spatial latent processes are expressed as a 
linear combination of $J$ zero mean univariate Gaussian processes having the
covariance functions
$k^{(\text{e})}_{j}(\mathbf{s}_{i},\mathbf{s}_{i'})$.
We collect all spatial random effects into vector
$\boldsymbol{\epsilon}(\mathbf{S})=(\boldsymbol{\epsilon}_{1}(\mathbf{S})^{\top},\dots,\boldsymbol{\epsilon}_{J}(\mathbf{S})^{\top})^{\top}$,
where $\boldsymbol{\epsilon}_{j}(\mathbf{S})$ denotes the vector having latent variables for species $j$ at spatial locations $\mathbf{S}$. The LMC model induces a multivariate Gaussian prior \citep{Gelfand+Schmidth+Banerjee+Sirmans:2004,Vanhatalo2020}
\begin{equation}
\boldsymbol{\epsilon}(\mathbf{S})\sim N\left(\mathbf{0},\sum_{j=1}^{J}
\mathbf{B}_{j}\otimes \bm{K}^{(\text{e})}_{j}\right)
\label{eq:multivariateGP}
\end{equation}
where $\otimes$ denotes Kronecker product of the covariance matrices that are constructed with $[\bm{K}^{(\text{e})}_{j}]_{i,i'} = k^{(\text{e})}_{j}(\mathbf{s}_{i},\mathbf{s}_{i'}; l_{j},\sigma_{j}^{2}=1)$ and the matrix $\mathbf{B}_{j}=\mathbf{L}_j\mathbf{L}_j^{\top}$. Matrix $\mathbf{L}_j$ is the $j^{\text{th}}$ column of the Cholesky decomposition of the coregionalization covariance matrix $\boldsymbol{\Sigma}_{\boldsymbol{\epsilon}}=\sum_{j=1}^{J}\mathbf{L}_j\mathbf{L}_j^{\top}$, which models the interspecific dependencies between species niche preferences. 
The variance parameter $\sigma_{j}^{2}$ of $k^{(\text{e})}_{j}(\mathbf{s}_{i},\mathbf{s}_{i'})$ is set to 1 to ensure identifiability.

The model \eqref{eq:multivariateGP} is the most flexible version of the LMC models where each species has its own process characteristics encoded by $k_j$. 
However, we can reduce the flexibility by reducing the number of unique covariance functions in the model. 
For example, if all species share the same covariance function $k^{(\text{e})}_{j}(\mathbf{s}_{i},\mathbf{s}_{i'})=k^{(\text{e})}(\mathbf{s}_{i},\mathbf{s}_{i'}) , \forall j=1,\dots J$, we obtain LMC(1) model, which is also called \textit{intrinsic model for coregionalization} \citep{Gelfand+Schmidth+Banerjee+Sirmans:2004}.
In this model the species specific latent processes are (correlated) random draws from the same underlying Gaussian process whereas in \eqref{eq:multivariateGP} the latent processes are (correlated) random draws from $J$ different Gaussian processes. 
We tested also models in between these two extremes and denote by LMC($k$) a model where we have $k$ distinct covariance functions so that $k^{(\text{e})}_{j}(\mathbf{s}_{i},\mathbf{s}_{i'})=k^{(\text{e})}_{k}(\mathbf{s}_{i},\mathbf{s}_{i'}), \forall j\geq k$ when $k$ is less than $J$.
Note that the LMC($k$) models always have $J$ unique spatial latent processes and the coregionalization matrix $\boldsymbol{\Sigma}_{\boldsymbol{\epsilon}}$ is positive definite in all these models. 

Our approach for modeling the multivariate spatial random effect is reasonable with our current application that has only moderate number of species. 
If extended for larger number of species, it would, however, run into trouble through rapid increase of the coregionalization covariance matrix $\mathbf{\Sigma}_{\boldsymbol{\epsilon}}$. Hence, with more species it would be reasonable to replace the LMC model with a latent factor model where we would have only $p<J$ spatial Gaussian processes so that the interspecific covariance matrix would be of low rank \citep[see, e.g.,][]{Taylor-Rodriguezetal:2017,Ovaskainen+Abrego:2020}.

\subsubsection{Non-stationary spatial Gaussian processes}
\label{sec:non-stationary}

Stationary Gaussian processes work typically well if the region of interest is sampled relatively sparsely, so that data does not allow inference for non-stationarity in spatial correlation, or if environmental covariates are accurate enough to describe potential non-stationarities in the latent process $f(s_i)$ \citep{Schmidt+Rodriguez:2011}. 
However, we do not have environmental covariates for the case study area, the case study data is sampled with high resolution (see Section~\ref{sec:case-study}), and the properties affecting peatland vegetation can vary considerably within the study region. Hence, we consider also non-stationary spatial random effects. 
We extended the stationary spatial random effect model by setting non-stationary covariance function for Gaussian process. We include non-stationarity into the model by changing stationary covariance function in equation \eqref{eq:exp-cov} to a non-stationary Mat{\'e}rn $(\nu=3/2)$ covariance function with spatially varying length scale parameter \citep[e.g.][]{Paciorek2006} 
\begin{equation}
k^{(\text{m})}_{j}(\mathbf{s}_{i},\mathbf{s}_{i'})=|\boldsymbol{\Sigma}_{ij}|^{1/4}|\boldsymbol{\Sigma}_{i',j}|^{1/4}\bigg\rvert \frac{\boldsymbol{\Sigma}_{ij}+\boldsymbol{\Sigma}_{i',j}}{2}\bigg\rvert^{-1/2}(1+\sqrt{3}\sqrt{Q_{i,i',j}})\exp(-\sqrt{3}\sqrt{Q_{i,i',j}})
\label{eq:nonstatmatern}
\end{equation} 
where $Q_{i,i',j}=(\mathbf{s}_{i}-\mathbf{s}_{i'})^{\top}\left(\frac{\boldsymbol{\Sigma}_{ij}+\boldsymbol{\Sigma}_{i',j}}{2}\right)^{-1}(\mathbf{s}_{i}-\mathbf{s}_{i'})$ is the Mahalanobis spatial distance between locations $\mathbf{s}_{i}$ and $\mathbf{s}_{i'}$ for species $j$ and $\boldsymbol{\Sigma}_{ij}=l_{ij}^{2}\begin{bmatrix}
1 & 0 \\
0 & 1
\end{bmatrix}$ so that $l_{ij}$ varies spatially. The superscript m in the above equation stands for Matern covariance. We model the spatially varying length scale parameter by giving a Gaussian process prior for its logarithm
\begin{equation}
\log(l_{ij}) \sim GP(\mu_{l_{j}},k^{(\text{e})}_{{j}}(\mathbf{s}_{i},\mathbf{s}_{i'})),
\label{eq:logGP}
\end{equation} 
where the mean function $\mu_{l_{j}}$ specifies the expected value of $\log(l_{ij})$ and $k^{(\text{e})}_{{j}}(\mathbf{s}_{i},\mathbf{s}_{i'})$ is the stationary exponential covariance function \eqref{eq:exp-cov}. Gaussian process prior gives smoothly varying length scale and modeling logarithm of the length scale $l_{ij}$ ensures positivity. 
Since the matrix $\boldsymbol{\Sigma}_{ij}$ is diagonal and all its diagonal elements are the same, the spatial correlation is isotropic at each location but the strength of the correlation decay varies within the area. 
The Non-stationary covariance function in equation \eqref{eq:nonstatmatern} reduces to stationary Mat{\'e}rn $(\nu=3/2)$ covariance function if the length scale parameter, $l_{ij}$, is set to be the same at each location.
Non-stationary multivariate spatial random effects have earlier been used in the context of species distribution modeling by \citet{Schmidt+Rodriguez:2011}. 
They applied also LMC framework but their non-stationary covariance kernels were constructed differently. 
We applied the non-stationary GP to models with and without interspecific dependence between the spatial latent processes. In the former case, each latent process was given an independent non-stationary GP prior. In the latter case, the latent processes were modeled jointly with LMC (equation \eqref{eq:multivariateGP}) where $\bm{K}^{(\text{e})}_{j}$ is replaced by $[\bm{K}^{(\text{m})}_{j}]_{i,i'} = k^{(\text{m})}_{j}(\mathbf{s}_{i},\mathbf{s}_{i'})$.

\subsection{Hyperpriors}\label{sec:Priors}

The model specification is completed by assigning prior distributions to the model hyperparameters. In the stationary GP models, we gave weakly informative half-inverse-Student-$t$ prior for the length scale parameters, $1/l_{j} \sim \text{Student}-t_{+}(\mu,s^2,\nu)$, which gives \emph{a priori} more weight for the larger length scales. The parameter values of the prior were chosen according to the size of the study area (see Section \ref{sec:case-study}). The location and scale parameters of the half-inverse-Student-$t$ prior for the length scale parameters of the stationary GP models were selected such that the length scale is less than 400 meters with probability 0.99. This gives $1/l_{j} \sim \text{Student}-t_{+}(0,0.19^2,5)$ which corresponds to preferring smooth over small scale variability in vegetation composition. 
In the non-stationary GP models, the prior for the mean of the log length-scale was $\mu_{l_{j}}\sim N(4.5,\sqrt{2}^{2})$, which favors relatively large $l_{ij}$ values ($>$ 100 m). 
For the length scale and variance parameters of the GP prior for $\log l_{ij}$ \eqref{eq:logGP} we gave weakly informative priors, $1/l_{j}\sim Student-t_{+}(0,2^2,4)$ and $\sigma_{j}^{2}\sim Student-t_{+}(0,1,4)$. 
The location and scale parameters in these priors were selected such that $l_{j}$ is less than 38 meters with probability 0.99. 

We gave a Gamma prior, $\gamma_{g} \sim \text{Gamma}(3/2,2/3)$, for the process parameters governing the level of randomness. 
It is \emph{a priori} likely that randomness in the vegetation composition is high for which reason we assigned scale and rate parameters of the Gamma distributions so that they give more weight for the smaller values of the process parameters. In order to model coregionalization matrix $\boldsymbol{\Sigma}_{\boldsymbol{\epsilon}}$ efficiently we use a separation strategy \citep[e.g.][]{Barnard2000} where coregionalization matrix is decomposed into correlation matrix $\boldsymbol{\Omega}$ and vector of standard deviations $\boldsymbol{\omega}$ such that $\boldsymbol{\Sigma}_{\boldsymbol{\epsilon}}=\text{diag}(\boldsymbol{\omega})\boldsymbol{\Omega}\text{diag}(\boldsymbol{\omega})$. 
The correlation matrix $\boldsymbol{\Omega}$ was given LKJ-prior with unit shape \citep{Lewandowski2009} that defines a prior distribution which is marginally uniform over all correlation parameters \citep{Vanhatalo2020}. 
We gave half-Student-$t$ prior for the standard deviations, $\boldsymbol{\omega} \sim \text{Student}-t_{+}(0,4^2,4)$. 
In the simulation study, we assigned environmental covariate effects $\bm{\beta}$ a multivariate normal prior $\bm{\beta}\sim N(\bm{0},\boldsymbol{\Sigma_\beta})$ where the covariance matrix was decomposed into correlation matrix $\boldsymbol{\Omega_\beta}$ and vector of standard deviations $\boldsymbol{\omega_\beta}$ such that $\boldsymbol{\Sigma_\beta}=\text{diag}(\boldsymbol{\omega}_\beta)\boldsymbol{\Omega_\beta}\text{diag}(\boldsymbol{\omega_\beta})$. The correlation matrix $\boldsymbol{\Omega_\beta}$ was given LKJ-prior with unit shape and for the standard deviations we gave half-Student-$t$ prior, $\boldsymbol{\omega_\beta} \sim \text{Student}-t_{+}(0,4^2,4)$. For each intercept term $\beta_{0,j}$ we assigned weakly informative Student-t priors $\beta_{0,j}\sim \text{Student}-t(0,2.5^2,4)$ in the simulation and case studies.
In principle, we could also give an informative prior for $N_{ig}$ to account for the fact that the estimate for experts' accuracy is not exact. 
However, we do not consider this option here.

\subsection{Posterior inference}\label{sec:inference}

All the alternative species distribution models were implemented using Stan via Rstan \citep{STAN} with which we conducted posterior sampling for all the model parameters and latent variables. 
We ran four parallel Markov chains of 2000 iterations such that first 1000 iterations of each chain were discarded as warmup. Posterior sampling was done using dynamic Hamiltonian Monte Carlo Sampler as coded in Stan version 2.18.1. The convergence and effective sample sizes were checked using trace- and autocorrelation plots and through potential scale reduction factor and Geyer's initial monotone sequence criterion.
After conducting the posterior sampling, we drew posterior predictive samples of the species specific percentage covers at prediction locations covering the study area \citep[see, e.g.,][]{Vanhatalo2020}.

\section{Model comparison and validation}
\label{sec:validation}

The central aims of our study are to provide posterior predictive maps for the vegetation cover within the study area, and to provide posterior predictive distributions for the total vegetation cover over the study area. 
Hence, we compare alternative models with the goodness of their out-of-sample posterior predictive distributions using cross-validation (CV) log posterior predictive density diagnostics \citep{Vehtari+Ojanen:2012}.
Model comparison indicates the best model among the alternatives but does not tell whether any of the models has actually good fit for the purpose. Hence, after choosing the best model we assess the goodness of its predictive distributions using probability integral transform (PIT) statistics \citep{Gneiting2007}. 
All the model comparison and model validation methods are summarized in Table~\ref{table:model_comparison_and_validation} and we explain them in detail next.

\begin{table}\caption{Summary of the model comparison and validation methods that are described in detail in Section~\ref{sec:validation}. \\ }\label{table:model_comparison_and_validation}

\small
\begin{tabular}{p{2.4cm}|p{4.6cm}|p{5.0cm}|}
Predictive task & Model comparison & Model validation \\
\hline
Species specific percentage cover maps & $CV_1$: An average of species- and location-wise log predictive densities. Equation \eqref{eq:CV1}. & $PIT_1$: Species specific PIT histograms of location-wise predictive distributions. Equations \eqref{eq:PIT} and \eqref{eq:PIT1}. \\
Total percentage cover maps & $CV_2$: An average of location-wise log joint over species predictive densities. Equation \eqref{eq:CV2}. & $PIT_2$: PIT histogram of location-wise total over species predictive distributions. Equations \eqref{eq:PIT} and \eqref{eq:PIT2}. \\
Species specific total vegetation cover over the study area & $CV_3$: An average of species- and CV-fold-wise log joint over locations (within a CV-fold) predictive densities. Equation \eqref{eq:CV3}. & $PIT_3$: Species specific PIT histograms of total over locations (within a CV-fold) predictive distributions. Equation \eqref{eq:PIT} and CDF is estimated analogously to \eqref{eq:PIT2}. \\
Total vegetation cover over the study area & $CV_4$: An average of CV-fold-wise log joint over species and locations (within a CV-fold) predictive densities. Equation \eqref{eq:CV4}. & $PIT_4$: PIT histogram of CV-fold-wise total over species and locations (within a CV-fold) predictive distributions. Equation \eqref{eq:PIT} and CDF is estimated analogously to \eqref{eq:PIT2}. \\
\end{tabular}
\end{table}

\subsection{Predictive model comparison with cross-validation}

Species specific percentage cover maps (so called, species distribution maps) are produced by forming a lattice mesh over the study area and calculating point-wise posterior predictive distributions for all the mesh cells. Summary statistics (mean, variance, quantiles, etc.) of these predictive distributions can then be drawn as maps. On the other hand, when predicting the total vegetation cover, we need to calculate the joint predictive distribution over all the mesh cells.
Moreover, we want to compare models' predictive performance both by species and for the total cover by all species.
Hence, we constructed own CV splitting strategy for each of these tasks.

In order to compare models in terms of producing species-specific percentage cover maps, we divided the observed dataset randomly into $K$ distinct subsets, indexed by sets of locations $\mathbf{S}_{1},...,\mathbf{S}_{K}$ such that the full data set is given by $\cup_{k=1}^{K} \mathbf{S}_{k}=\mathbf{S}$. The single species, point-wise, predictive performance was then
\begin{equation} \label{eq:CV1}
CV_{1}=\frac{1}{Jn} \sum_{i=1}^{n}\sum_{j=1}^{J}\text{log}\left( \pi(y_{ij}|\mathbf{Y}(\mathbf{S}_{\setminus k(i)}), \mathbf{S}_{\setminus k(i)}, \mathbf{s}_i)\right),
\end{equation}
where $\pi(y_{ij}|\mathbf{Y}(\mathbf{S}_{\setminus k(i)}), \mathbf{S}_{\setminus k(i)}, \mathbf{s}_i)$ is the CV posterior predictive density for $y_{ij}$. The set $k(i)$ is the CV set that contains location $\mathbf{s}_{i}$, and $\mathbf{Y}(\mathbf{S}_{\setminus k(i)})$ includes all other species observations except the observations in the CV set $k(i)$.
The comparison of models in the task of producing total vegetation cover maps was done analogously so that we calculated the average log (spatially) point-wise joint over species posterior predictive density. This leads to $K$-fold CV criterion 
\begin{equation} \label{eq:CV2}
CV_{2}=\frac{1}{n}\sum_{i=1}^{n}\text{log} \left(\pi(\mathbf{y}_{i}|\mathbf{Y}(\mathbf{S}_{\setminus k(i)}), \mathbf{S}_{\setminus k(i)}, \mathbf{s}_i)\right)
\end{equation}
where $\pi(\mathbf{y}_{i}|\mathbf{Y}(\mathbf{S}_{\setminus k(i)}), \mathbf{S}_{\setminus k(i)}, \mathbf{s}_i)$ is the joint posterior predictive probability mass function of all the species at location $i$.
We compared models in predicting per species total percentage cover over an area by first calculating for each CV-fold and species the log joint posterior predictive density of observations in that CV-fold and then taking the average over the CV-folds and species. That is, 
\begin{equation} \label{eq:CV3}
CV_{3}=\frac{1}{KJ}\sum_{k=1}^{K}\sum_{j=1}^{J}\text{log} \left(
\pi(\mathbf{Y}_j(\mathbf{S}_{k})|\mathbf{Y}(\mathbf{S}_{\setminus k}), \mathbf{S}_{\setminus k}, \mathbf{S}_k)\right),
\end{equation}
where $\mathbf{Y}_j(\mathbf{S}_{k})$ collects all the observations of the $j$th species at locations $\mathbf{S}_{k}$. Similarly, to evaluate models' performance in predicting the total percentage cover over all species and an area we calculated the average of CV-fold-wise log joint predictive densities with 
\begin{equation} \label{eq:CV4}
CV_{4}=\frac{1}{K}\sum_{k=1}^{K}\text{log} \left(
\pi(\mathbf{Y}(\mathbf{S}_{k})|\mathbf{Y}(\mathbf{S}_{\setminus k}), \mathbf{S}_{\setminus k}, \mathbf{S}_k)\right),
\end{equation}
where $\mathbf{Y}(\mathbf{S}_{k})$ collects observations of all species at locations $\mathbf{S}_{k}$. Note that in all above CV criterion we have calculated the average over individual log predictive densities. Hence, the statistics are in comparable scale.

We used $K=10$ in all CV metrics. Moreover, we estimated the posterior predictive densities using Monte Carlo over Markov chain samples from the posterior distribution. For example, in $CV_1$ the posterior predictive density is approximated with $\pi(y_{ij}|\mathbf{Y}(\mathbf{S}_{\setminus k(i)}), \mathbf{S}_{\setminus k(i)}, \mathbf{s}_i) \approx \frac{1}{M}\sum_{m=1}^{M}\pi(y_{ij}|f_{ij}^{(m)},\gamma^{(m)})$ where $f_{ij}^{(m)}$ and $\gamma^{(m)}$ denote the $m^{\text{th}}$ posterior sample from the joint posterior (predictive) distribution of the latent variable and Dirichlet model parameters respectively; that is $f_{ij}^{(m)},\gamma^{(m)}\sim p(f_{ij},\gamma|\mathbf{Y}(\mathbf{S}_{\setminus k(i)}), \mathbf{S}_{\setminus k(i)}, \mathbf{s}_i)$. 
In $CV_2$ the posterior predictive density is approximated with $\pi(\mathbf{y}_{i}|\mathbf{Y}(\mathbf{S}_{\setminus k(i)}), \mathbf{S}_{\setminus k(i)}, \mathbf{s}_i)\approx \frac{1}{M}\sum_{m=1}^{M}\pi(\mathbf{y}_{i}|\mathbf{f}(\mathbf{s}_{i})^{(m)},\gamma^{(m)})$ and analogously in $CV_3$ and $CV_4$.
After obtaining the posterior samples for the model parameters and latent variables at data locations, the posterior predictive samples can be constructed in Gibbs style using the full Gaussian conditional distribution of the latent variables \citep[see][Section~4.1]{Vanhatalo2020}.
If the samples of posterior predictive densities, such as $\pi(y_{ij}|f_{ij}^{(m)},\gamma^{(m)})$, have large variance, the Monte Carlo approximations for log predictive densities may become unreliable. 
This is likely, especially in case of $CV_2$ - $CV_4$ where we estimate multivariate densities. 
We used bootstrapping with 1000 replicates to estimate the uncertainty in the CV criterion induced by the Monte Carlo approximation for the log predictive densities. Each bootstrap replicate of the CV criterion was based on a random sample of size $M$ with replacement from the posterior sample of model parameters and latent variables.

\subsection{Model validation with PIT}

A key property of a predictive distribution is its calibration \citep{Gneiting2007} 
which we evaluated using randomized probability integral transform \citep[PIT][]{Denuit2005}. Uniform Q-Q plots of the randomized PIT is a graphical method to evaluate whether data, $y_{ij}$, can be considered as a random sample from the discrete predictive distribution that is given by the fitted model or not. To construct Q-Q plots of the randomized PIT for species- and location-wise predictive distributions we define 
\begin{equation} \label{eq:PIT}
u_{ij}=F(y_{ij}-r_{min})+v_{ij}\left(F(y_{ij})-F(y_{ij}-r_{min})\right),
\end{equation}
where $v_{ij}$ is a draw from standard uniform distribution, $F({y_{ij}})$ is the posterior predictive cumulative distribution function (CDF) for observation $y_{ij}$, and $r_{min}$ is the minimum gap between any possible adjacent values of $y$. We set $F({y_{ij}-r_{min}})=0$ if ${y_{ij}-r_{min}}<0$ since percentage cover cannot be negative. 
Calibration is evaluated graphically by plotting uniform Q-Q plot with point-wise 95$\%$ confidence interval of the values $u_{ij}$ since they follow standard uniform distribution if $F({y_{ij}})$ corresponds to the true data generating process \citep{Gneiting2007}.

Q-Q plots of the PIT were drawn for CV predictive distributions using the same splitting strategy as in the CV tests. Hence, $PIT_1$ corresponds to species- and location-wise predictions for $y_{ij}$, $PIT_2$ to location-wise predictions for the sum over species, $\bar{y}_i=\sum_j y_{ij}$, $PIT_3$ to species-wise predictions for the sum over locations, $\bar{y}_j=\sum_i y_{ij}$, and $PIT_4$ to predictions for the sum over species and locations, $\bar{y}=\sum_{ji} y_{ij}$ (Table~\ref{table:model_comparison_and_validation}). 
We used Monte Carlo approximation for the posterior predictive CDFs. For $PIT_1$ we approximated the posterior predictive CDF directly as
\begin{equation} \label{eq:PIT1}
F_1(y_{ij}|\mathbf{Y}(\mathbf{S}_{\setminus k(i)}), \mathbf{S}_{\setminus k(i)}, \mathbf{s}_i)=\sum_{z=0}^{y_{ij}}\frac{1}{M}\sum_{m=1}^{M}\pi(z|f_{ij}^{(m)},\gamma^{(m)}).
\end{equation}
In $PIT_2$--$PIT_4$ we first constructed Monte Carlo estimator for the posterior predictive densities for the sums over species and/or locations conditional on latent variables and parameters. These were calculated for each posterior sample so that $\pi(\bar{y}_i|\mathbf{f}_{i}^{(m)},\gamma^{(m)}) = \frac{1}{B} \sum_{b=1}^B \mathbf{1}_{\bar{y}_i}\left(\sum_{j=1}^J \tilde{y}_{ij}^{(b,m)}\right)$
where $[y_{i,1}^{b,m},\dots,y_{i,J}^{b,m}]\sim \pi(\mathbf{y}_i|\mathbf{f}_{i}^{(m)},\gamma^{(m)})$. After this we calculated CDF as 
\begin{equation} \label{eq:PIT2}
F_2(\bar{y}_{i}|\mathbf{Y}(\mathbf{S}_{\setminus k(i)}), \mathbf{S}_{\setminus k(i)}, \mathbf{s}_i)=\sum_{z=0}^{\bar{y}_{i}}\frac{1}{M}\sum_{m=1}^{M}\pi(z|\mathbf{f}_{i}^{(m)},\gamma^{(m)}).
\end{equation}
The CDFs for $PIT_3$ and $PIT_4$ were estimated analogously.

\section{Experiments}\label{sec:experiments}

\subsection{Simulation study}

We conducted a simulation study to demonstrate and test the applicability of the proposed model. 
To mimic the design of our case study, the simulated data included 200 plots that were selected at random from the inventory plots of the case study. 
We constructed a regular mesh, $\mathcal{D}_M$, over the study area, $\mathcal{D}$, which was used for posterior predictive checks. 
We included four species competing for space (i.e., mutually exclusive species) and four species that do not compete for space with other species. 
First, we simulated two environmental covariates, $x$ and $z$, in the inventory plots and in the mesh nodes. 
We drew covariate $x$ as a random realization from a non-stationary Gaussian process \eqref{eq:nonstatmatern} where logarithm of a spatially varying length scale was given a Gaussian process prior \eqref{eq:logGP} with mean function $\mu=4.5$ and exponential covariance function with length scale $l=80$ and variance $\sigma^{2}=1$. 
The second covariate, $z$, was sampled from a stationary Gaussian process \eqref{eq:independentGP} having exponential covariance function with length scale $l=80$ and variance $\sigma^{2}=1$.
We formed a latent Gaussian variable for all species following $f_{j}(\bm{s}_i)=\beta_{0,j}+\beta_{1,j}x_i+\beta_{2,j}z_i$, where the species specific intercept $\beta_{0,j}$ and covariate effects $\beta_{1,j}$ and $\beta_{2,j}$ were sampled from independent Gaussian distributions. We sampled species observations, $y_{ij}$, for the competing species group from a Dirichlet-Multinomial distribution and for the non-competing species from a Beta-Binomial distributions. The scale parameter of the Dirichlet-multinomial was set to $\gamma=10$ and the scale parameters of the Beta-binomial distributions were sampled from a log-normal distribution $\gamma_{g} \sim \text{log-N}(\mu=2,\sigma^2=0.5^2)$. 

We fitted three models to the simulated data. In all of them, the conditional distribution for species observations given the latent Gaussian variables corresponded to the true data generating process with Dirichlet-multinomial and Beta-Binomial components whose scale parameters were given priors as in Section~\ref{sec:Priors}. 
In the first model, to be denoted Cov+LMC(1)$_\text{S}$, the Gaussian latent variable model included covariate $x$ and a spatial random effect with LMC(1) prior. The covariance function of the LMC(1) prior was the stationary exponential with hyperpriors as defined in Section~\ref{sec:Priors}. Hence, this model corresponds to a situation where the (non-stationary) environmental covariate is included in the model and the (multivariate) stationary spatial random effect captures the effects of the missing (stationary) covariate. 
In the two other models, the Gaussian latent variable model included only the intercept and spatial random effects, that is $f(s_i)=\beta_{0,j} + \epsilon_{i,j}$. 
In the second model, the spatial random effects were given LMC(1) prior with the same non-stationary covariance function that was used to generate the covariate $z$ (model LMC(1)$_\text{NS}$). In the third model, the spatial random effects were given LMC(1) prior with the same stationary covariance function that was used to generate the covariate $x$ (model LMC(1)$_\text{S}$). After fitting the three models to the data, we compared them using the CV criteria and Q-Q plots of the randomized PIT as well as with respect to the true underlying Gaussian latent variable model over the grid cells.

\subsection{Case study analyses}
\label{sec:case-study}

In the plant community modeling case study (Section~\ref{sec:site}), we used one species group for all the Sphagnum moss species and each vascular plant formed its own species group. 
Since Sphagnum mosses had larger than 1\% cover throughout the study region we set $N_{i,a}=100$ for them corresponding to the used 1 percentage unit measurement accuracy.
Vascular plant percentage cover measurements ranged from below to above 1\% so we assumed in average 0.5 percentage unit accuracy for them leading to $N_{i,a}=200$ for vascular plants.
To evaluate the effect of model structures to leaf area predictions, and to test if modeling competition for space improves model predictions, we compared 12 alternative models with different structures for the Gaussian latent variable model and the observation model.
Since we did not have detailed covariate information from the study region, we did not incorporate covariate effects to the latent Gaussian variable model \eqref{eq:LGVM} but tested seven alternative spatial GPs for it: a constant, $f_{j}(\mathbf{s}_{i})=\beta_{0,j}$, or a spatial GP, $f_{j}(\mathbf{s}_{i})=\beta_{0,j}+\epsilon_{ij}$, where Gaussian random effects $\epsilon_{ij}$ were either independent or dependent with either a stationary or non-stationary covariance function.
Each of the latent variable models was combined with either an observation model containing Dirichlet-multinomial for the group of mosses or an observation model where all species had independent Beta-Binomial distributions.
The compared models are summarized in table~\ref{table:modelproperties} and their DAGs are summarized in Figure~\ref{fig:DAG_LMC+DM} and in Figures~B1--B3.

\begin{table}[t]
\footnotesize
\centering

\caption{Summary of the alternative models compared in the case study. The abbreviations in the model names are: C for constant latent function ($f_{j}(\mathbf{s}_{i})=\beta_{0,j}$); IGP$_\mathrm{S}$ and IGP$_\mathrm{NS}$ denote $f_{j}(\mathbf{s}_{i})=\beta_{0,j}+\epsilon_{ij}$ where spatial GPs are respectively independent stationary and non-stationary; LMC($k$)$_\mathrm{S}$ and LMC($k$)$_\mathrm{NS}$ denote respectively stationary and non-stationary linear model of coregionalization with $k$ distinct covariance functions; BB and DM denote respectively Beta-Binomial and Dirichlet-Multinomial processes. }
\renewcommand*{\arraystretch}{0.5}
\begin{tabular}[t]{lccccccc}

\toprule
\textbf{Model} & \multicolumn{5}{c}{Species niche preference} & \multicolumn{2}{c}{Competition}   \\
\cmidrule(lr){2-6}\cmidrule(lr){7-8}
& Constant & \multicolumn{2}{c}{Independent heterogeneous} & \multicolumn{2}{c}{Dependent heterogeneous} & No & Yes \\
\cmidrule(lr){3-4}\cmidrule(lr){5-6}
 &  & stationary & non-stationary & stationary & non-stationary & & \\
\midrule
C+BB & \hfil$\bullet$ & & & & & \hfil$\bullet$ & \\
C+DM & \hfil$\bullet$ & & & & & & \hfil$\bullet$ \\
IGP$_\mathrm{S}$+BB & & \hfil$\bullet$ & & & & \hfil$\bullet$ & \\
IGP$_\mathrm{NS}$+BB & & & \hfil$\bullet$ & & & \hfil$\bullet$ & \\
IGP$_\mathrm{S}$+DM & & \hfil$\bullet$ & & & & & \hfil$\bullet$  \\
IGP$_\mathrm{NS}$+DM  & & & \hfil$\bullet$ & & & & \hfil$\bullet$  \\
LMC(1)$_\mathrm{S}$+BB & & & & \hfil$\bullet$ & & \hfil$\bullet$ & \\
LMC(1)$_\mathrm{NS}$+BB & & & & & \hfil$\bullet$ & \hfil$\bullet$ & \\
LMC(1)$_\mathrm{S}$+DM & & & & \hfil$\bullet$ & & & \hfil$\bullet$ \\
LMC(1)$_\mathrm{NS}$+DM & & & & & \hfil$\bullet$ & & \hfil$\bullet$ \\
LMC(2)$_\mathrm{S}$+DM & & & & \hfil$\bullet$ & & & \hfil$\bullet$ \\
LMC(2)$_\mathrm{NS}$+DM & & & & & \hfil$\bullet$ & & \hfil$\bullet$ \\
\midrule

\end{tabular}
\label{table:modelproperties}
\normalsize
\end{table}

We compared and assessed the alternative models with the CV criteria and Q-Q plots with point-wise 95$\%$ confidence intervals of the randomized PIT summarized in Table~\ref{table:model_comparison_and_validation}.
For computational reasons, these were done using a random subset of 200 inventory plots.
We then trained the best model with all data to predict the vegetation cover for each species separately, across all species and across both species groups (mosses and vascular plants). 
The predictions were done over a regular mesh, $\mathcal{D}_M$, covering the study area, $\mathcal{D}$. 
The resolution of the mesh was chosen so that further decreasing the size of the grid cells ($A=4$ m$^{2}$) did not affect the posterior predictive variance of the total vegetation cover $Var(\tilde{\phi})$.

\section{Results} \label{sec:results}

\subsection{Simulation study}

The results from the simulation study are collected in Appendix A and here we summarize them. As expected, the model closest to the data generating model (Cov+LMC(1)$_{\text{S}}$) had the best and the non-stationary random effect model (LMC(1)$_{\text{NS}}$) the second best overall performance when measured with the 10-fold CV log predictive density estimates (Table~A1). 
In terms of Q-Q plots of the PIT (Figures~A5--A7, A14--A16 and A23--A25) and the total percentage cover predictions over the grid cells (Figures~A8, A17 and A26) the differences between the models were small. 
The main purpose of the simulation study was, however, to critically assess the identifiability and posterior sampling of model parameters. 
For all models, the MCMC sampling of model parameters and Gaussian latent variables converged and mixed reasonably well. 
For model LMC(1)$_{\text{NS}}$ the convergence was slower and the mixing of the sample chain not as good as for the other two models though.
In Cov+LMC(1)$_{\text{S}}$ practically all parameters and latent variables were well identified. 
The 95\% posterior credible intervals for model parameters included the true data generating parameter values for all cases except the length-scale of the LMC(1) covariance function, which was slightly underestimated (Figure~A1). 
The mismatch between the posterior distribution and the true value of the lengthscale most likely results from randomness in the realization of the hidden covariate, $z$, which the LMC(1) term adapts to.
Since Cov+LMC(1)$_{\text{S}}$ did not include the hidden covariates its spatial random effect should adapt to the species specific $\beta_{2,j}z$ terms and its coregionalization covariance matrix, $\bm{\Sigma}_{\boldsymbol{\epsilon}}$, estimates the true interspecific covariance matrix induced by $\beta_{2,j}$ parameters; that is, $\bm{\beta}_{2}\bm{\beta}_{2}^{\top}$ where $\bm{\beta}_{2}^{\top}=[\beta_{2,1},\dots,\beta_{2,J}]$. 
However, since we have sampled the process only in 200 locations the coregionalization covariance matrix should estimate better the sample covariances between by $\beta_{2,j}z$ terms.
The 95\% posterior credible intervals of the elements of the coregionalization matrix included the true interspecific covariance in most cases and the sample covariance in all cases (Figure~A2). 
The posterior mean of the interspecific exclusive competition measure \eqref{eq:competition_measure} predicted over the study area matched also well the true simulated competition measure (Figure~A9). 

The models LMC(1)$_{\text{S}}$ and LMC(1)$_{\text{NS}}$ did not include any covariates so the only parameters having a counterpart in the data generating model are the covariance function length-scales, $\mu_l$ in the non-stationary covariance function, the scale parameters of the Dirichlet and Beta distributions, and the coregionalization covariance matrix. 
All these parameters were well identified since the most of their 95\% posterior credible intervals included the true data generating values (Figures~A10, A19). The only exception was the posterior distribution of the length-scale of the LMC(1)$_{\text{S}}$ model, for which the 95\% posterior credible interval did not include the true data generating value.
In general, the coregionalization matrices $\bm{\Sigma}_{\epsilon}$ of LMC(1)$_{\text{S}}$ and LMC(1)$_{\text{NS}}$ estimated well the interspecific sample covariances but underestimated the true interspecific covariances, induced by the two covariate effects in the simulated data, given by $\bm{BB}^{\top}$ where $\bm{B}$ is a $J\times2$ matrix with $j$'th row $\bm{B}_{j\cdot}=[\beta_{1,j},\beta_{2,j}]$ (Figures~A11, A20). 
Also, the posterior means of the interspecific competition measures \eqref{eq:competition_measure} predicted over the study area matched well the true simulated competition measure (Figures~A18 and A27). These interspecific exclusive space competition measure predictions of LMC(1)$_{\text{S}}$ and LMC(1)$_{\text{NS}}$ were smoother and less accurate than the corresponding prediction with Cov+LMC(1)$_{\text{S}}$ which is reasonable since the former two lack the information from $x_1$ at the prediction locations.

\subsection{Case study}

\subsubsection{Model comparison}

Table \ref{table:10fCV_comparison} summarizes the model comparison results for the alternative models in our case study. 
In all predictive tasks, the models with constant latent function (C+BB and C+DM) had the lowest CV predictive performance. This gives strong support for spatial heterogeneity in vegetation covers. 
However, when predicting the species-wise total over the region (CV$_3$), the difference to the best model was not significant if measured relative to standard error of the CV estimates. 
In general, models with Dirichlet-Multinomial observation model outperform the otherwise same models with Beta-Binomial in joint species predictions (CV$_2$ and CV$_4$) and perform practically equally well in single species predictions (CV$_1$ and CV$_3$). This indicates that accounting for species competition is more important for community predictions than for single species predictions.

The best performing model in terms of $CV_1$ and $CV_3$ was LMC(1)$_\mathrm{S}$+BB with practically equal performance by LMC(1)$_\mathrm{S}$+DM and LMC(2)$_\mathrm{S}$+DM. 
However, the $CV_1$ and $CV_3$ estimates of all spatially heterogeneous models except IGP$_S$+DM in $CV_1$ were within one standard error of the best CV estimate indicating that these differences were practically negligible. 
When looking at $CV_1$ and $CV_3$ for each species separately (Table~B1), the relative differences between alternative models were small but the best model for all species always had either stationary or non-stationary LMC latent function prior. 
In terms of, $CV_2$ and $CV_4$ the best performing models were LMC(1)$_\mathrm{S}$+DM, LMC(2)$_\mathrm{S}$+DM and LMC(1)$_\mathrm{NS}$+DM. 
The former two models had practically equal $CV_2$ and $CV_4$ estimates and those of the latter were within one standard error estimate from the former two. 
The relatively larger difference between LMC(1)$_\mathrm{NS}$+DM and the best model might be, at least partly, explained by worse mixing of its Markov chain compared to the Markov chains of LMC(1)$_\mathrm{S}$+DM and LMC(2)$_\mathrm{S}$+DM which induces more weight to occasional low log predictive density values. 
The $CV_2$ and $CV_4$ estimates of all other models were significantly worse than those of the best three models. 
The Markov chains for IGP$_\mathrm{NS}$+DM and LMC(2)$_\mathrm{NS}$+DM converged slowly and their mixing was poor resulting into considerably smaller effective sample size for the log predictive densities compared to other models so we excluded them from the CV comparison.
Time needed for posterior sampling of JSDMs with stationary spatial random effects ranged from 3 to 11 hours with a regular desktop computer. 
The sampling for single species SDMs (stationary or non-stationary) was considerably faster whereas the sampling for JSDMs with non-stationary spatial random effects was considerably slower (6.5 days for LMC(1)).

\begin{table}[ht]
\footnotesize
\centering
\caption{Model comparison results with 10-fold CV using log predictive density utilities ($CV_1,\dots,CV_4$; see Table~\ref{table:model_comparison_and_validation}) together with the standard error estimates (se) and the Monte Carlo error estimate (me) for the CV estimates (see Section~\ref{sec:validation}). The best CV estimate is indicated by bold font and the CV estimates that are within one standard error from the best model are indicated by italic font. Rows with -- indicate models whose effective sample size for log predictive density was considerably smaller than for the rest of the models and, hence, were excluded from the comparison. 
\\
}
\begin{tabular}[ht]{lcccc}

\toprule
\textbf{Model} & $CV_{1}$ (se/me) &  $CV_{2}$ (se/me) & $CV_{3}$ (se/me) & $CV_{4}$ (se/me) \\
\midrule
C+BB & -1.80 (4e-2/3e-5) & -25.2 (3e-1/5e-4) & \emph{-35.9} (2/6e-4) & -503.0 (8/2e-2) \\
C+DM & -1.80 ( 4e-2/3e-5) & -23.8 (4e-1/5e-4 ) & \emph{-36.1} (2/7e-4) & -474.9 (9/2e-2) \\
\hline
IGP$_\mathrm{S}$+BB & \emph{-1.69} (4e-2/1e-5) & -23.6 (4e-1/3e-3) & \emph{-33.5} (2/6e-3) & -474.5 (10/4e-1) \\
IGP$_\mathrm{NS}$+BB & \emph{-1.71} (4e-2/3e-4) & -23.9 (4e-1/7e-3) & \emph{-34.1} (2/2e-2) & -487.1 (10/6e-1) \\
IGP$_\mathrm{S}$+DM & -1.75 (4e-2/5e-4) & -23.1 (4e-1/1e-2) & \emph{-34.1} (2/2e-2) & -460.3 (10/9e-1) \\
IGP$_\mathrm{NS}$+DM & -- & -- & -- & -- \\
\hline
LMC(1)$_\mathrm{S}$+BB & \textbf{-1.67} (4e-2/1e-4) & -23.3 (4e-1/4e-3) & \textbf{-33.4} (2/6e-3) & -470.3 (10/4e-1) \\
LMC(1)$_\mathrm{NS}$+BB & \emph{-1.70} (4e-2/7e-4) &  -23.6 (5e-1/1e-2) &  \emph{-34.5} (2/3e-2) &  -495.8 (14/9e-1) \\
LMC(1)$_\mathrm{S}$+DM & \emph{-1.68} (4e-2/1e-4) & \textbf{-21.9} (4e-1/3e-3) & \emph{-33.5} (2/6e-3) & \emph{-440.1} (10/4e-1) \\
LMC(1)$_\mathrm{NS}$+DM & \emph{-1.71} (4e-2/4e-4) & \emph{-22.1} (4e-1/9e-3) & \emph{-34.2} (2/2e-2) & \emph{-450.4} (12/8e-1) \\
\hline
LMC(2)$_\mathrm{S}$+DM & \emph{-1.68} (4e-2/1e-4) & \textbf{-21.9} (4e-1/3e-3) & \emph{-33.5} (2/5e-3) & \textbf{-439.2} (10/4e-1) \\
LMC(2)$_\mathrm{NS}$+DM & -- & -- & -- & -- \\
\bottomrule
\end{tabular}
\label{table:10fCV_comparison}
\normalsize
\end{table}

\subsubsection{Model validation}

\begin{figure}
\begin{center}
\includegraphics[]{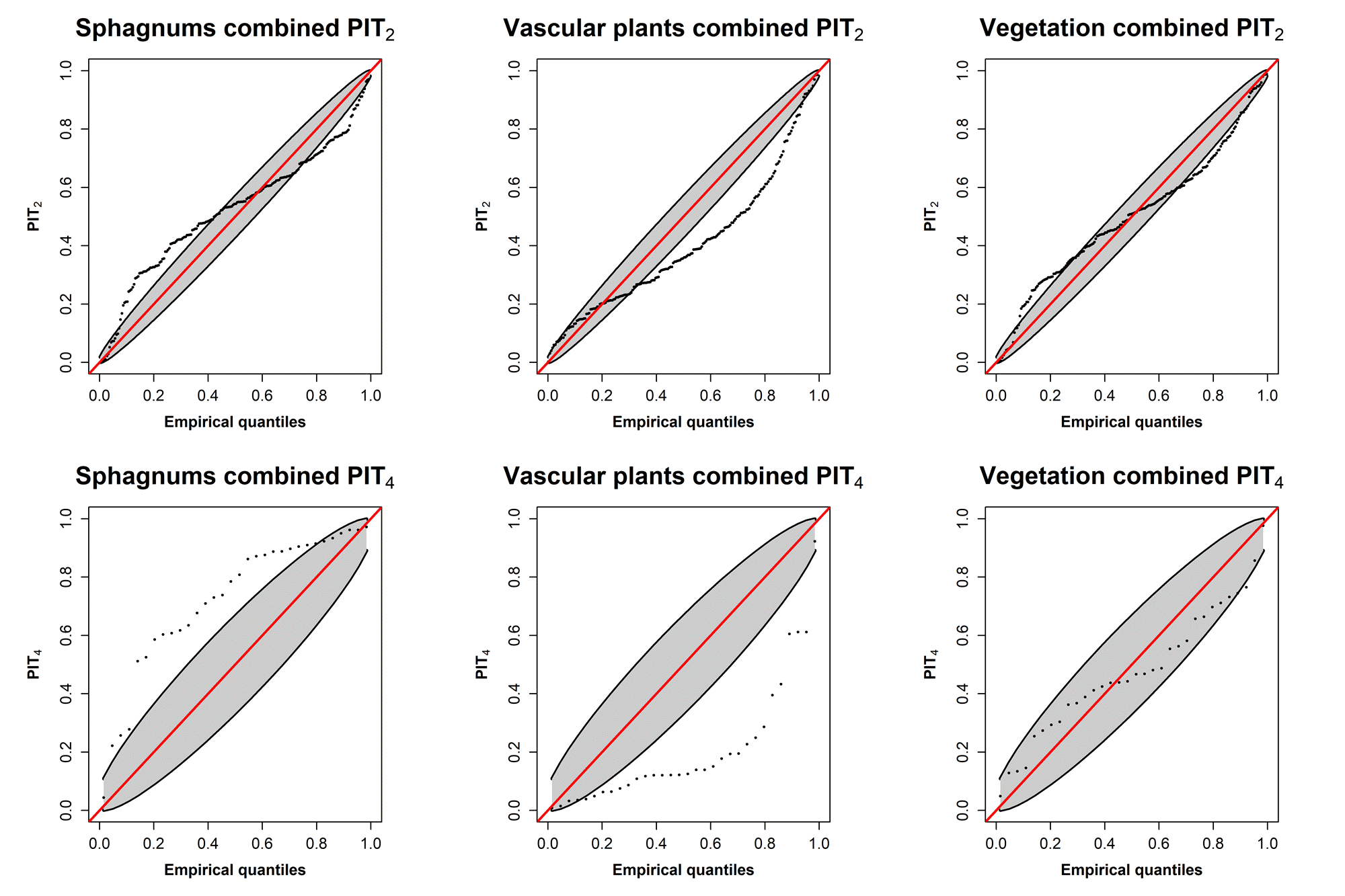}
\captionof{figure}{Randomized PIT$_{2}$ and PIT$_{4}$ histograms for the LMC(1)$_\mathrm{NS}$+DM model separately for sphagnums, vascular plants and combined vegetation. }
\label{PIT1}
\end{center}
\end{figure}

Model comparison with CV did not indicate clear differences between the best models: LMC(1)$_\mathrm{S}$+DM, LMC(2)$_\mathrm{S}$+DM and LMC(1)$_\mathrm{NS}$+DM. Moreover, the former two were practically the same model since LMC(2) effectively reduced to LMC(1). Hence, we conducted model validation for the latter two. 
Q-Q plots with point-wise 95$\%$ confidence intervals of the randomized PIT$_2$ and PIT$_4$ for the LMC(1)$_\mathrm{NS}$+DM and LMC(2)$_\mathrm{S}$+DM models (Figure~\ref{PIT1} and B4) 
show moderate departure from uniformity and they indicate that the total percentage cover is to some extent underestimated in average over the CV folds. 
However, it should be noted that Q-Q plots of the PIT$_2$ are formed from only 200 and Q-Q plots of the PIT$_4$ from 32 posterior predictive distributions. 
Hence, the deviation from uniformity is likely also due to randomness. 
Similar deviations from uniformity were observed also in the simulation study with the Cov+LMC(1)$_{\text{S}}$ model, which was corresponded to the true data generating process in the simulation study (Figures~A5-A7).
The species wise Q-Q plots of the PIT$_1$ for LMC(2)$_\mathrm{S}$+DM and LMC(1)$_\mathrm{NS}$+DM models (Figures B5 and B6) 
are close to uniform for all other species except S. \emph{papillosum} for which Q-Q plots of the PIT$_1$ is slightly $\cup$-shaped indicating underestimation of predictive uncertainty. This deviation from uniformity was more distinct for LMC(2)$_\mathrm{S}$+DM than LMC(1)$_\mathrm{NS}$+DM. Out of all species, S. \emph{papillosum} was the one showing the most clear non-stationarity in the latent function of LMC(1)$_\mathrm{NS}$+DM model indicating that non-stationary Gaussian process prior might have been benefitial in this case. 
For most species the species wise Q-Q plots of the PIT$_3$ for LMC(2)$_\mathrm{S}$+DM and LMC(1)$_\mathrm{NS}$+DM models (Figures B7 and B8) 
do not show clear deviations from uniformity even though the Q-Q plots are noisier than the Q-Q plots of the PIT$_1$.

\subsubsection{Percentage cover predictions and species interactions}

\begin{figure}
\begin{center}
\includegraphics{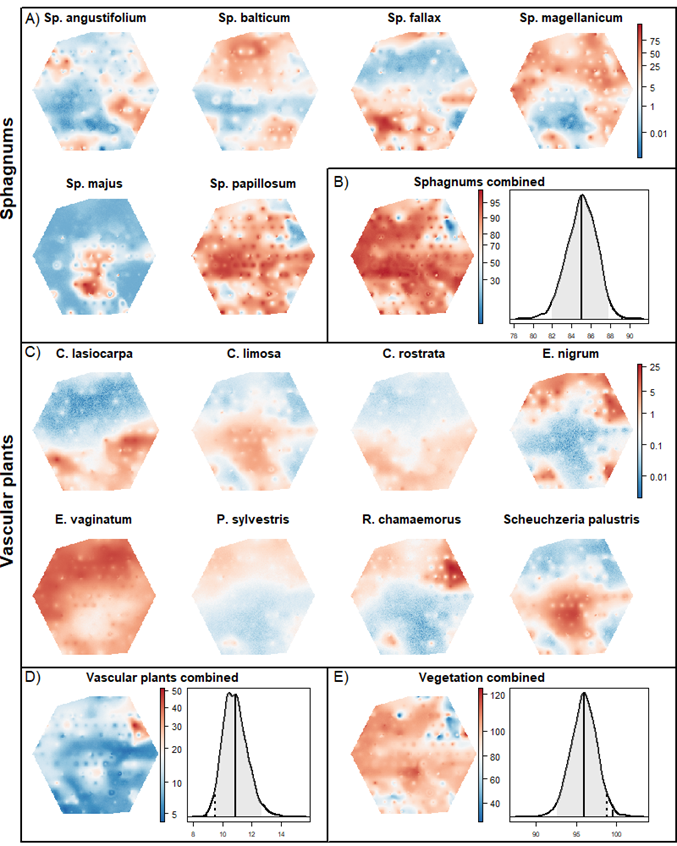}
\captionof{figure}{The posterior predictive means of the percentage covers and probability distributions of the combined percentage covers over the study area as predicted by LMC(1)$_\mathrm{NS}$+DM model. (A) species-wise percentage covers of the sphagnums, (B) total percentage cover of all sphagnum species combined and the total combined cover over the study area, (C) species-wise percentage covers of vascular plants, (D) total percentage cover of vascular plants combined and the total combined cover over the study area and (E) combined total cover of sphagnums and vascular plants. For combined totals over the area probability distributions are estimated using kernel density estimation on MCMC samples. Solid vertical lines show the posterior means for the combined covers, dashed lines show the sample means computed from the training data. }
\label{species_covers}
\end{center}
\end{figure}

Based on the model comparison and validation we selected LMC(1)$_\mathrm{NS}$+DM for the final inference. The percentage covers of the species have clearly different spatial patterns (Figure~\ref{species_covers}). Spatial distribution of the vegetation is typical for minerotrophic fen where center is lower than the edges and therefore also water table is on average higher (wetter) at the center. As spatial composition of sphagna and vascular plants follows the variation in the water table (WT), plants adapted to grow in drier areas are concentrated to edges of the study area and flarks are inhabitated by species that can withstand waterlogging.

Hummock species like \textit{S. angustifolium} and \textit{E. nigrum} favoring drier habitat grow mostly at the edges of the study area. Lawn species \textit{S. papillosum} dominating the composition of sphagna grows fairly evenly throughout the area. Lawn community type can be further divided into three subgroups specified by vascular plants (from drier to wetter) \textit{E. vaginatum lawn}, \textit{C. rostrata lawn} and \textit{C. lasiocarpa lawn}. \textit{C. rostrata} and \textit{C. lasiocarpa} grow mainly on the southern edges of the study area while percentage cover of \textit{E. vaginatum} tends to be higher on the northern edges. Species (\textit{S. majus}, \textit{C. limosa} and \textit{S. palustris}) associated to wetter growing conditions (hollows) are concentrated in smaller hotspots at the center of the study region. 
Combined sphagnum cover (sum of sphagnum species covers) is relatively stable (85-95\%) over the study area although species-wise distributions have distinct spatial patterns. Contrarily, spatial distribution of combined vascular plants cover has high variation such that hotspots of higher combined covers are located on both wetter center areas and drier parts of the study area.

\begin{figure}  
\begin{center}  
\includegraphics{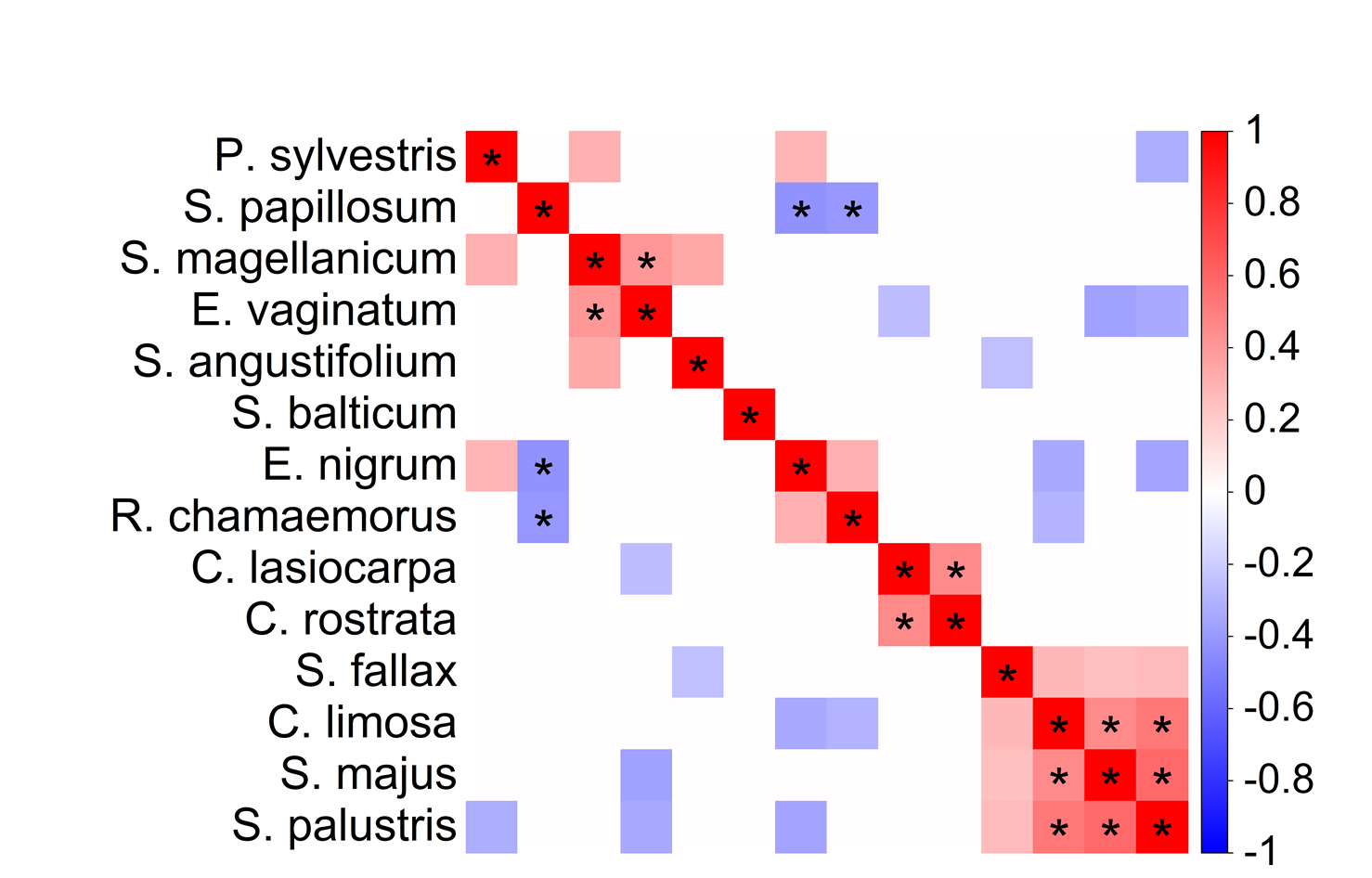}
\captionof{figure}{Posterior mean estimates of the interspecific correlations in the species niche preference in the LMC(1)$_\mathrm{NS}$+DM model. White cells indicate that the 80\% posterior credible interval of the correlation overlapped zero (i.e., weak support for interspecific correlation) and stars indicate that the 95\% posterior credible interval did not overlap zero (i.e., strong support for interspecific correlation).}
\label{fig:spatial-correlations}
\end{center}
\end{figure}

\begin{figure}
\begin{center}
\includegraphics{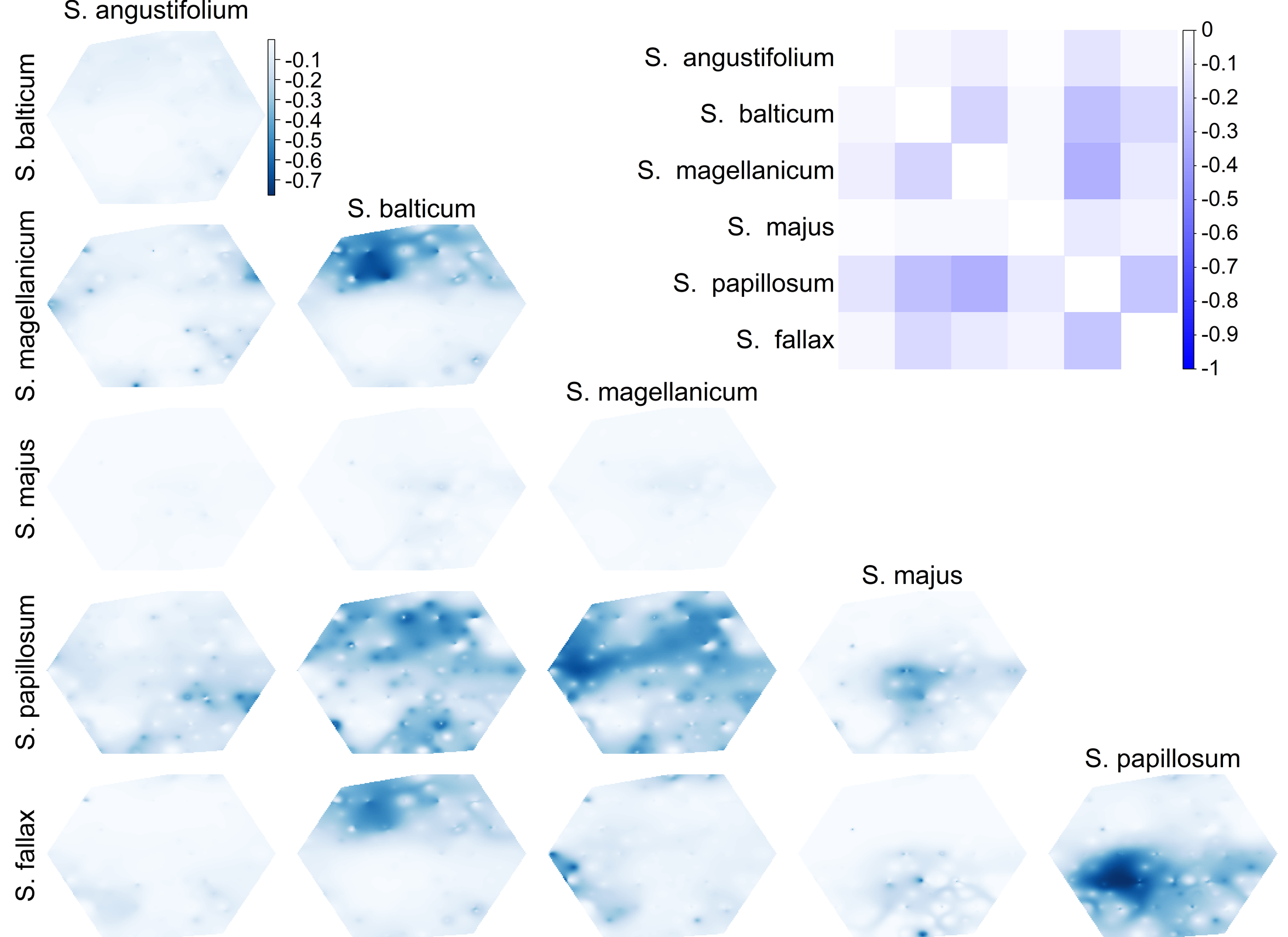}
\captionof{figure}{The exclusive competition for space between \textit{sphagnum} species. The maps on the lower left corner show the spatial distribution of the (negative) interspecific correlation in percentage covers for all pairs of \textit{sphagnum} species. The matrix on the upper right corner shows the average of these correlations over the whole study area. }
\label{fig:species_competition}
\end{center}
\end{figure}

The estimated interspecific correlations in niche preferences are summarized in Figure~\ref{fig:spatial-correlations} and the exclusive competition for space between \textit{sphagnum} species is summarized in Figure~\ref{fig:species_competition} in terms of the (negative) correlation in their percentage cover \eqref{eq:competition_measure}. 
Species can be differentiated into co-occuring groups according to the interspecific dependencies between species niche preferences. Sphagnums \textit{S. majus} and \textit{S. fallax} with vascular plants \textit{S. palustris} and \textit{C. limosa} adapted to grow in wetter conditions form one cluster. Lawn species \textit{C. lasiocarpa} and \textit{C. rostrata} can be classified into the second co-occuring cluster. High lawn-low hummock species \textit{S. angustifolium}, \textit{S. magellanicum} and \textit{P. sylvestris} form the third cluster. Dwarf shrubs \textit{E. nigrum} and \textit{R. chamaemorus} form the fourth cluster (both species were also positively correlated with \textit{P. sylvestris} but 80\% credibility intervals for the correlation between \textit{R. chamaemorus} and \textit{P. sylvestris} overlapped zero). Lawn species \textit{S. papillosum}, \textit{S. balticum} and \textit{E. vaginatum} were also positively correlated but 80\% credibility intervals for correlations overlapped zero. The interspecific correlations in niche preference followed qualitatively similar pattern in all LMC models. However, models with Beta-Binomial observation model found more negative interspecific correlations than models with Dirichlet-Multinomial observation model as illustrated by comparison of Figures~\ref{fig:spatial-correlations} and B10.

Estimated total vegetation cover distributions for each model presented in Figure B9 
show that non-stationary JSDM models predicted combined covers that are close to the observed sample means. All stationary models tend to predict combined cover to be smaller than their non-stationary counterparts. 
In each case, the 95\% prediction intervals for non-stationary JSDM models overlap observed sample mean but for stationary models 95 \% prediction intervals do not overlap observed sample mean. 
This suggests that there is non-stationarity in the observation process that should be taken into account or received vegetation cover estimates might be biased. 
Stacked species distribution models tend to overpredict combined vascular plants cover but underpredict combined sphagnum and total vegetation covers.

\section{Discussion}
\label{sec:discussion}

Our proposed model performed well in the simulation experiments in terms of the identifiability of the model parameters and (out-of-sample) predictive performance. 
Importantly, the proposed model was able to reproduce the interspecific correlations in the residual latent Gaussian variables (the coregionalization matrices $\Sigma_{\epsilon}$) (Figures~A2, A11, A20) and the parameters for exclusive competition for space \eqref{eq:competition_measure} (Figures~A9, A18, A27). 
The latter were equally accurately inferred in the presence of (simulated) environmental covariates (Figure~A9) and in the absence of them (Figure~A18). The computational time for posterior inference ranged from hours to several days but, to our experience, it was comparable to other JSDMs employing spatial random effects. 

Our case study results (table~\ref{table:10fCV_comparison}) support previous empirical findings that modeling interspecific correlations improves JSDM predictions through information sharing between species \citep{Ovaskainen2011,Vanhatalo2020,Nordberg2019}. 
Our case study results show additionally that explicit modeling of exclusive competition for space improves models' predictive performance. 
This improvement is more important in predictions concerning community structure and total percentage cover ($CV_2$ and $CV_4$ in table~\ref{table:10fCV_comparison}) than in single species predictions ($CV_1$ and $CV_3$ in table~\ref{table:10fCV_comparison}).
The models did not differ only in their predictive properties, though, but they led to different inference results as well. For example, models with Beta-Binomial observation model found more negative interspecific correlations in species niche preferences than models with Dirichlet-Multinomial observation model (figures~\ref{fig:spatial-correlations} and B10). Even though the differences between the best models were small, the proposed non-stationary JSDM model worked slightly better than its stationary counterpart. The reason for this was that some species, most clearly S. \textit{papillosum}, had non-stationary latent Gaussian field (see Figure~\ref{species_covers}). 
However, we want to emphasize that the more complex non-stationary spatial structure is an overkill in JSDMs in general. 
As mentioned already by \cite{Schmidt+Rodriguez:2011}, non-stationary spatial random effects are more likely needed when environmental covariates are not available than when they are available. To our experience, the environmental covariates typically capture non-stationary aspects in the data. Moreover, often spatial data are collected with so low resolution that inferring non-stationarity from them is not possible. Non-stationary LMC was justified for our case study, though, since our data were sampled with high spatial resolution and we did not have environmental covariates from the region.
 
Our results demonstrate also that the relative performances of alternative models can be different in different model comparison metrics. 
This highlights the importance of tailoring the model comparison method to the specific task at hand. 
The best models (LMC(1)$_\mathrm{S}$+DM, LMC(2)$_\mathrm{S}$+DM and LMC(1)$_\mathrm{NS}$+DM) were among the best in all predictive tasks whereas other models performed equally well to them only in some of the predictive tasks. 
Namely, all spatially heterogeneous models were practically equally good in producing species distribution maps and per species total cover estimates ($CV_1$ and $CV_3$ see table~\ref{table:10fCV_comparison}). However, in joint species predictions the best models stood out as clearly the best. 
Notably, the spatially homogeneous models C+BB and C+DM did not differ significantly from the other models in per-species total over area predictions ($CV_3$ in table~\ref{table:10fCV_comparison} and Figure~B9), which is reasonable since, in theory, sample mean over uniform random locations is an unbiased estimate for species-wise total covers. 
However, \citet{Fosteretal:2021} give an example on how sample mean over uniform random locations may fail in practice if the underlying field is very heterogeneous. 
We expect that estimating areal vegetation composition more precisely can be beneficial in areal greenhouse gas emission models since vegetation is important explanatory factor in carbon dioxide and methane flux models. 
The case study results demonstrate that presented model offers valid method to estimate areal vegetation cover, and uncertainty associated to it, for those purposes.
The lack of covariates is a shortcoming in our case study though, but information on relevant covariates was not available for the study area.
For example, height of water table and covariates related to soil properties could improve the model fit further as they are a commonly known drivers of vegetation patterns in peatlands \citep{Andersen2011}. 

Drawing conclusions about species associations and competition from observational data is challenging since we cannot control for all the covariates and processes shaping the community.
As demonstrated and discussed by, for example, \citet{Clark+etal:2014}, \citet{Blanchet+etal:2020} and \citet{Poggiato+etal:2021} latent factor models offer limited insight into realized dependence behavior between species at sites and cannot, in reality, separate environmental effects from biotic interactions since these two components are functionally confounded. 
In this respect, our inference is an improvement compared to other JSDMs where interspecific interactions are modelled only with latent factors. 
Our models with Dirichlet-Multinomial observation model describe exclusive competition for space functionally differently from the effects of environmental covariates, and these two descriptions are located in different hierarchical levels of our probabilistic model. 
Hence, the negative interspecific dependencies arising from the exclusive space competition are filtered from the other interspecific dependencies captured by the latent Gaussian model. 
A clear example of this is provided by the different inference results for the coregionalization matrices of the Beta-Binomial model compared to that of the Dirichlet-Multinomial observation model (Figures~\ref{fig:spatial-correlations} and B10).
Our case study results are also highly congruent with the earlier ecological knowledge of the species in the study area.
However, the Dirichlet-Multinomial model does not contain mechanistic description on biological processes (such as abiotic filtering and biotic competition for nutrients) underlying this exclusion process.
It is a probabilistic description for the end result of those processes.

Even though Dirichlet process has previously been used in JSDM literature our approach differs from these earlier models significantly. \citet{Taylor-Rodriguezetal:2017} and \citet{Shirota2019} use Dirichlet processes to cluster species in their responses to latent factors and \citet{Johnson+Sinclar:2017} and \citet{Sollmann+etal:2021} use them to cluster species in their responses to environmental covariates. We use Dirichlet distribution in conceptually different manner and in technically different part of the model compared to sample species specific percentage covers. 
Moreover, Dirichlet-multinomial model has been used to model vegetation cover data, for example, in ordination settings \citep{Damgaard+Hansen+Hui:2020} but our model is the first one combining it with formal joint species distribution modeling framework. 

We believe that the proposed modeling framework can be broadly used for vegetation distribution studies. 
The only \emph{a priori} information on interspecific dependence between species required in our model is whether species inhabit same vegetation layer or not. 
In our case study, this boils down to grouping together the moss species. 
However, such information is often available also in other plant studies. 
In principle, the model could be extended also to cases where the space dependencies between species are not clear so that we make the group identifiers of species random and infer them alongside other model parameters. 
Our model could also be applied to other cases where species can be assigned to exclusive species groups according to a limited resource, and potentially also to cases where competition arises from priority effects where a species arriving first excludes a species arriving later.
For example, coral species occupying same depth layer behave similarly to the mosses in our study and colonial breeding bird communities show competition for nesting space that could potentially be formulated through Dirichlet process.
Another application domain for our model are compositional data where species are not necessarily ecologically exclusive but due to sampling process they are so in the observations. 
For example, sequencing based microbial community data are compositional by nature due to the limited volume of sample even if the microbial species were not competing for space. 
Similarly, \citet{Juntunen+etal:2012} modeled (fixed volume) trawl catches from a fish community with multinomial distribution without assumptions for exclusive competition for space. 
Hence, we believe that combining the Dirichlet-Multinomial layer with the hierarchical latent Gaussian layer of contemporary JSDMs can improve their predictive and inference performance considerably in applications where interspecific competition for space is present or where sampling process induces Multinomial structure among species observations.

\section*{Acknowledgements}

This work was funded by the Doctoral Programme in Science, Technology and Computing of the University of Eastern Finland through support to JK, by University of Eastern Finland, Faculty of Science and Forestry, through support to AK, and by the Academy of Finland through a grant 317255 to JV, grants 337550 and 337549 to E-ST and grant 338980 to AK, and by Tiina and Antti Herlin Foundation through a grant to JK.

\bibliographystyle{Chicago}
\bibliography{Bibliography}

\appendix

\section{Simulation study analyses}

\setcounter{figure}{0} 
\setcounter{table}{0} 
\renewcommand{\thefigure}{A\arabic{figure}}
\renewcommand{\thetable}{A\arabic{table}}

In this supplementary material we present the results for the simulation study analyses. The model comparison results are summarized in Table~\ref{table:10fCV_comparison_simul} and the model specific inference results are given in the subsequent sections.

\begin{table}[ht]
\footnotesize
\centering
\caption{Model comparison results with 10-fold cross-validation using log predictive density utilities ($CV_1,\dots,CV_4$) together with the standard error estimates (se) and the Monte Carlo error estimate (me) for the CV estimates. The best CV estimate is indicated by bold font and the CV estimates that are within one standard error from the best model are indicated by italic font.
\\
}
\begin{tabular}[ht]{lcccc}
\toprule
\textbf{Model} & $CV_{1}$ (se/me) &  $CV_{2}$ (se/me) & $CV_{3}$ (se/me) & $CV_{4}$ (se/me) \\
\midrule
Cov+LMC(1)$_\mathrm{S}$ & \textbf{-2.30} (5.0e-2/1.0e-7) & \textbf{-16.8} (3.1e-1/8.6e-6) & \textbf{-45.9} (1.9/5.4e-3) & \textbf{-337.5} (4.7/1.3e-1) \\
LMC(1)$_\mathrm{NS}$ & -2.53 (5.7e-2/5.7e-7) & -18.3 (3.8e-1/2.0e-4) & -50.6 (2.1/4.4e-2) & -389.9 (11.2/6.3e-1) \\
LMC(1)$_\mathrm{S}$ & -2.64 (6.5e-2/4.1e-7) &  -20.0 (5.9e-1/6.1e-5) &  -52.2 (2.3/2.5e-2) &  -404.2 (10.3/4.6e-1) \\
\bottomrule
\end{tabular}
\label{table:10fCV_comparison_simul}
\normalsize
\end{table}

\subsection{Inference using Cov+LMC(1) model}

Latent process for species $j$ at location $\mathbf{s}_{i}$ was modelled as
\setlength{\abovedisplayskip}{15pt}
\setlength{\belowdisplayskip}{15pt}
\begin{equation}
f_{ij}=\beta_{0,j}+\beta_{1,j}x_{i}+\epsilon_{ij},
\end{equation}
where $\beta_{0,j}$ is an intercept for species $j$, $x_{i}$ is (environmental) covariate value at location $\mathbf{s}_{i}$, $\beta_{1,j}$ is a covariate effect for species $j$ and $\epsilon_{ij}$ is a stationary Gaussian random effect for species $j$ at location $\mathbf{s}_{i}$. Dependencies are modeled through linear model of coregionalization with one distinct covariance function (k=1).

\begin{figure}[H]
\begin{center}
\includegraphics[]{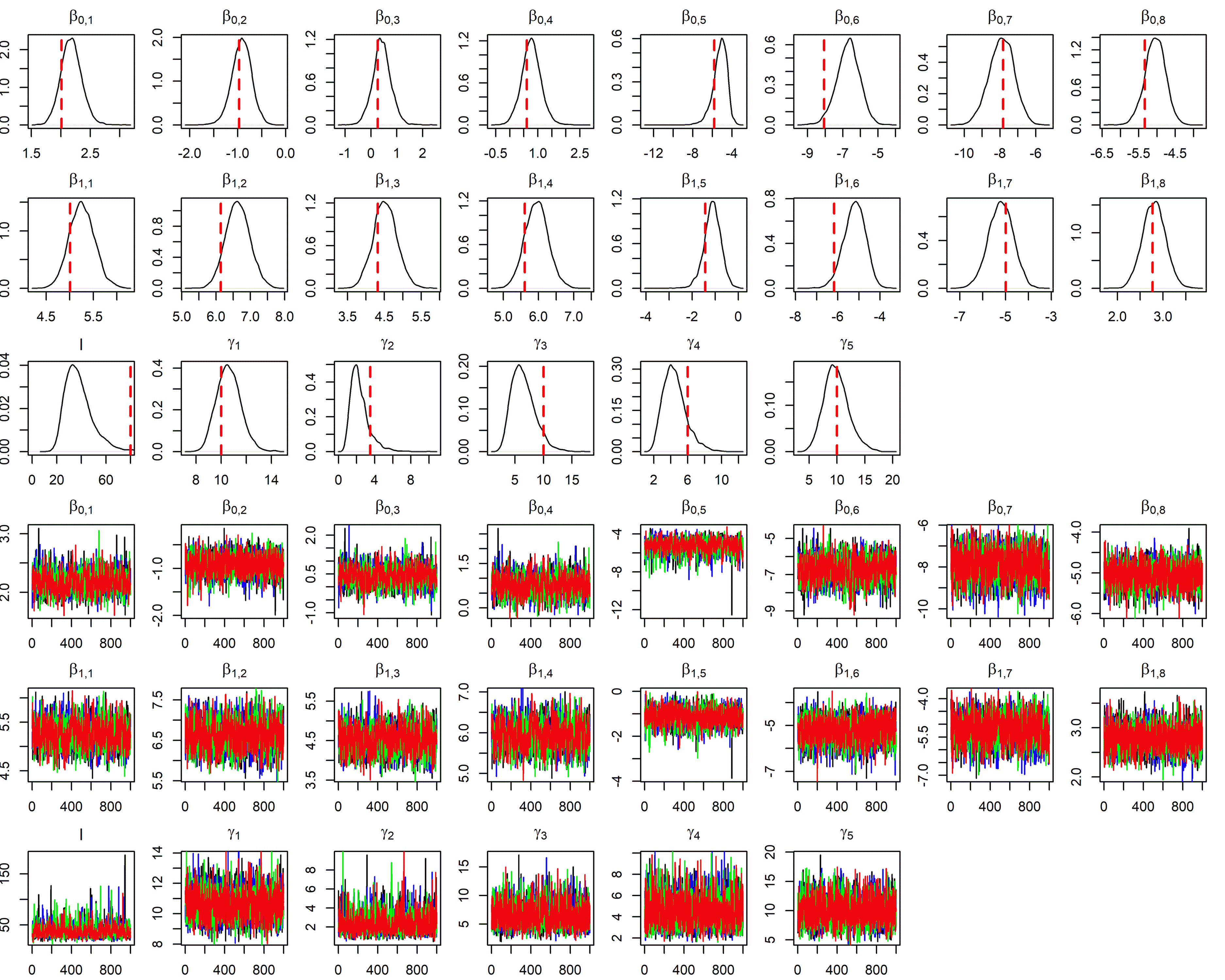}
\end{center}
\caption{The posterior distributions and trace plots as estimated by Cov+LMC(1) model for intercept terms, covariate effects, the hyperparameters of the covariance function, the hyperparameter of the Dirichlet distribution and the hyperparameters of the Beta distributions. The red dashed line shows the true parameter used in simulating the data. Trace plots show 1000 posterior samples after 1000 warmup steps for four chains.} \label{fig:cov_stat_post}
\end{figure}

\begin{figure}[H]
\begin{center}
\includegraphics[]{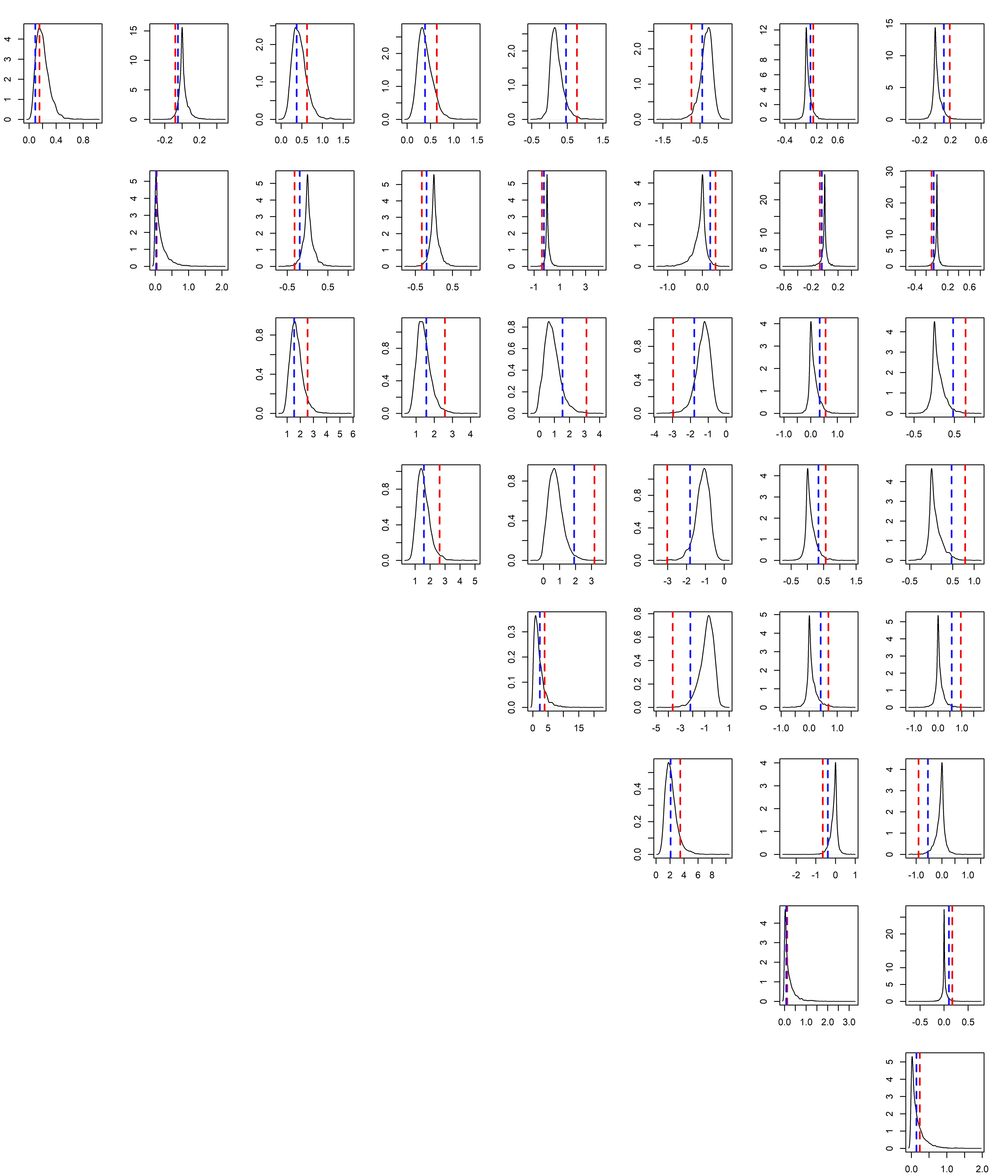}
\end{center}
\caption{Posterior distributions for variances and covariances of the coregionalization matrix $\bm{\Sigma}_{\epsilon}$ as estimated by Cov+LMC(1) model. The red dashed lines show the covariances induced by the unobserved covariate, z, that is elements of $\boldsymbol{\beta}_2\bm{\beta}_2^{\top}$. The blue dashed lines show the sample covariances between the 200 simulated species specific covariate effects $\beta_{2,j}z$ and $\beta_{2,j'}z$.} \label{fig:cov_stat_covariance}
\end{figure}

\begin{figure}[H]
\begin{center}
\includegraphics[]{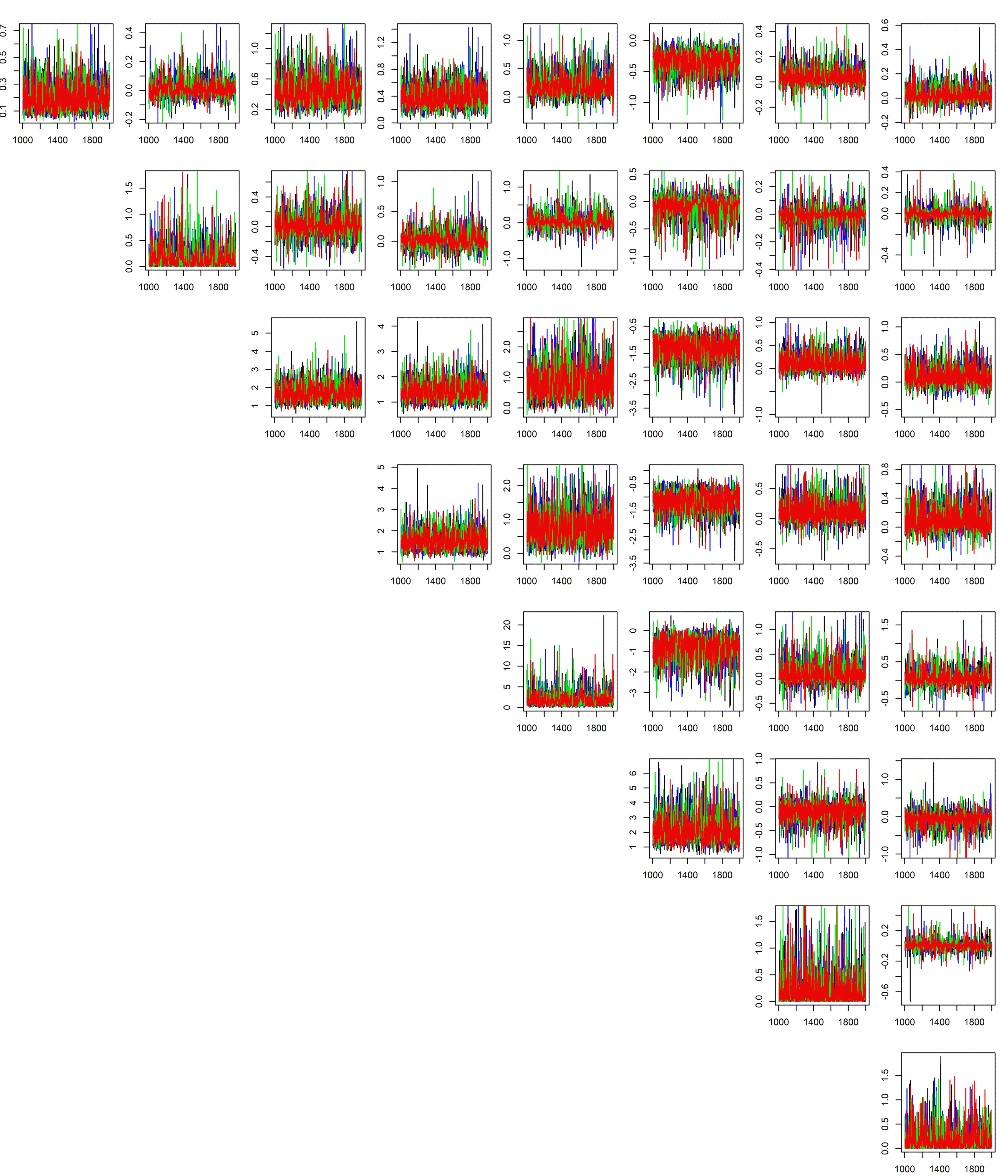}
\end{center}
\caption{Trace plots for the variances and covariances of the coregionalization matrix $\bm{\Sigma}_{\epsilon}$. Trace plots show 1000 posterior samples after 1000 warmup steps for four chains for the Cov+LMC(1) model.} \label{fig:cov_stat_trace2}
\end{figure}

\begin{figure}[H]
\begin{center}
\includegraphics[]{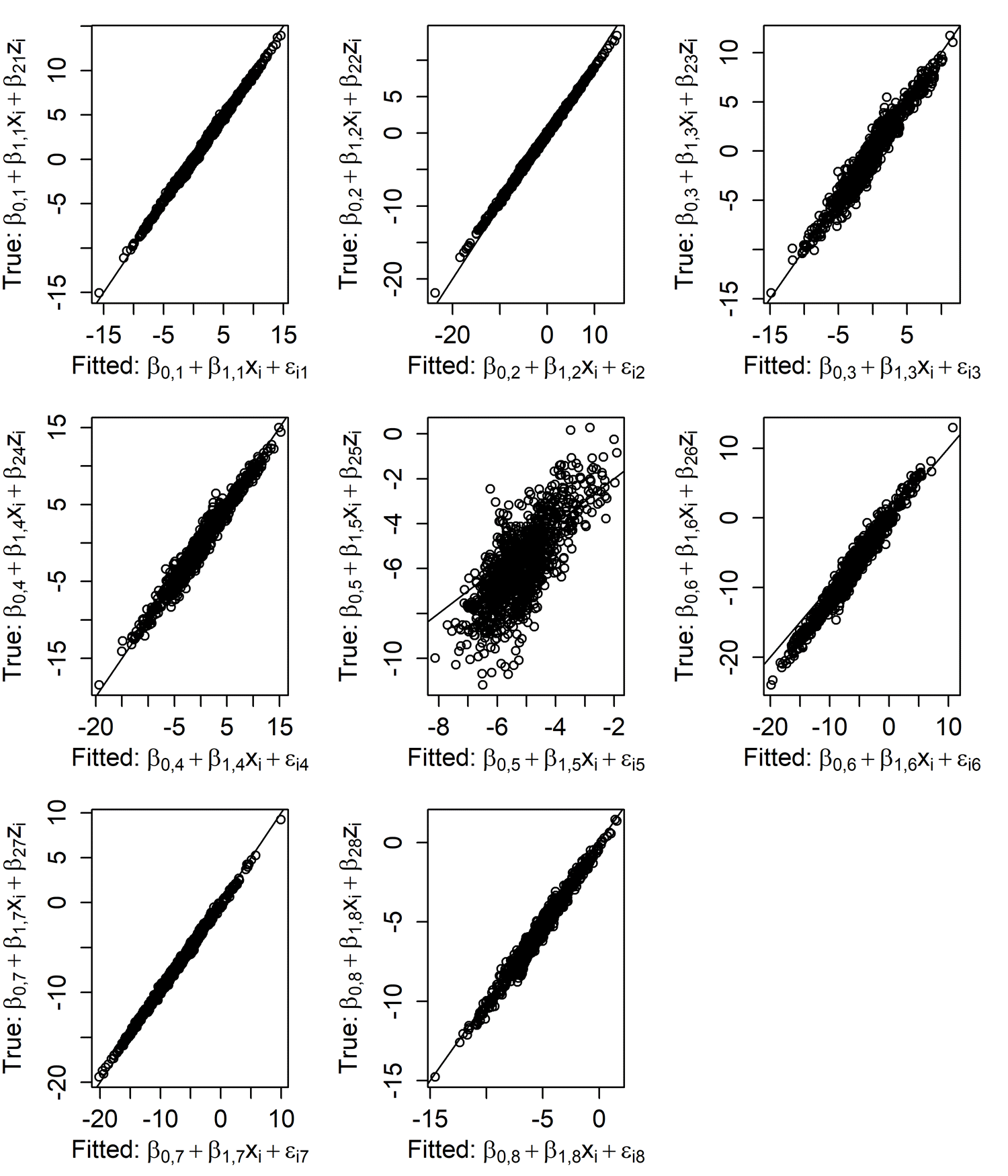}
\end{center}
\caption{Fitted latent variable as predicted by Cov+LMC(1) model vs. true data generating latent variable over the mesh grid locations.} \label{fig:cov_stat_fit_true}
\end{figure}

\begin{figure}[H]
\begin{center}
\includegraphics[]{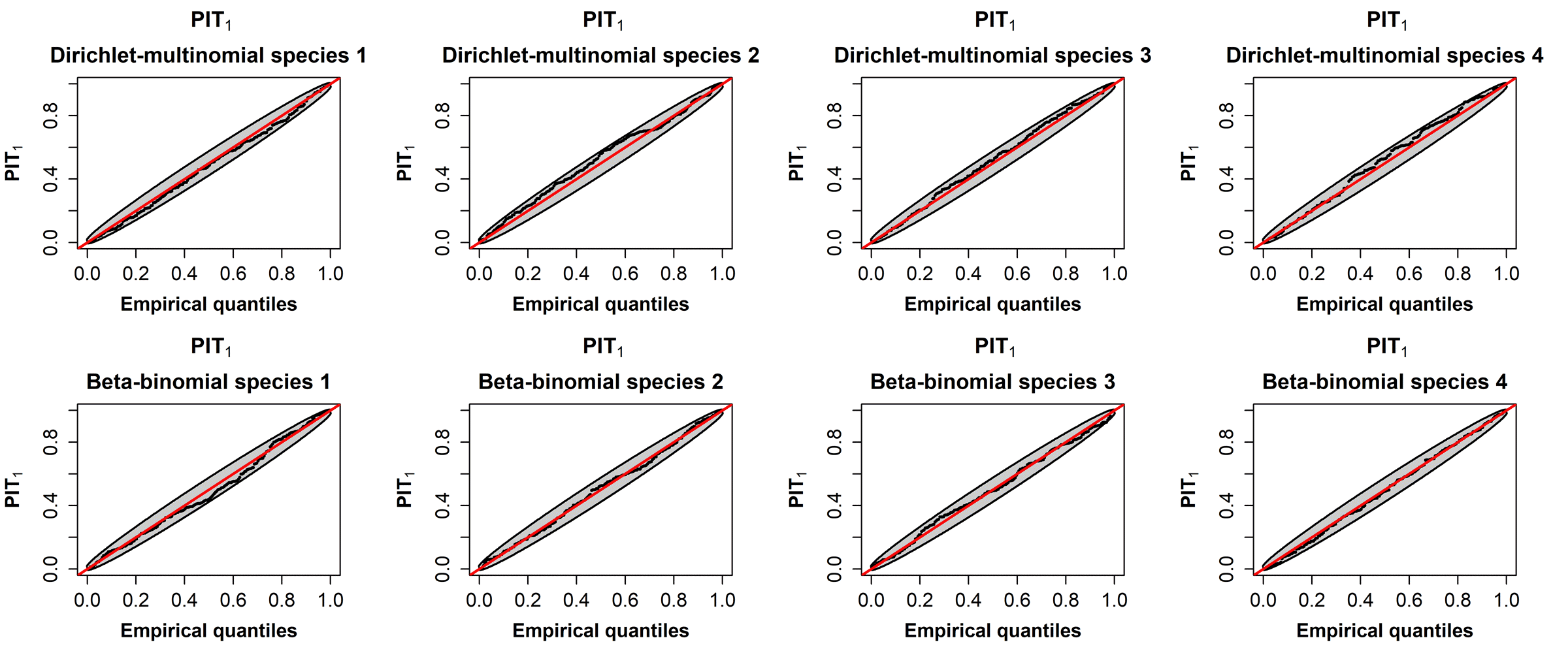}
\end{center}
\caption{Species specific Q-Q plots with point-wise 95$\%$ confidence intervals of the PIT of location-wise cross validation predictive distributions (PIT$_1$) for Cov+LMC(1) model.} \label{fig:cov_stat_PIT1}
\end{figure}

\begin{figure}[H]
\begin{center}
\includegraphics[scale=0.9]{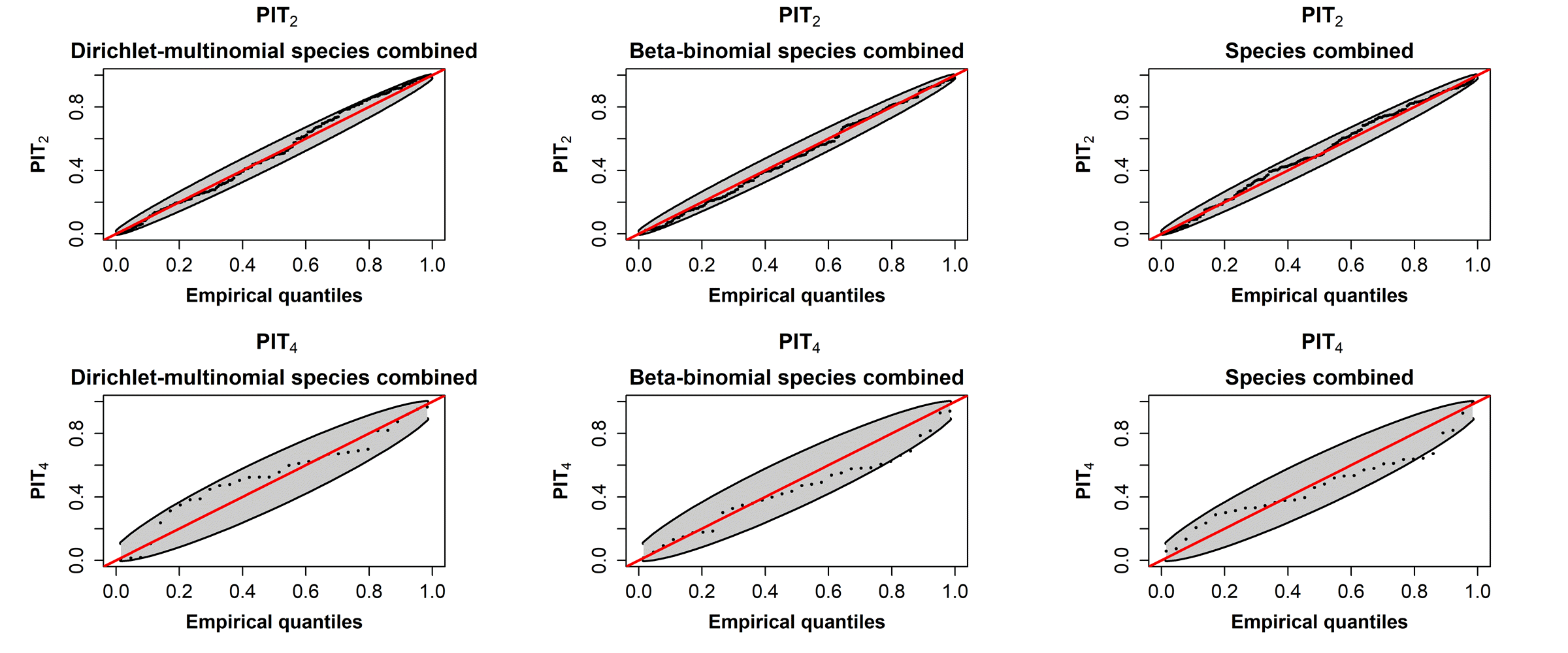}
\end{center}
\caption{Q-Q plots with point-wise 95$\%$ confidence intervals of the PIT of location-wise total over species (PIT$_2$) and CV-fold-wise total over species and locations predictive distributions (PIT$_4$) for Cov+LMC(1) model.} \label{fig:cov_stat_PIT24}
\end{figure}

\begin{figure}[H]
\begin{center}
\includegraphics[scale=0.9]{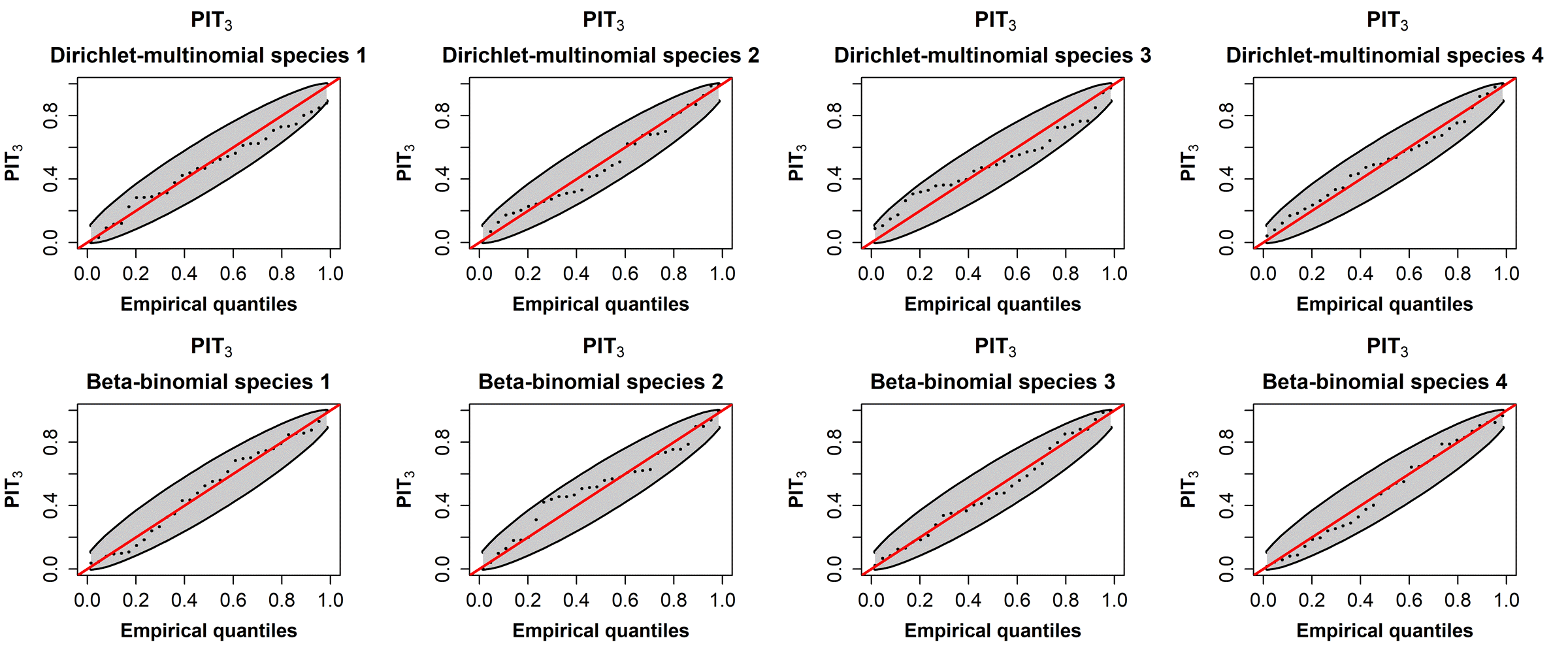}
\end{center}
\caption{Species specific Q-Q plots with point-wise 95$\%$ confidence intervals of the PIT of total over locations (within a CV-fold) predictive distributions (PIT$_3$) for Cov+LMC(1) model.} \label{fig:cov_stat_PIT3}
\end{figure}

\begin{figure}[H]
\begin{center}
\includegraphics[]{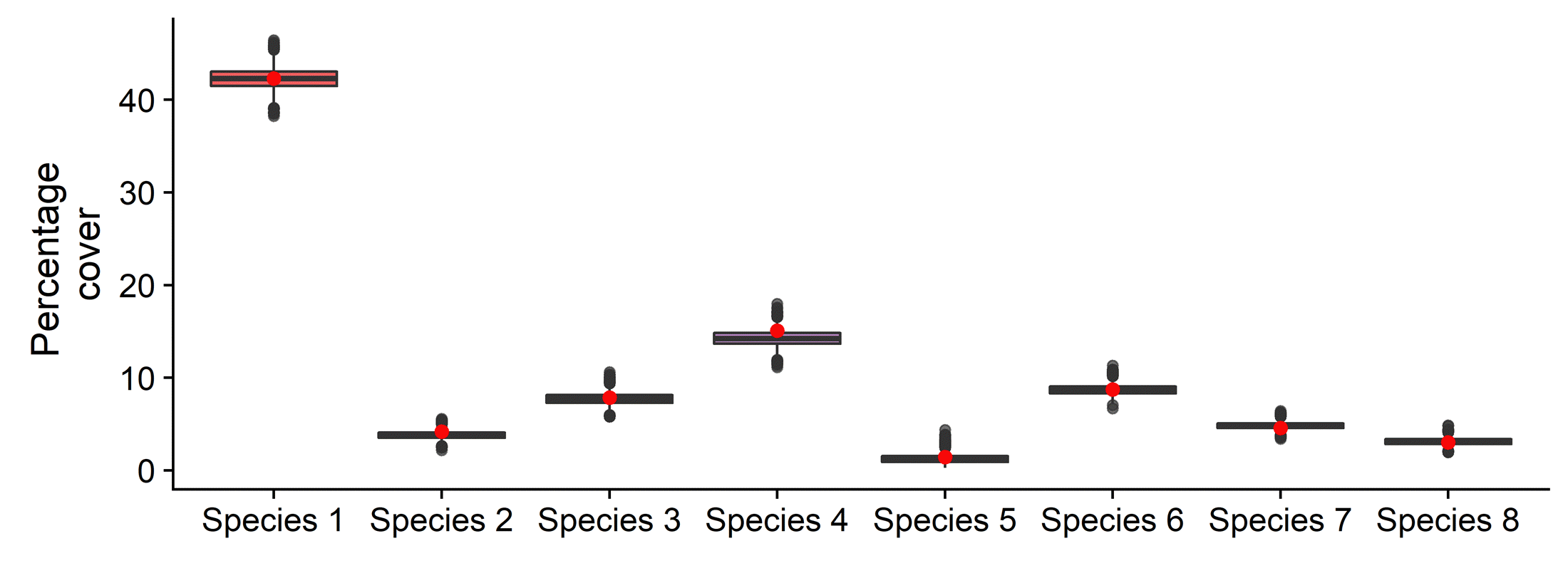}
\end{center}
\caption{Posterior distributions for total percentage covers over the study area as predicted by Cov+LMC(1) model. Red dots show the simulated true value.} \label{fig:cov_stat_pred}
\end{figure}

\begin{figure}[H]
\begin{center}
\includegraphics[]{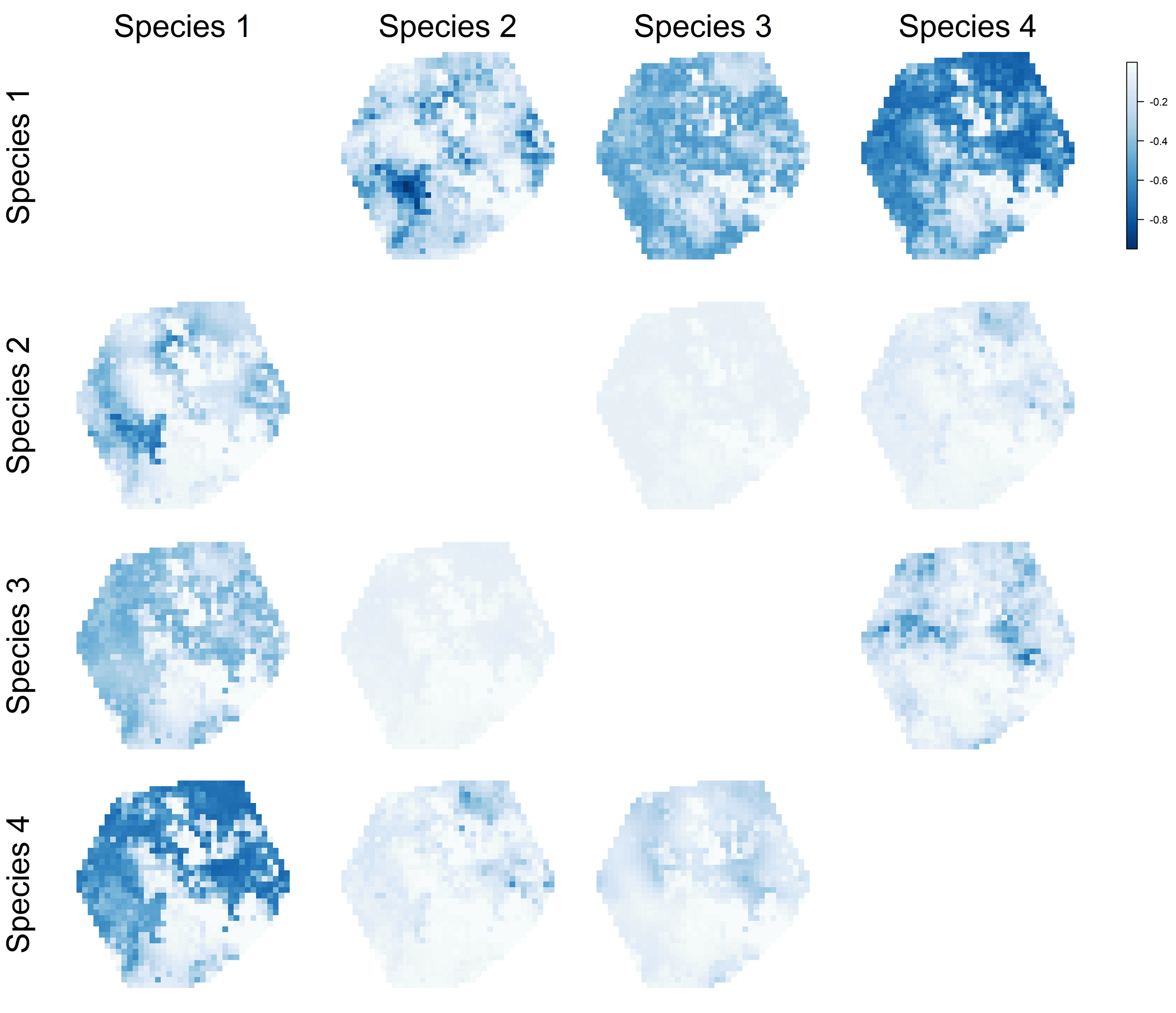}
\end{center}
\caption{The interspecific competition between mutually exclusive species as estimated by Cov+LMC(1) model. The maps on the lower left triangle show the spatial distribution of the estimated interspecific correlation in percentage covers for all pairs of the mutually exclusive species. The maps on the upper right triangle show the true spatial distribution of the interspecific correlation in percentage covers as calculated from the simulated $\alpha$ values.} \label{fig:cov_stat_cor}
\end{figure}


\subsection{Inference using LMC(1)$_{\text{NS}}$}

Latent process for species $j$ at location $\mathbf{s}_{i}$ was modelled as
\setlength{\abovedisplayskip}{15pt}
\setlength{\belowdisplayskip}{15pt}
\begin{equation}
f_{ij}=\beta_{0,j}+\epsilon_{ij},
\end{equation}
where $\beta_{0,j}$ is an intercept for species $j$ and $\epsilon_{ij}$ is a non-stationary Gaussian random effect for species $j$ at location $\mathbf{s}_{i}$. Dependencies are modelled through linear model of coregionalization with one distinct covariance function (k=1).

\begin{figure}[H]
\begin{center}
\includegraphics[]{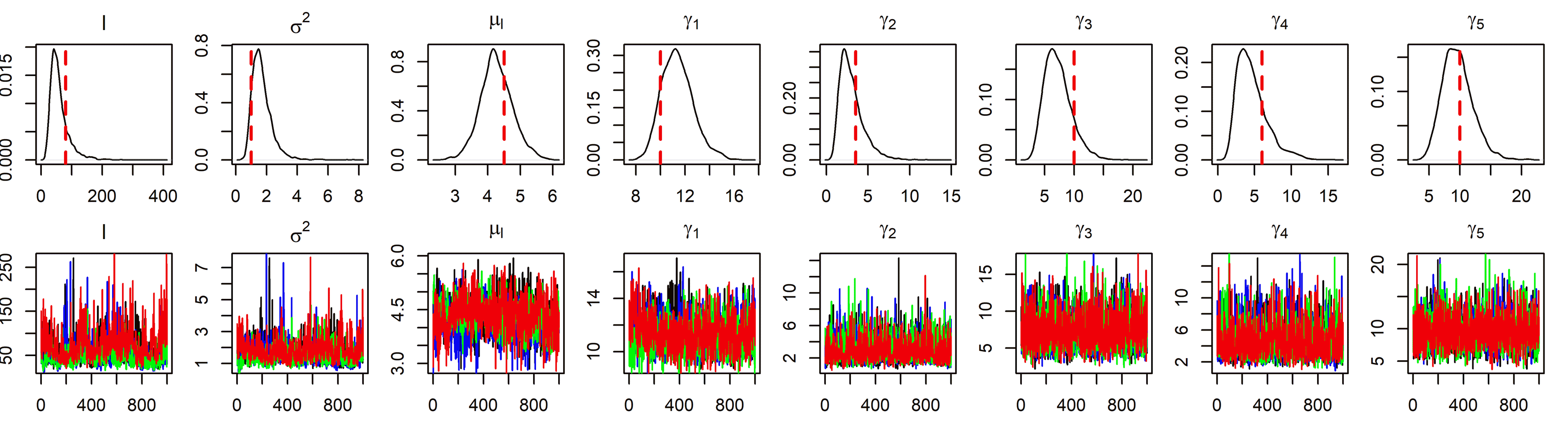}
\end{center}
\caption{Posterior distributions and trace plots as estimated by LMC(1)$_{\text{NS}}$ model for the hyperparameters of the covariance function, the hyperparameter of the Dirichlet distribution and the hyperparameters of the Beta distributions. The red dashed line shows the true parameter used in simulating the data.  Trace plots show 1000 posterior samples after 1000 warmup steps for four chains.} \label{fig:nonstat_ICM_post}
\end{figure}

\begin{figure}[H]
\begin{center}
\includegraphics[]{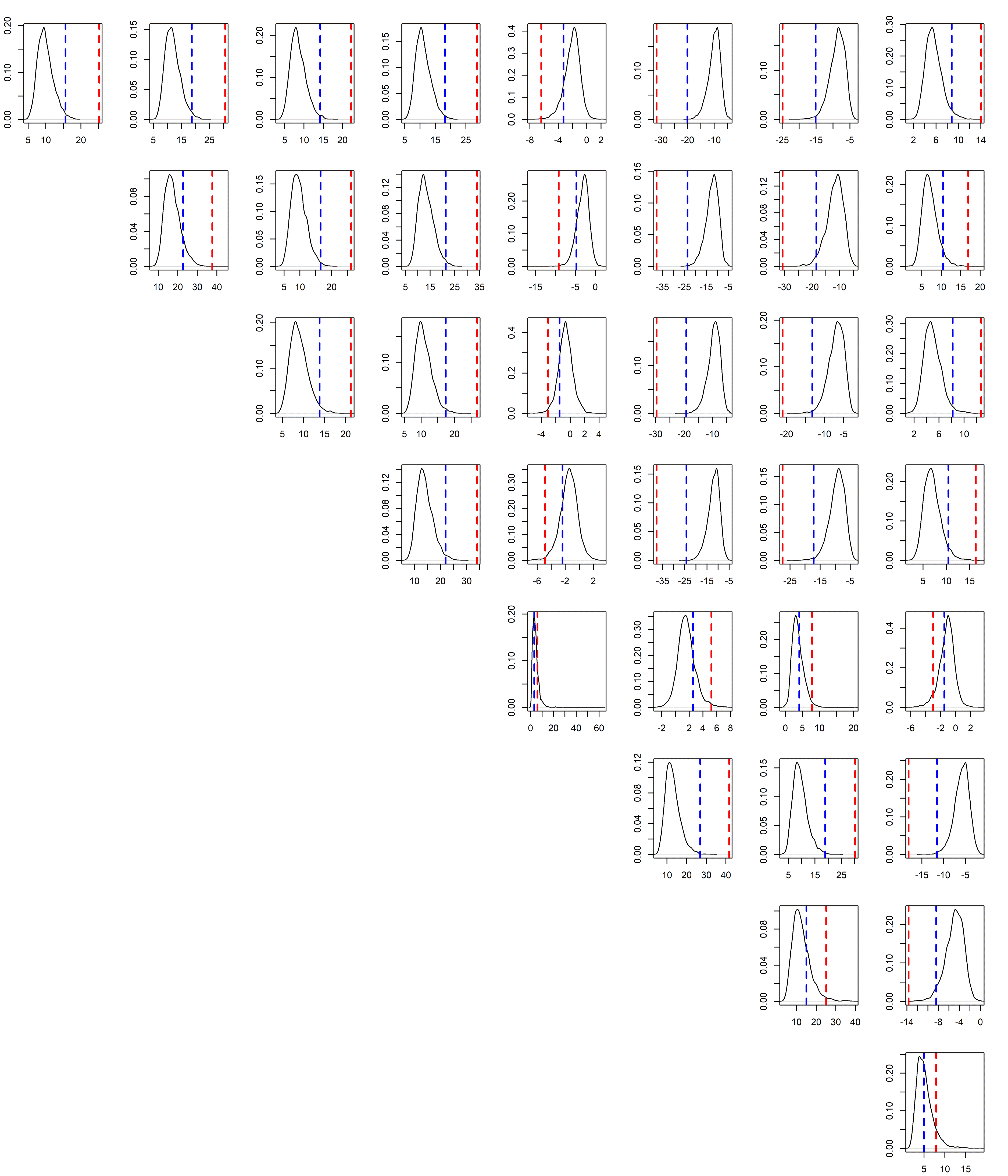}
\end{center}
\caption{Estimated variances and covariances of the coregionalization matrix $\bm{\Sigma}_{\epsilon}$ as estimated by LMC(1)$_{\text{NS}}$ model. The red dashed lines show the true interspecific covariances induced by the covariates $x$ and $z$; that is, the elements of $\bm{BB}^T$ where $\bm{B}$ is a $J\times2$ matrix with $j$'th row $\bm{B}_{j\cdot}=[\beta_{1,j},\beta_{2,j}]$. The blue dashed lines show the sample covariances between the 200 simulated species specific latent variables $\beta_{1,j}x+\beta_{2,j}z$ and $\beta_{1,j'}x+\beta_{2,j'}z$.}  \label{fig:nonstat_ICM_covariance}
\end{figure}

\begin{figure}[H]
\begin{center}
\includegraphics[]{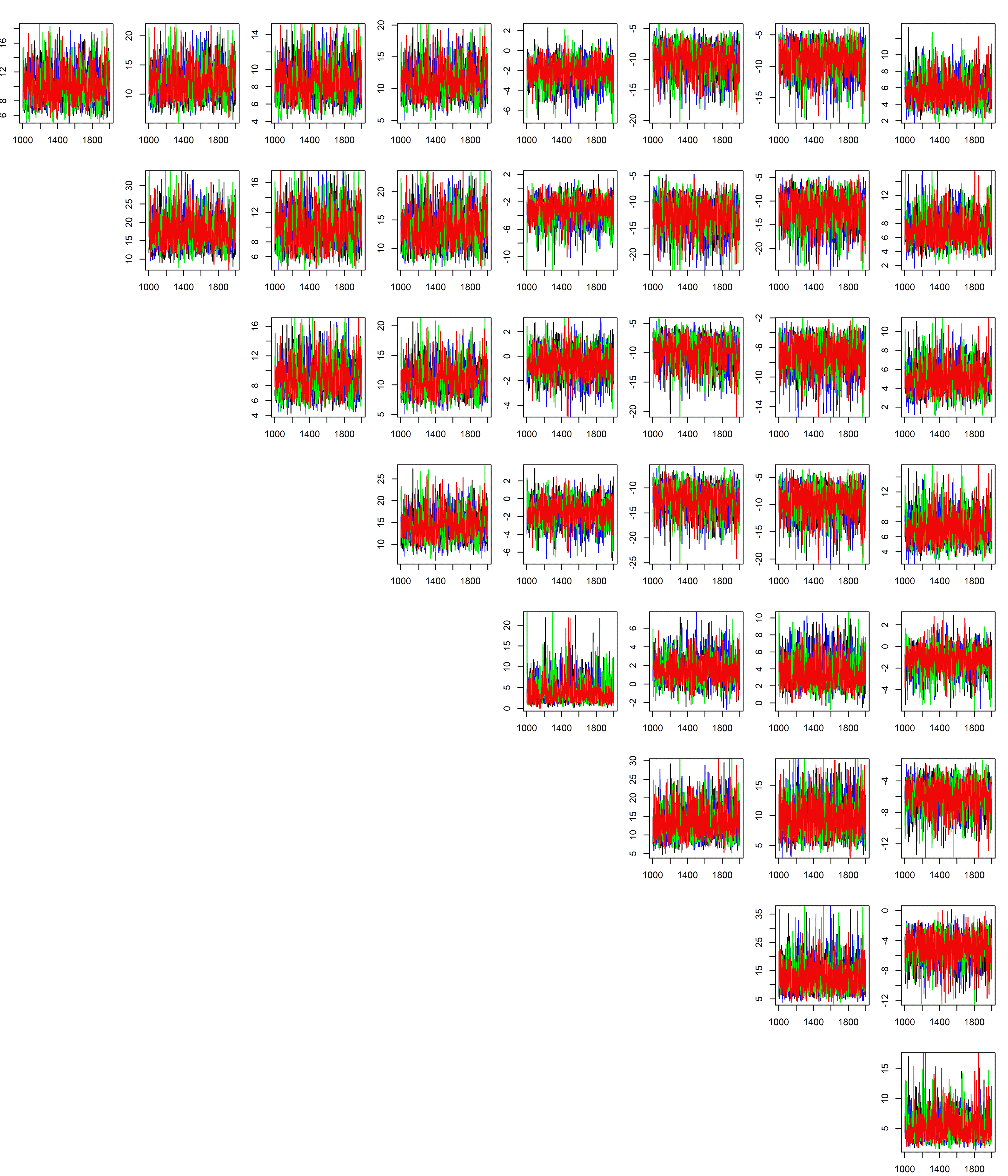}
\end{center}
\caption{Trace plots for the variances and covariances of the coregionalization matrix $\bm{\Sigma}_{\epsilon}$. Trace plots show 1000 posterior samples after 1000 warmup steps for four chains for the LMC(1)$_{\text{NS}}$ model.}  \label{fig:nonstat_ICM_trace2}
\end{figure}

\begin{figure}[H]
\begin{center}
\includegraphics[]{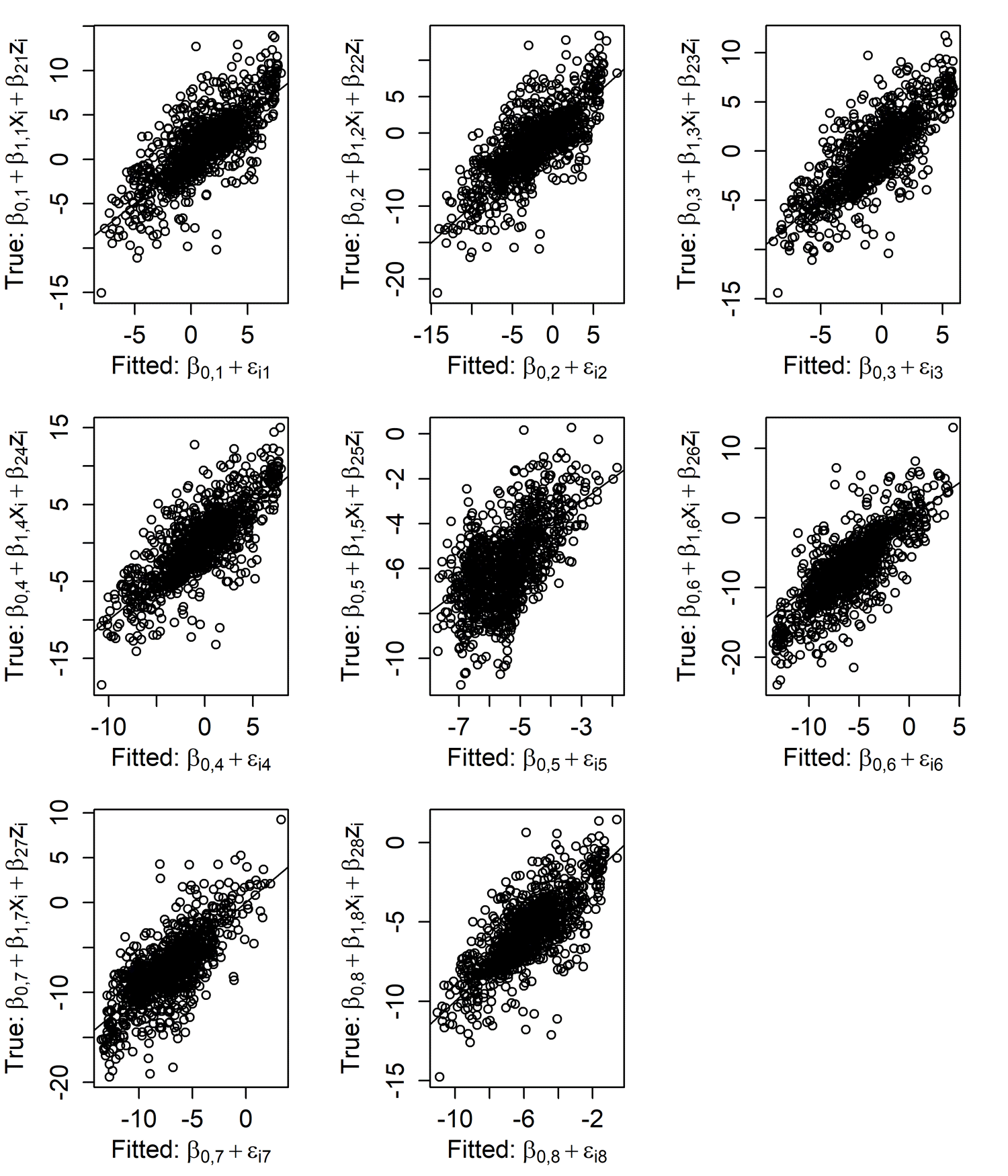}
\end{center}
\caption{Fitted latent variable as predicted by LMC(1)$_{\text{NS}}$ model vs. true data generating latent variable over the grid locations.} \label{fig:nonstat_ICM_fit_true}
\end{figure}

\begin{figure}[H]
\begin{center}
\includegraphics[]{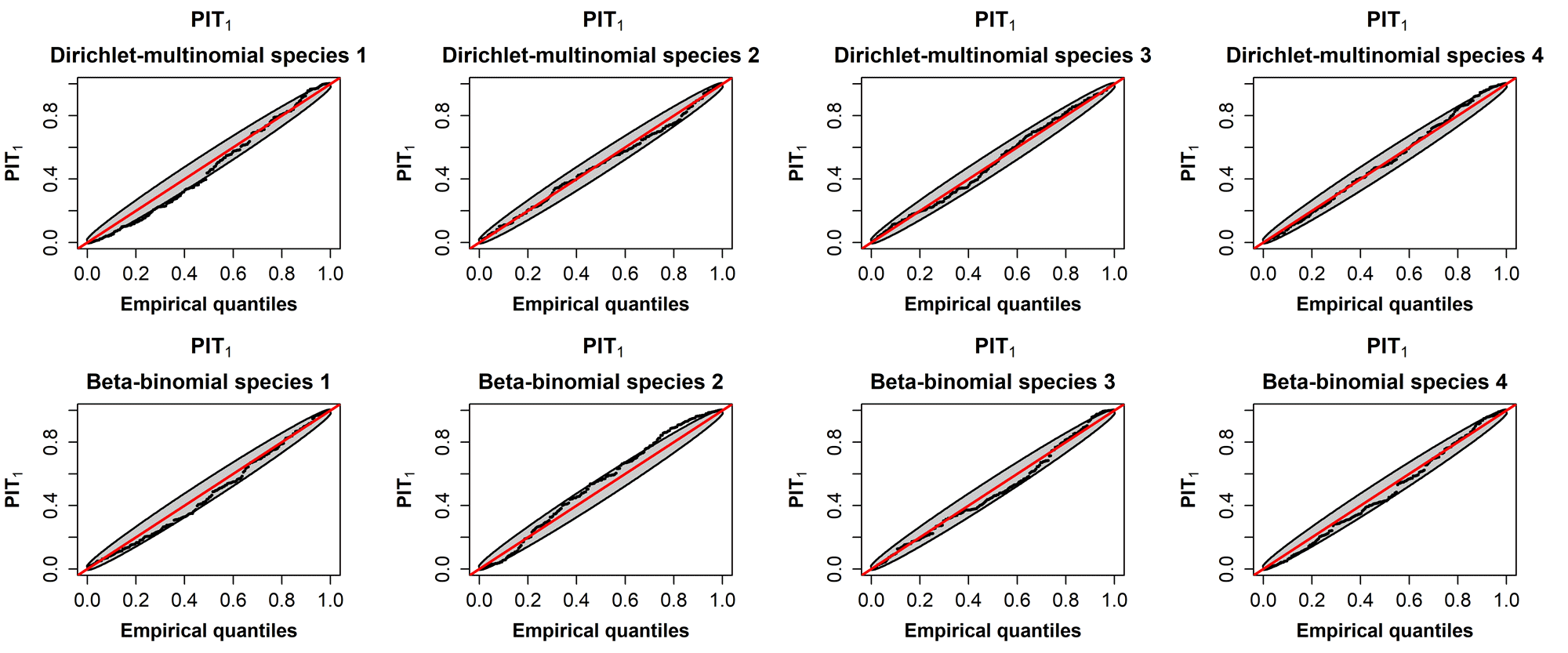}
\end{center}
\caption{Species specific Q-Q plots with point-wise 95$\%$ confidence intervals of the PIT of location-wise cross validation predictive distributions (PIT$_1$) for LMC(1)$_{\text{NS}}$ model.} \label{fig:nonstat_ICM_PIT1}
\end{figure}

\begin{figure}[H]
\begin{center}
\includegraphics[scale=0.90]{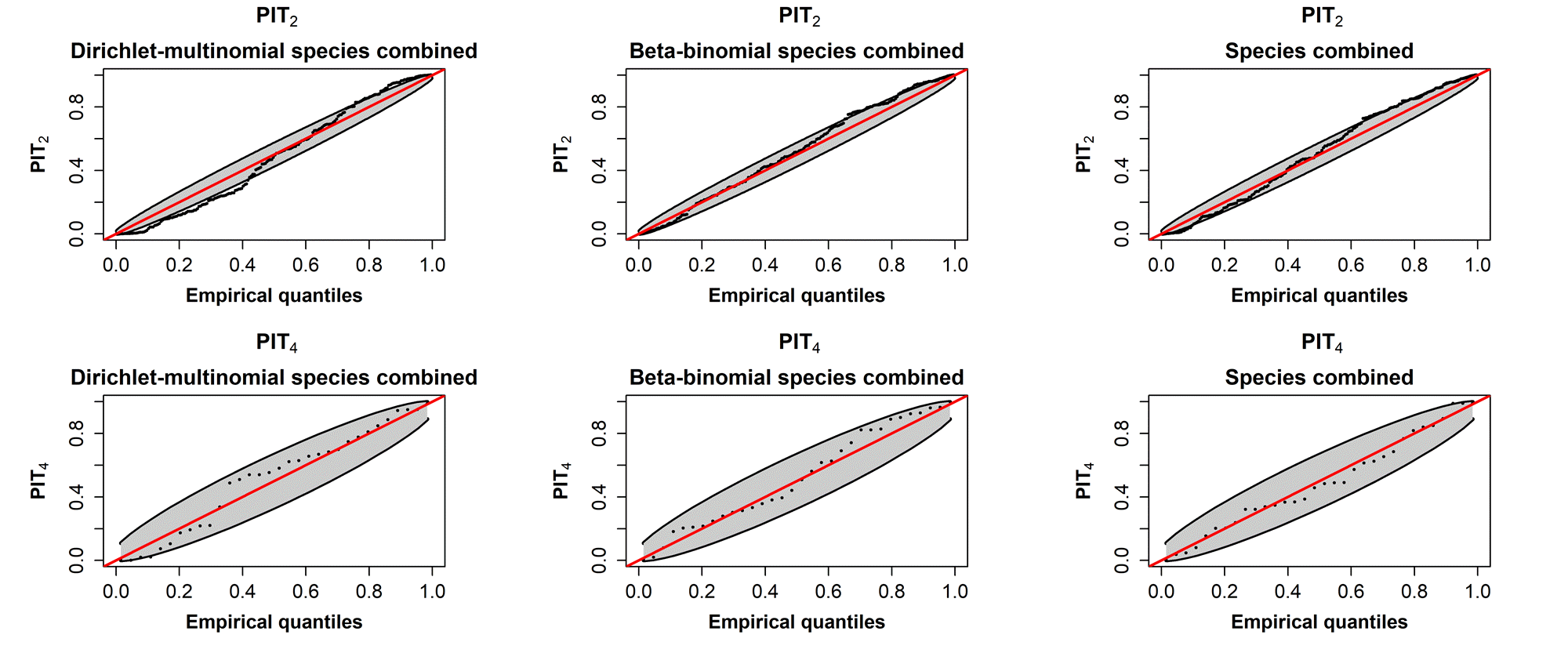}
\end{center}
\caption{Q-Q plots with point-wise 95$\%$ confidence intervals of the PIT of location-wise total over species predictive distributions $PIT_2$ and Q-Q plots with point-wise 95$\%$ confidence intervals of the PIT of CV-foldwise total over species and locations (within a CV-fold) predictive distributions ($PIT_4$) for LMC(1)$_{\text{NS}}$ model.} \label{fig:nonstat_ICM_PIT24}
\end{figure}

\begin{figure}[H]
\begin{center}
\includegraphics[scale=0.9]{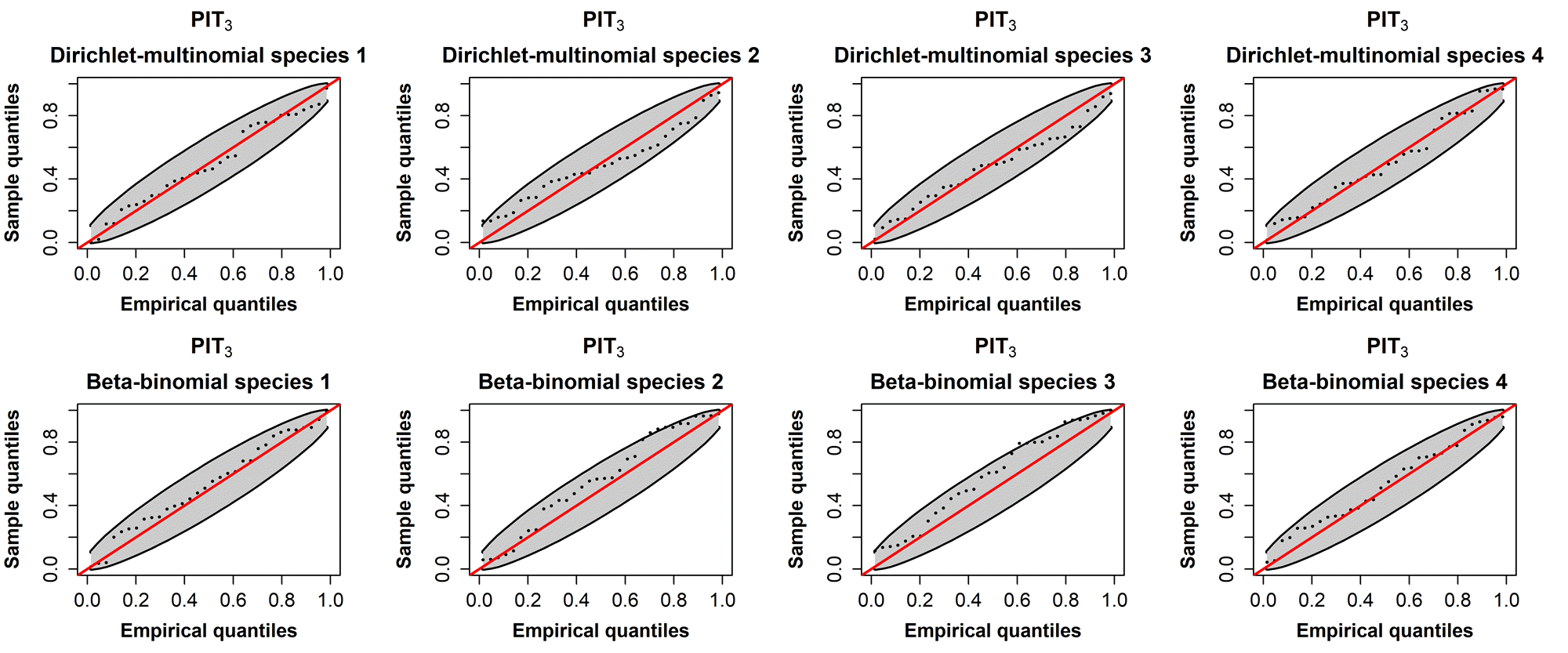}
\end{center}
\caption{Species specific Q-Q plots with point-wise 95$\%$ confidence intervals of the PIT of total over locations (within a CV-fold) predictive distributions $PIT_3$ for LMC(1)$_{\text{NS}}$ model.} \label{fig:nonstat_ICM_PIT3}
\end{figure}

\begin{figure}[H]
\begin{center}
\includegraphics[]{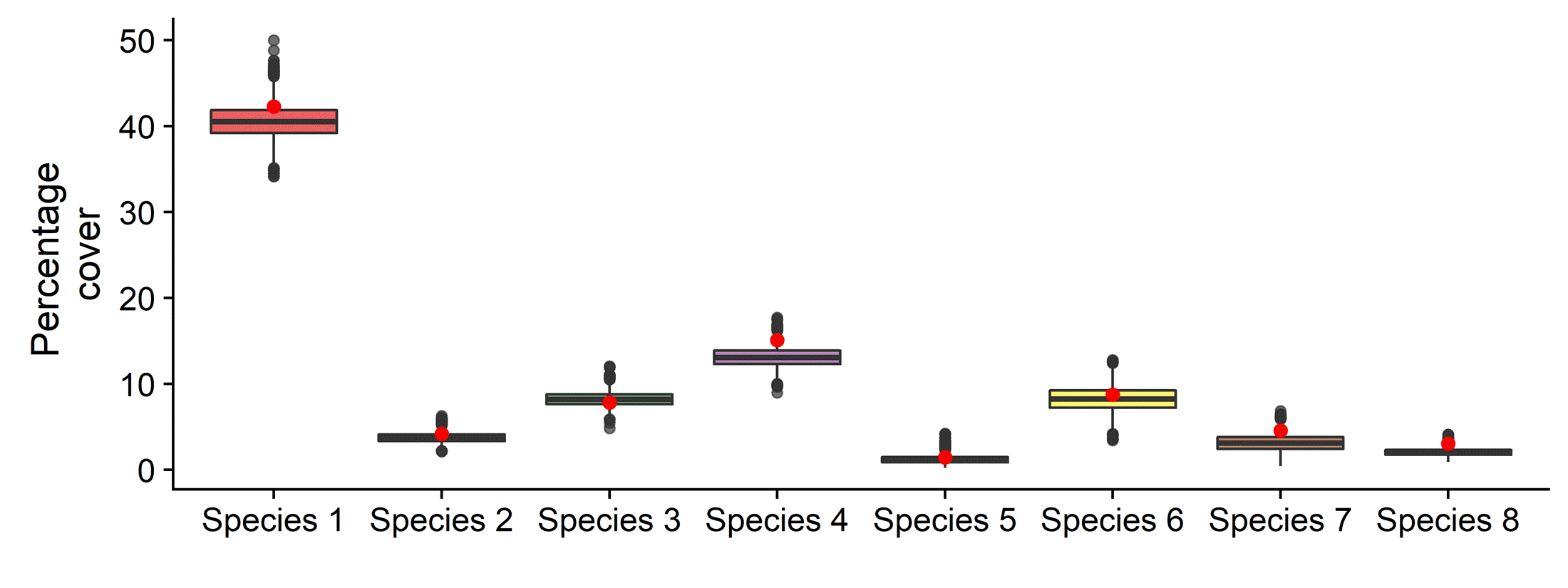}
\end{center}
\caption{Posterior distributions for total percentage covers over the study area as predicted by the LMC(1)$_{\text{NS}}$ model. Red dots show the simulated true value.} \label{fig:nonstat_ICM_pred}
\end{figure}

\begin{figure}[H]
\begin{center}
\includegraphics[]{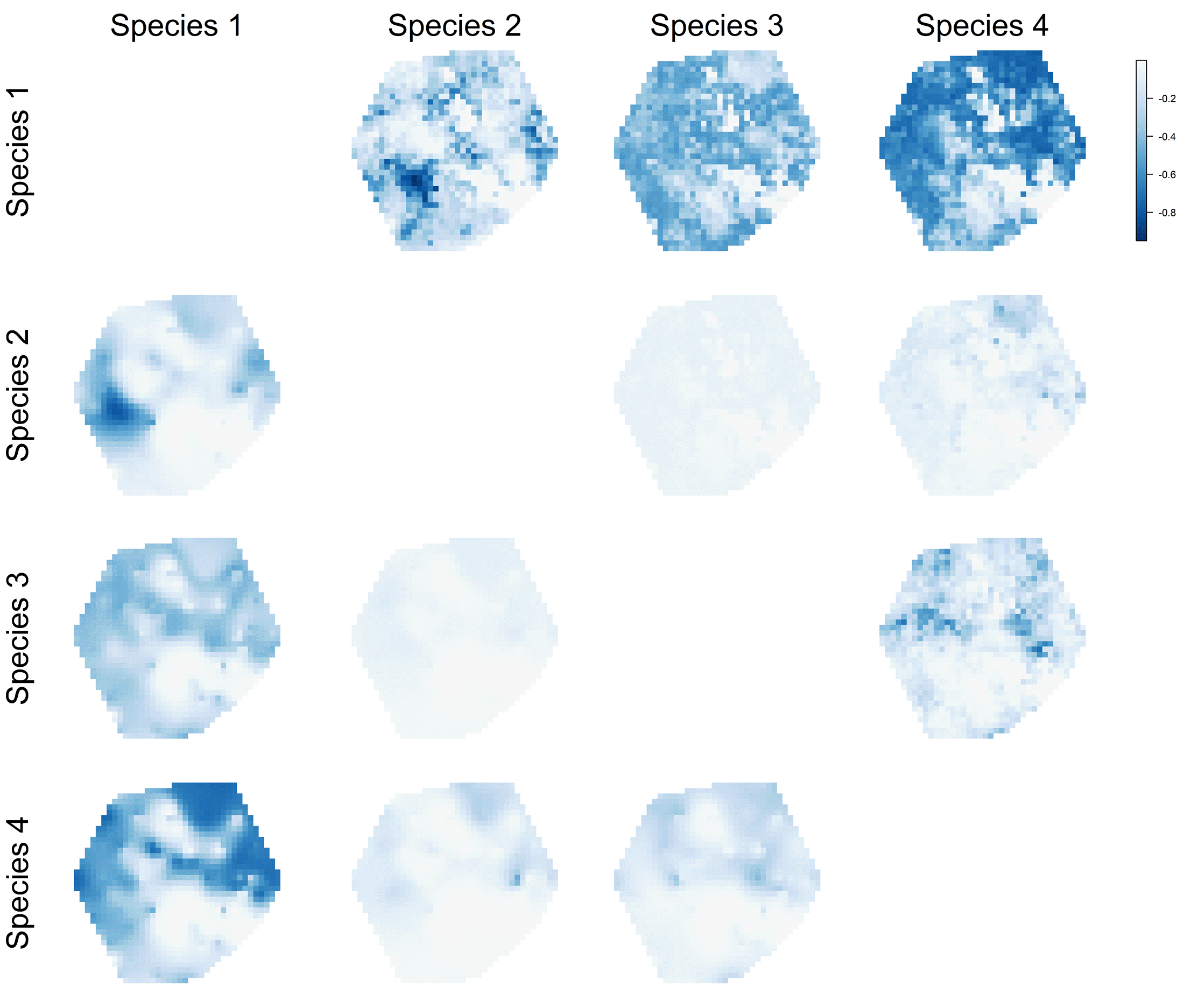}
\end{center}
\caption{The interspecific competition between mutually exclusive species as estimated by LMC(1)$_{\text{NS}}$ model. The maps on the lower left triangle show the spatial distribution of the estimated interspecific correlation in percentage covers for all pairs of the mutually exclusive species. The maps on the upper right triangle show the true spatial distribution of the interspecific correlation in percentage covers.} \label{fig:nonstat_ICM_cor}
\end{figure}

\subsection{Inference using $LMC(1)_{S}$}

Latent process for species $j$ at location $\mathbf{s}_{i}$ was modelled as
\setlength{\abovedisplayskip}{15pt}
\setlength{\belowdisplayskip}{15pt}
\begin{equation}
f_{ij}=\beta_{0,j}+\epsilon_{ij},
\end{equation}
where $\beta_{0,j}$ is an intercept for species $j$ and $\epsilon_{ij}$ is a stationary Gaussian random effect for species $j$ at location $\mathbf{s}_{i}$. Dependencies are modelled through linear model of
coregionalization with one distinct covariance function (k=1).

\begin{figure}[H]
\begin{center}
\includegraphics[]{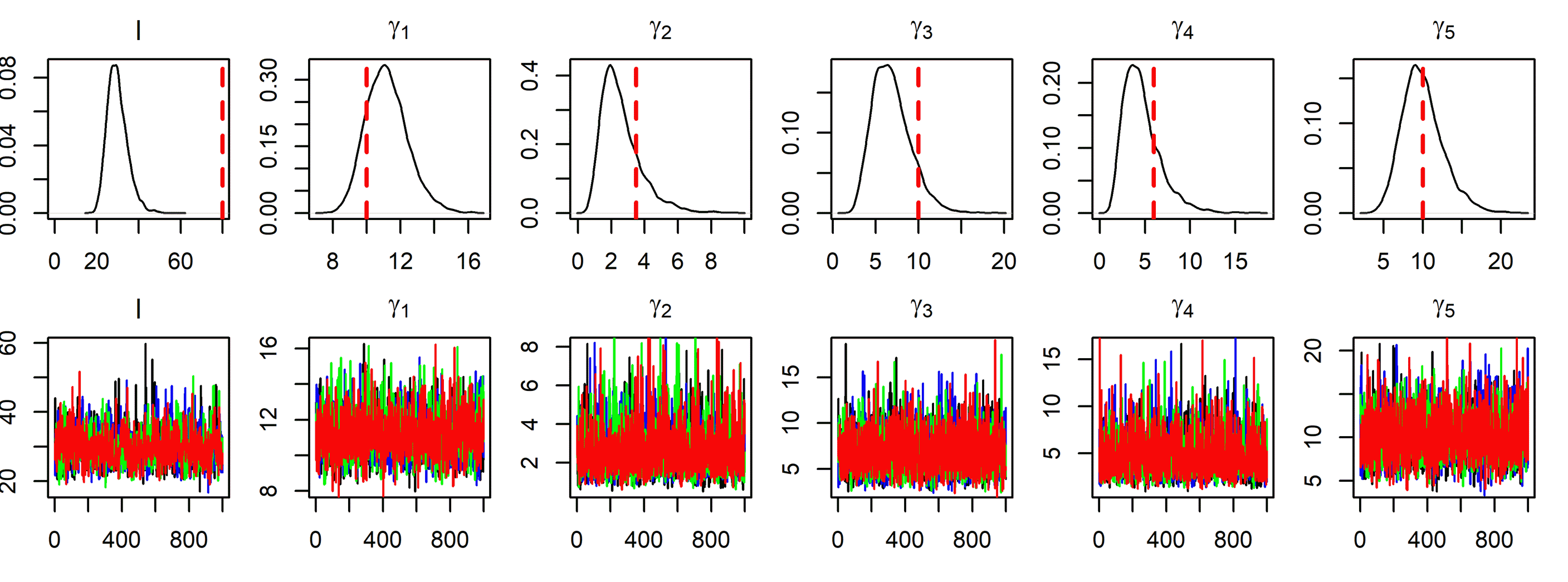}
\end{center}
\caption{Posterior distributions and trace plots as estimated by LMC(1)$_{\text{S}}$ model for the hyperparameters of the covariance function, the hyperparameter of the Dirichlet distribution and the hyperparameters of the Beta distributions. The red dashed line shows the true parameter used in simulating the data.  Trace plots show 1000 posterior samples after 1000 warmup steps for four chains.} \label{fig:stat_ICM_post}
\end{figure} 

\begin{figure}[H]
\begin{center}
\includegraphics[]{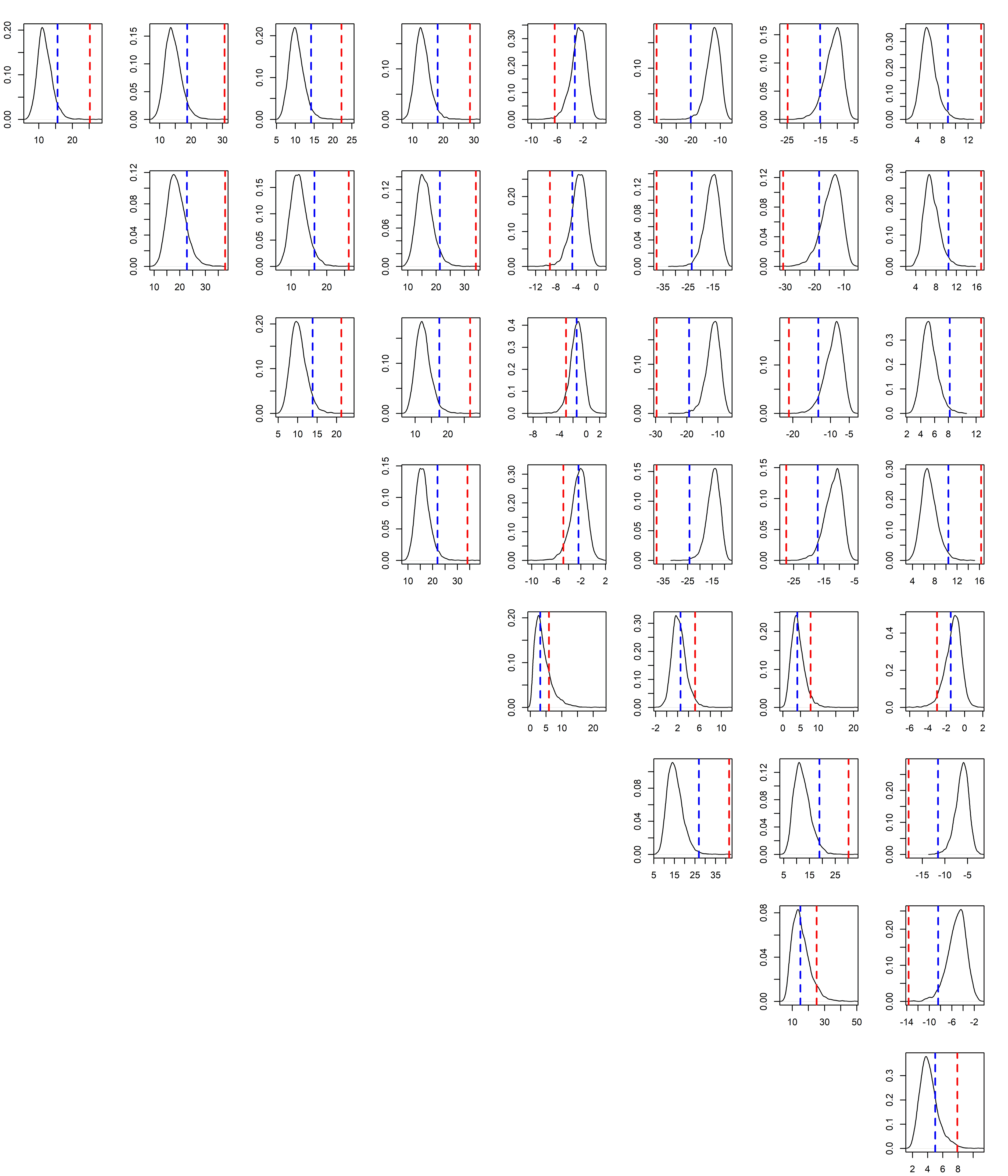}
\end{center}
\caption{Estimated variances and covariances of the coregionalization matrix $\bm{\Sigma}_{\epsilon}$ as estimated by LMC(1)$_{\text{S}}$ model. The red dashed lines show the true interspecific covariances induced by the covariates $x$ and $z$; that is, the elements of $\bm{BB}^T$ where $\bm{B}$ is a $J\times2$ matrix with $j$'th row $\bm{B}_{j\cdot}=[\beta_{1,j},\beta_{2,j}]$. The blue dashed lines show the sample covariances between the 200 simulated species specific latent variables $\beta_{1,j}x+\beta_{2,j}z$ and $\beta_{1,j'}x+\beta_{2,j'}z$. } \label{fig:stat_ICM_covariance}
\end{figure}

\begin{figure}[H]
\begin{center}
\includegraphics[]{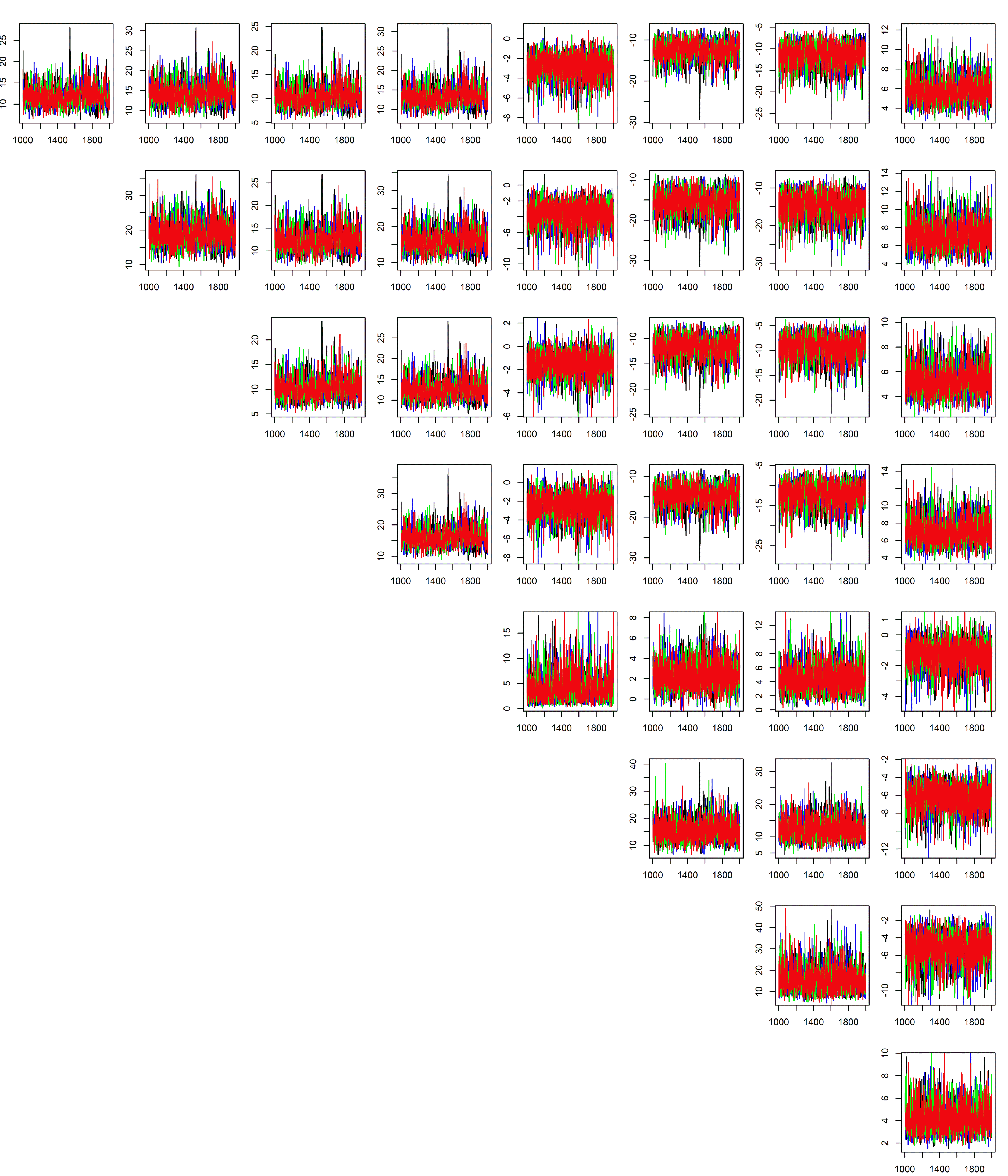}
\end{center}
\caption{Trace plots for the variances and covariances of the coregionalization matrix $\bm{\Sigma}_{\epsilon}$. Trace plots show 1000 posterior samples after 1000 warmup steps for four chains for the LMC(1)$_{\text{S}}$ model.}  \label{fig:stat_ICM_trace2}
\end{figure}

\begin{figure}[H]
\begin{center}
\includegraphics[]{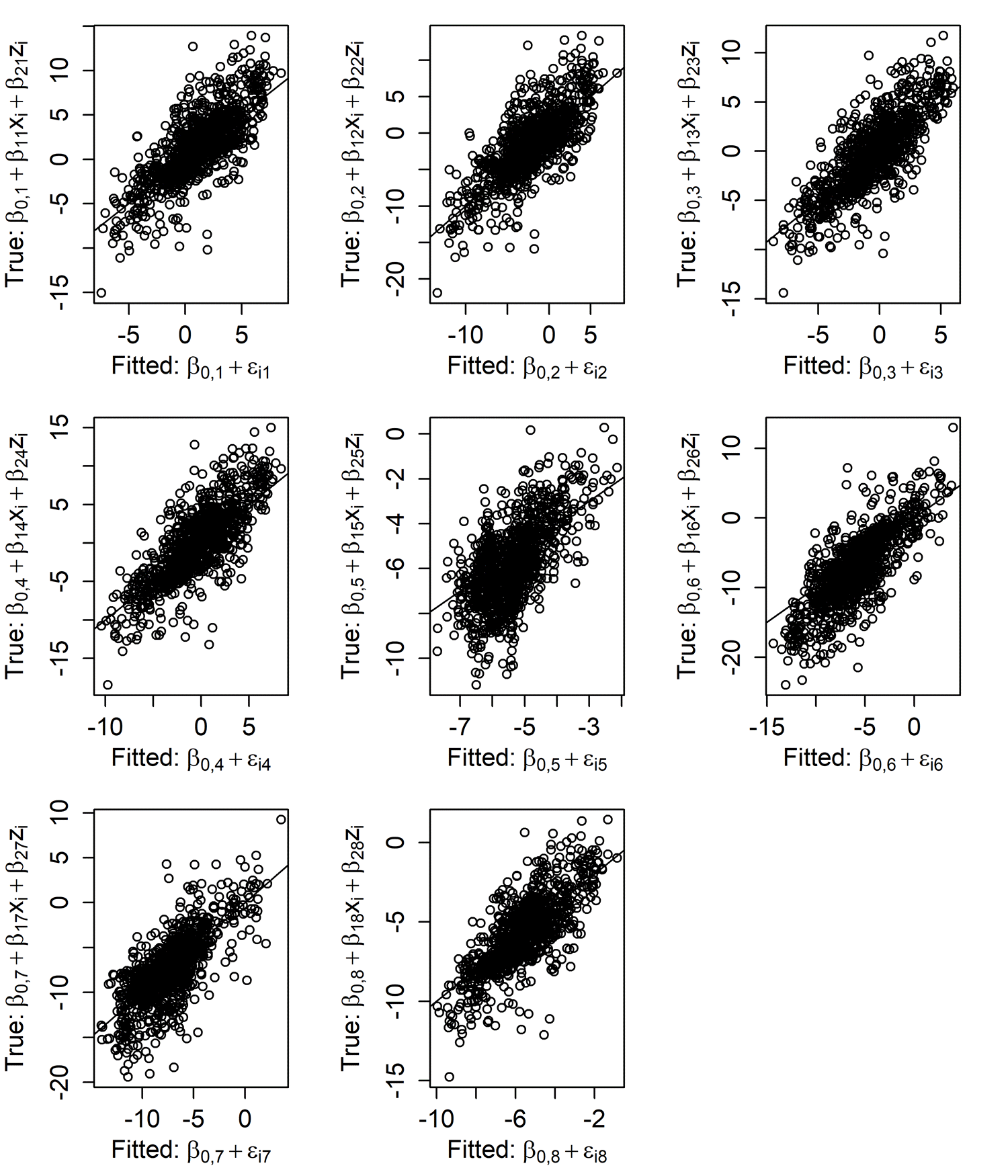}
\end{center}
\caption{Fitted latent variable as predicted by LMC(1)$_{\text{S}}$ model vs. true data generating latent variable over the grid locations.} \label{fig:stat_ICM_fit_true}
\end{figure}

\begin{figure}[H]
\begin{center}
\includegraphics[]{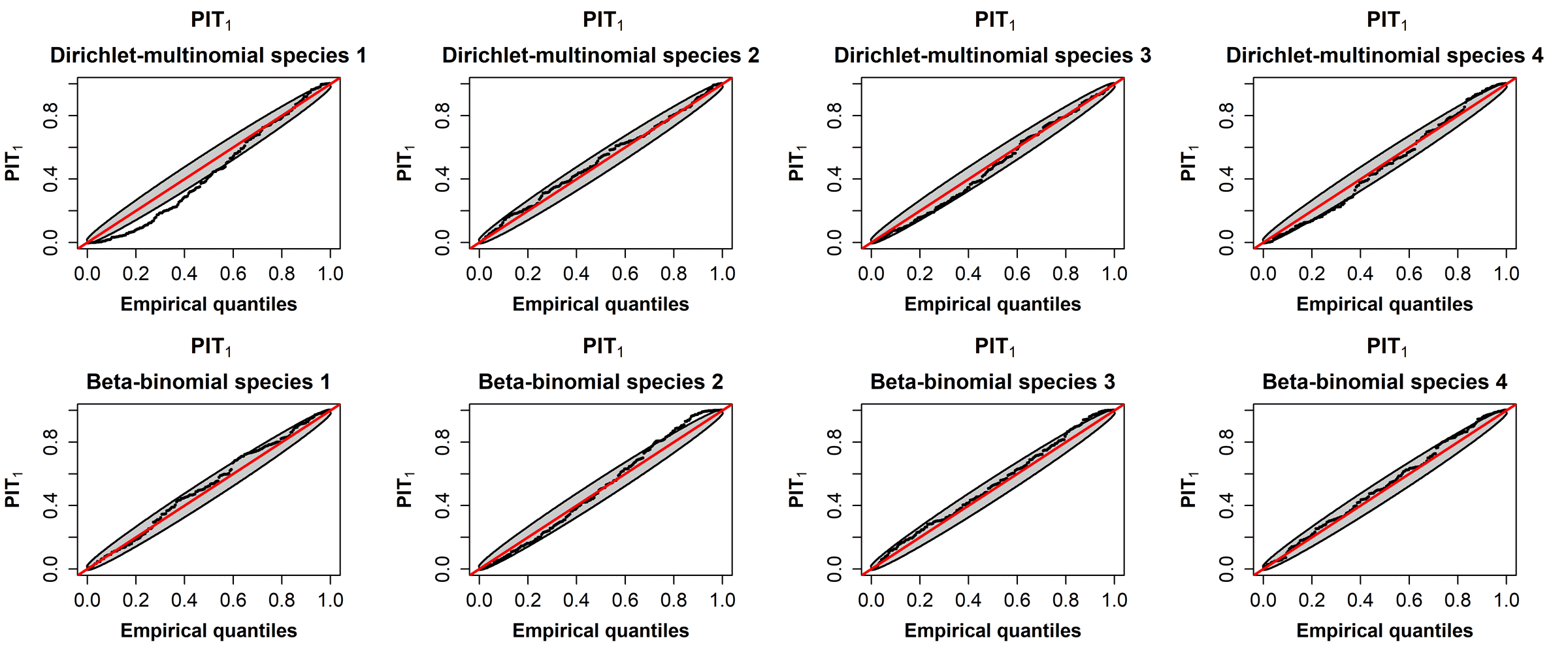}
\end{center}
\caption{Species specific Q-Q plots with point-wise 95$\%$ confidence intervals of the PIT of location-wise cross validation predictive distributions $PIT_1$ for LMC(1)$_{\text{S}}$ model.} \label{fig:stat_ICM_PIT1}
\end{figure}

\begin{figure}[H]
\begin{center}
\includegraphics[scale=0.9]{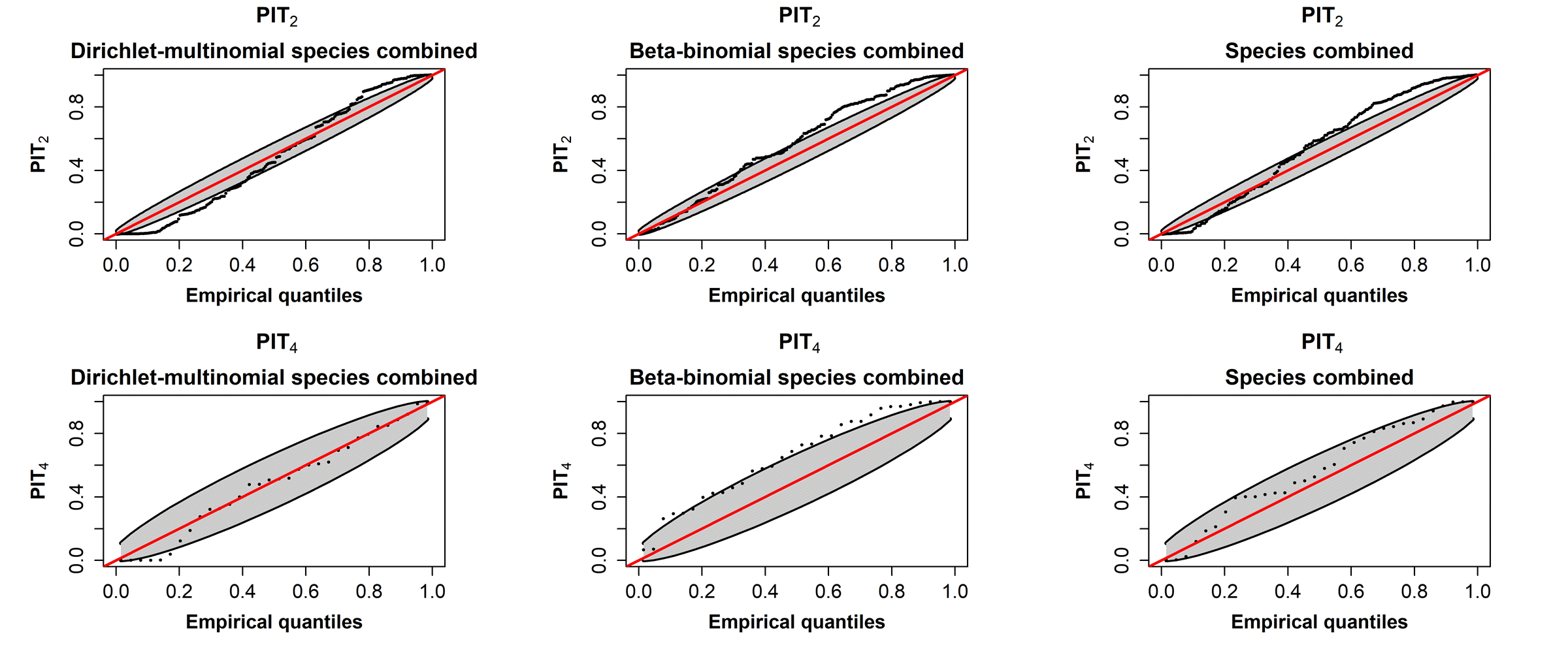}
\end{center}
\caption{Q-Q plots with point-wise 95$\%$ confidence intervals of the PIT of location-wise total over species predictive distributions $PIT_2$ and Q-Q plots with point-wise 95$\%$ confidence intervals of the PIT of CV-foldwise total over species and locations (within a CV-fold) predictive distributions ($PIT_4$) for LMC(1)$_{\text{S}}$ model.} \label{fig:stat_ICM_PIT24}
\end{figure}

\begin{figure}[H]
\begin{center}
\includegraphics[scale=0.9]{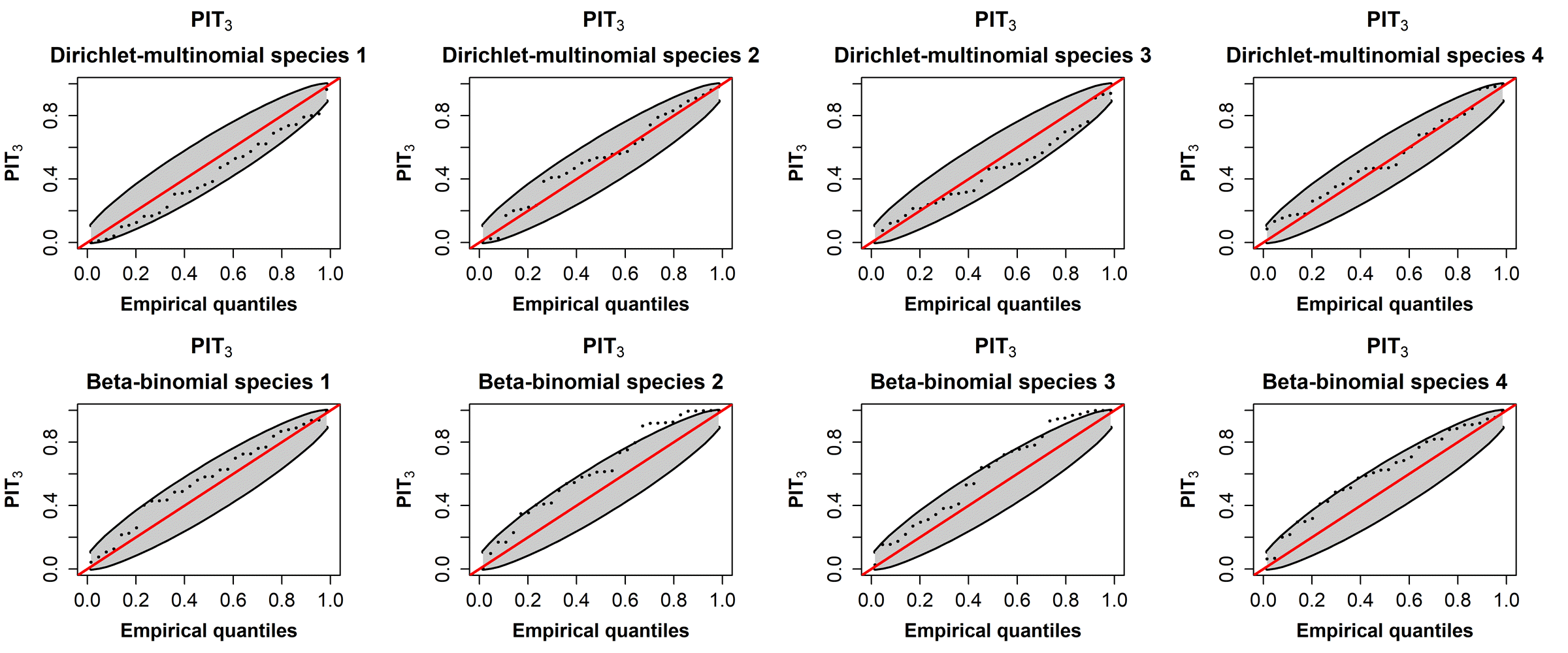}
\end{center}
\caption{Species specific Q-Q plots with point-wise 95$\%$ confidence intervals of the PIT of total over locations (within a CV-fold) predictive distributions $PIT_3$ for LMC(1)$_{\text{S}}$ model.} \label{fig:stat_ICM_PIT3}
\end{figure}

\begin{figure}[H]
\begin{center}
\includegraphics[]{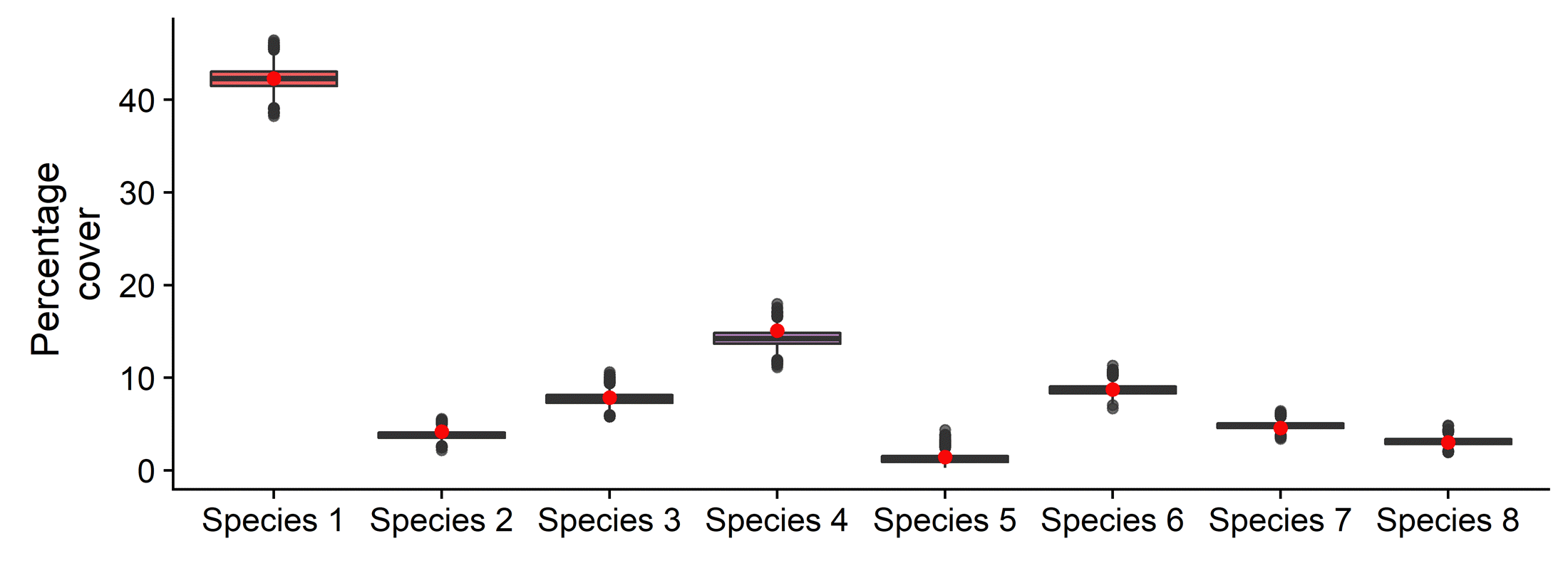}
\end{center}
\caption{Posterior distributions for total percentage covers over the study area as predicted by the LMC(1)$_{\text{S}}$ model. Red dots show the simulated true value.} \label{fig:stat_ICM_pred}
\end{figure}

\begin{figure}[H]
\begin{center}
\includegraphics[]{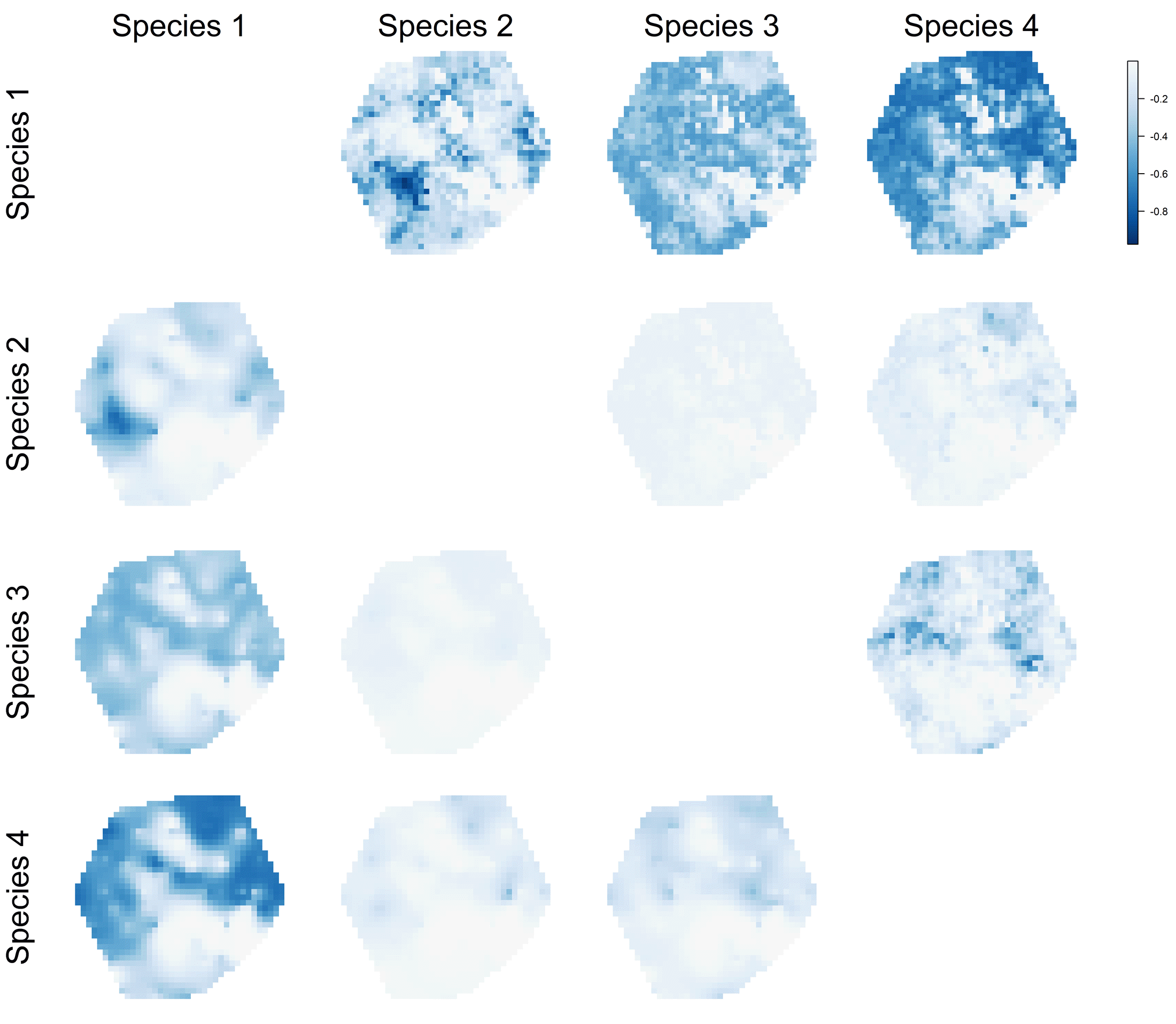}
\end{center}
\caption{The interspecific competition between mutually exclusive species as estimated by LMC(1)$_{\text{S}}$ model. The maps on the lower left triangle show the spatial distribution of the estimated interspecific correlation in percentage covers for all pairs of the mutually exclusive species. The maps on the upper right triangle show the true spatial distribution of the interspecific correlation in percentage covers.} \label{fig:stat_ICM_cor}
\end{figure}

\section{Case study analyses}

Appendix B contains supplementary results for the case study analyses. It includes the following tables and figures:

\begin{enumerate}
\item Figures \ref{fig:DAG_I+BB}-\ref{fig:DAG_LMC+BB}: DAG representations of the alternative models. 
\item Figures \ref{fig:LMC(2)_S+DM-PIT2+PIT4}-\ref{fig:LMC(2)_S+DM_PIT3}: Q-Q plots with point-wise 95$\%$ confidence intervals of the randomized PIT$_{2}$ and PIT$_{4}$ for the LMC(2)$_{S}$+DM model and species specific Q-Q plots with point-wise 95$\%$ confidence intervals of the randomized PIT$_{1}$ and PIT$_{3}$ for LMC(2)$_{S}$+DM and LMC(1)$_{NS}$+DM models.
\item Figure \ref{fig:combined_totals}: Estimated posterior distributions for total vegetation covers for each model.
\item Figure \ref{fig:spatial-correlations_ICM_BB}: Posterior mean estimates of the  interspecific correlations for LMC(1)$_{NS}$+BB model.
\item Tables \ref{table:species-specific-CV1} and \ref{table:species-specific-CV3}: Species specific model comparison results with 10-fold cross-validation using log predictive density utilities $CV_{1}$ and $CV_{3}$.
\end{enumerate}

\setcounter{figure}{0} 
\setcounter{table}{0} 
\renewcommand{\thefigure}{B\arabic{figure}}
\renewcommand{\thetable}{B\arabic{table}}

\begin{figure}[h]
\begin{center}
\includegraphics[]{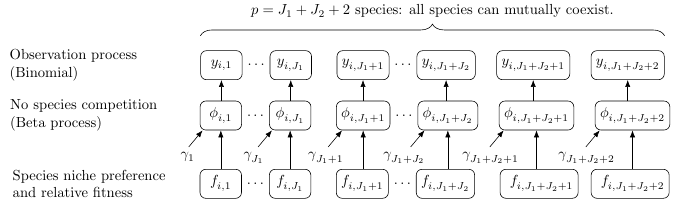}
\end{center}
\caption{DAG of the stacked single species distribution models: C+BB (with $f_{i,j}=\beta_{j}$) and IGP+BB. The model does not contain species interaction in any form. } \label{fig:DAG_I+BB}
\end{figure}

\begin{figure}[h] 
\begin{center}
\includegraphics[]{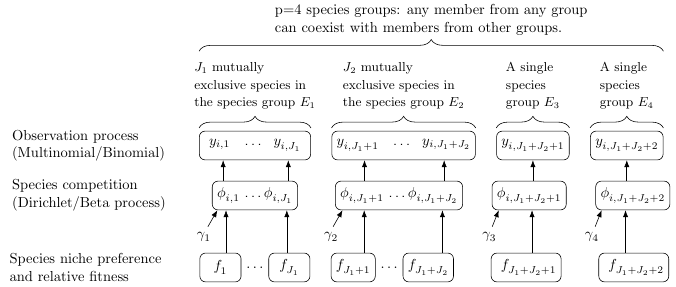}
\end{center}
\caption{DAG of joint species distribution models which include interspecific competition but not include interspecific correlations between site preferences: C+DM and IGP+DM. } \label{fig:DAG_I+DM}
\end{figure}

\begin{figure} 
\begin{center}
\includegraphics[]{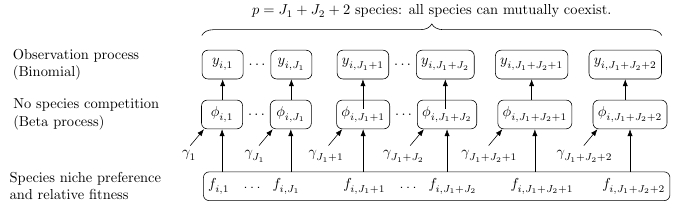}
\end{center}
\caption{The directed acyclic graph (DAG) representation of a joint species distribution model which includes interspecific correlation between site preferences of species but does not include interspecific competition: LMC($k$)+BB. This is the DAG that corresponds to many state-of-the-art JSDMs such as the Hierarchical model of Species Communities (Ovaskainen and Abrego, 2020) and the Additive Multivariate GP model (Vanhatalo et al., 2020). } 
\label{fig:DAG_LMC+BB}
\end{figure}

\begin{figure}
\begin{center}
\includegraphics[]{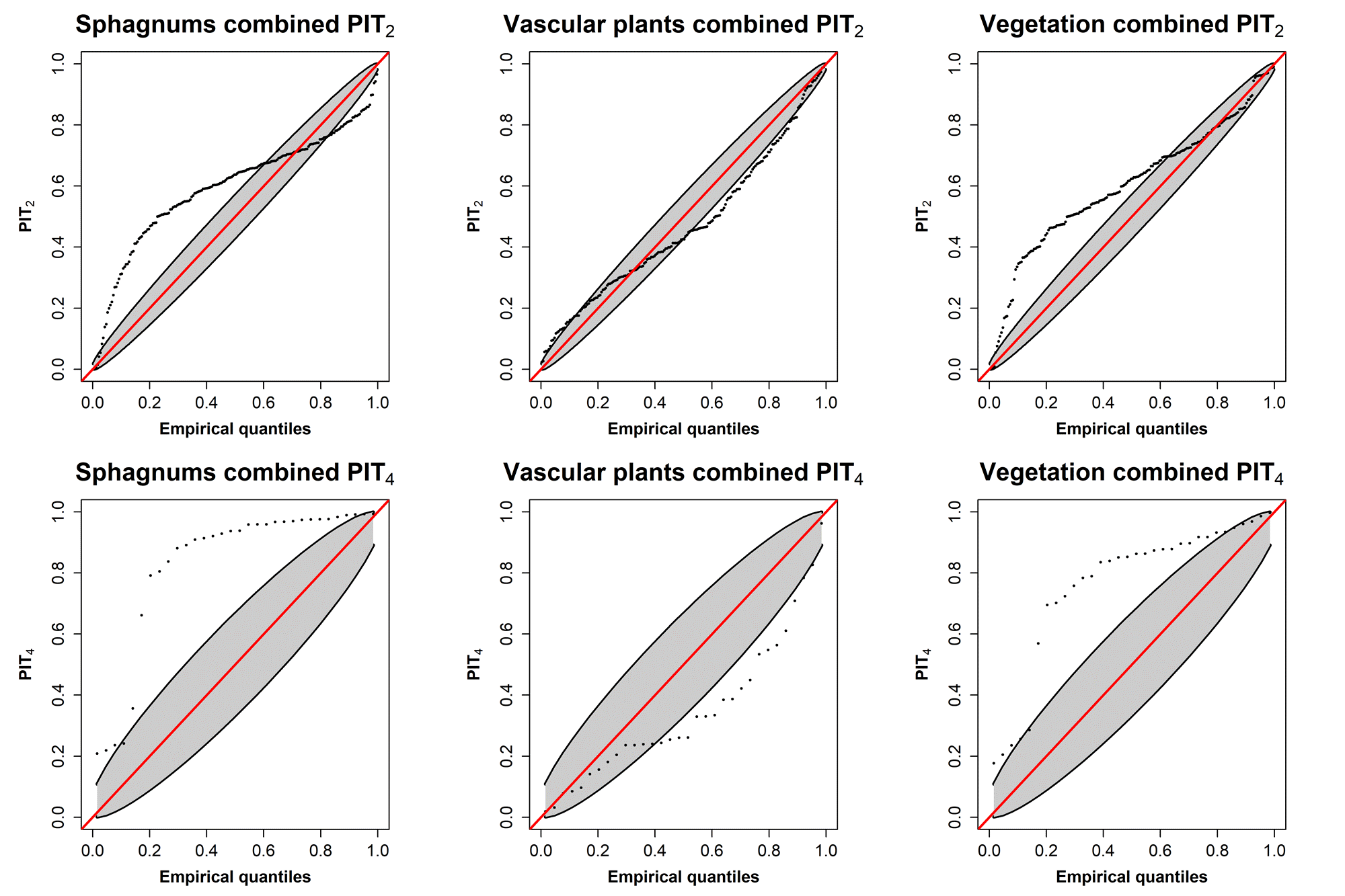}
\captionof{figure}{Q-Q plots with point-wise 95$\%$ confidence intervals of the randomized PIT$_{2}$ and PIT$_{4}$ for the LMC(2)$_\mathrm{S}$+DM model separately for sphagnums, vascular plants and combined vegetation.}
\label{fig:LMC(2)_S+DM-PIT2+PIT4}
\end{center}
\end{figure}

\begin{figure}
\begin{center}
\includegraphics[]{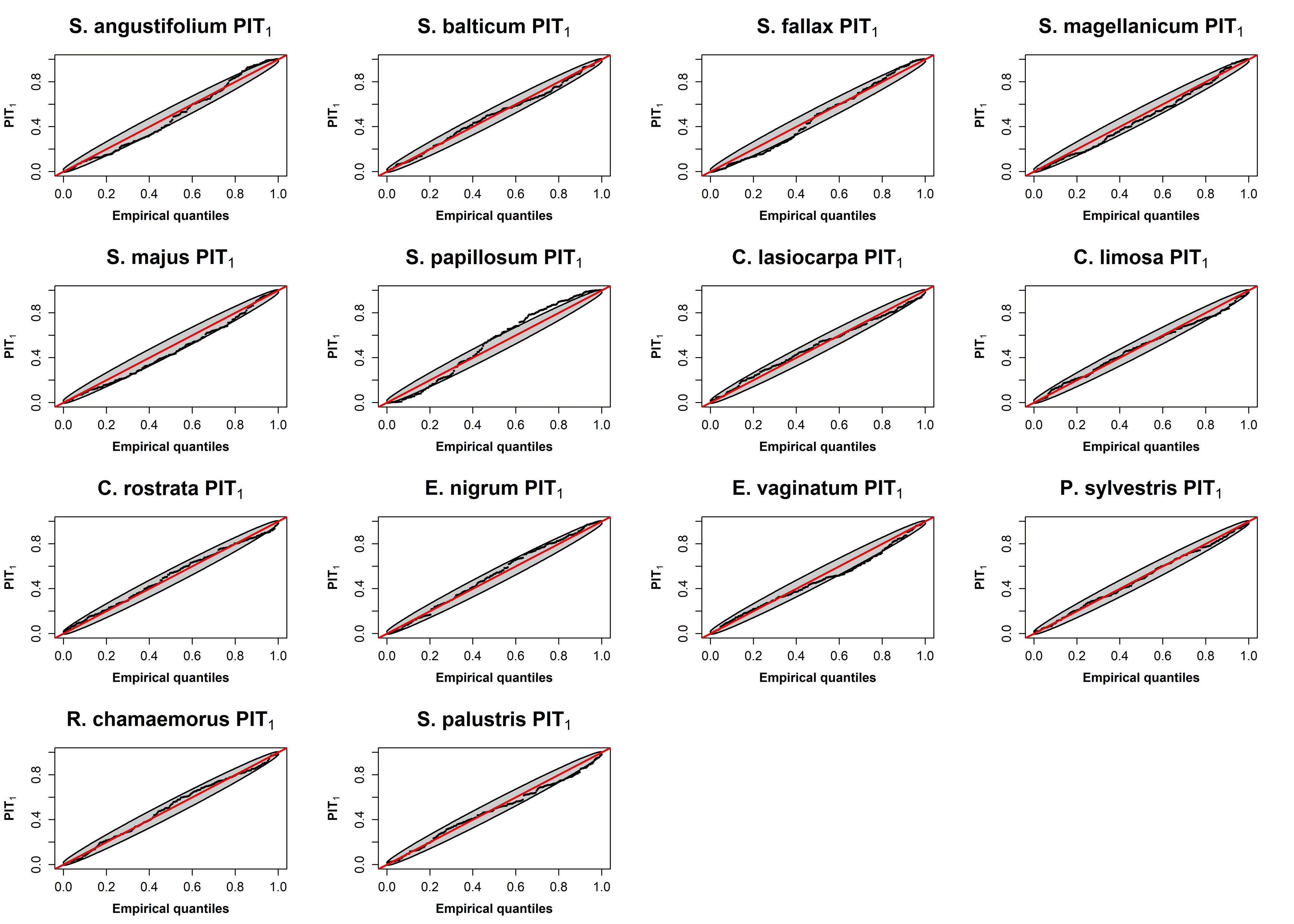}
\captionof{figure}{Species specific Q-Q plots with point-wise 95$\%$ confidence intervals of the randomized PIT$_{1}$ for the LMC(1)$_\mathrm{NS}$+DM model.}
\label{fig:PIT2}
\end{center}
\end{figure}

\begin{figure}
\begin{center}
\includegraphics[]{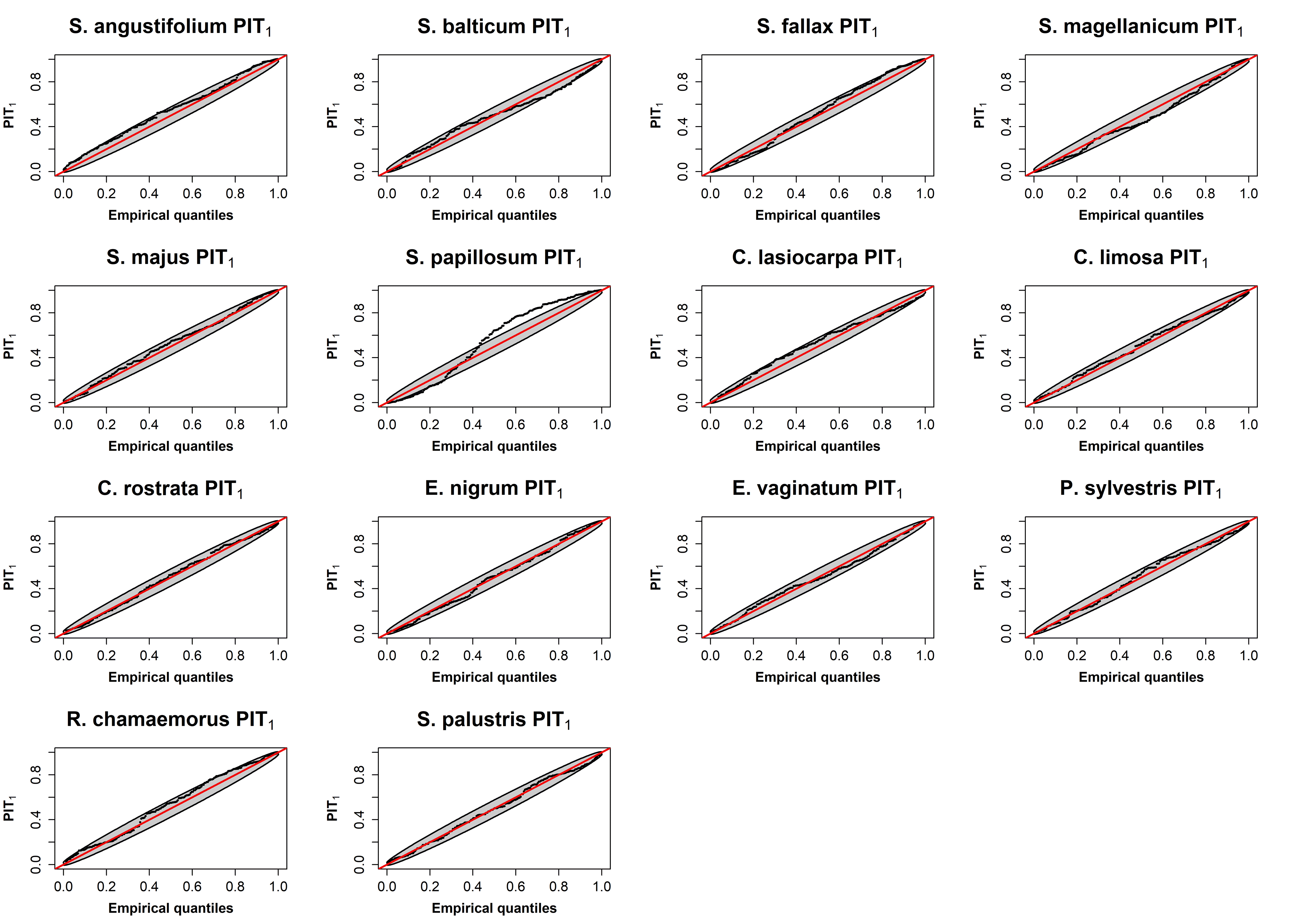}
\captionof{figure}{Species specific Q-Q plots with point-wise 95$\%$ confidence intervals of the randomized PIT$_{1}$ for the LMC(2)$_\mathrm{S}$+DM model.}
\label{fig:LMC(2)_S+DM-PIT1}
\end{center}
\end{figure}

\begin{figure}
\begin{center}
\includegraphics[]{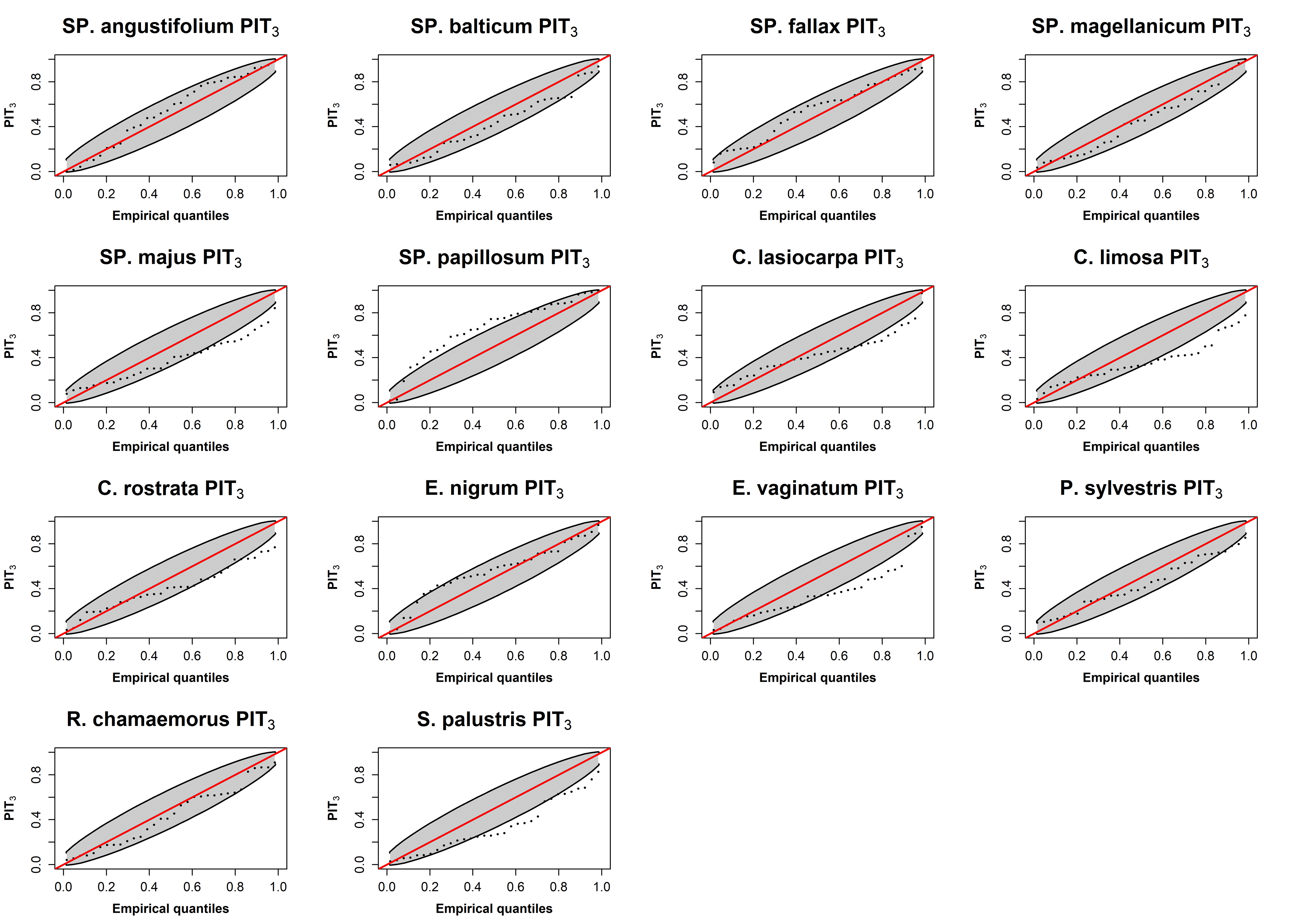}
\captionof{figure}{Species specific Q-Q plots with point-wise 95$\%$ confidence intervals of the randomized PIT$_{3}$ for the LMC(1)$_\mathrm{NS}$+DM model.}
\label{fig:PIT3}
\end{center}
\end{figure}

\begin{figure}
\begin{center}
\includegraphics[]{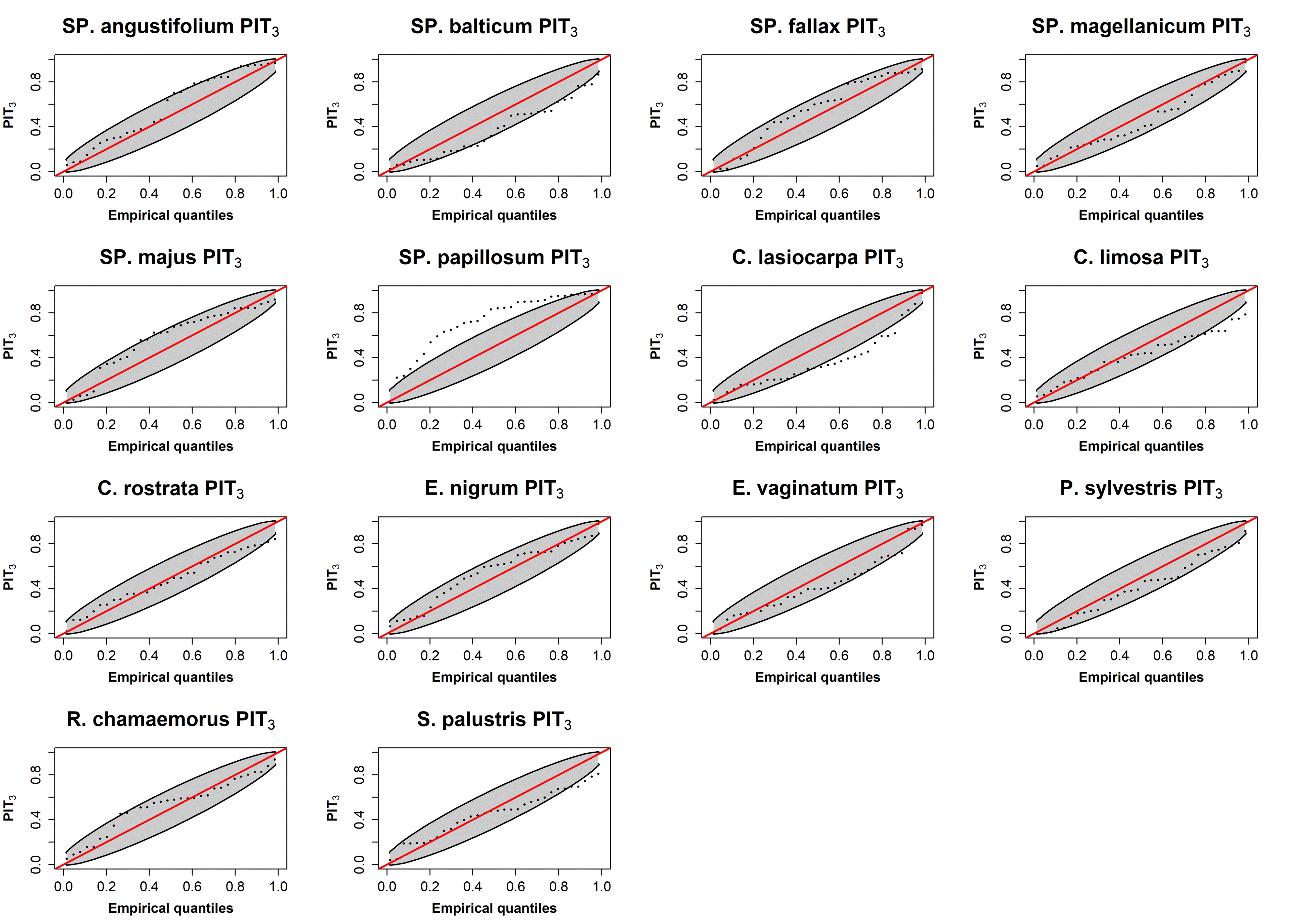}
\captionof{figure}{Species specific Q-Q plots with point-wise 95$\%$ confidence intervals of the randomized PIT$_{3}$ for the LMC(2)$_\mathrm{S}$+DM model.}
\label{fig:LMC(2)_S+DM_PIT3}
\end{center}
\end{figure}

\begin{figure}
\begin{center}
\includegraphics{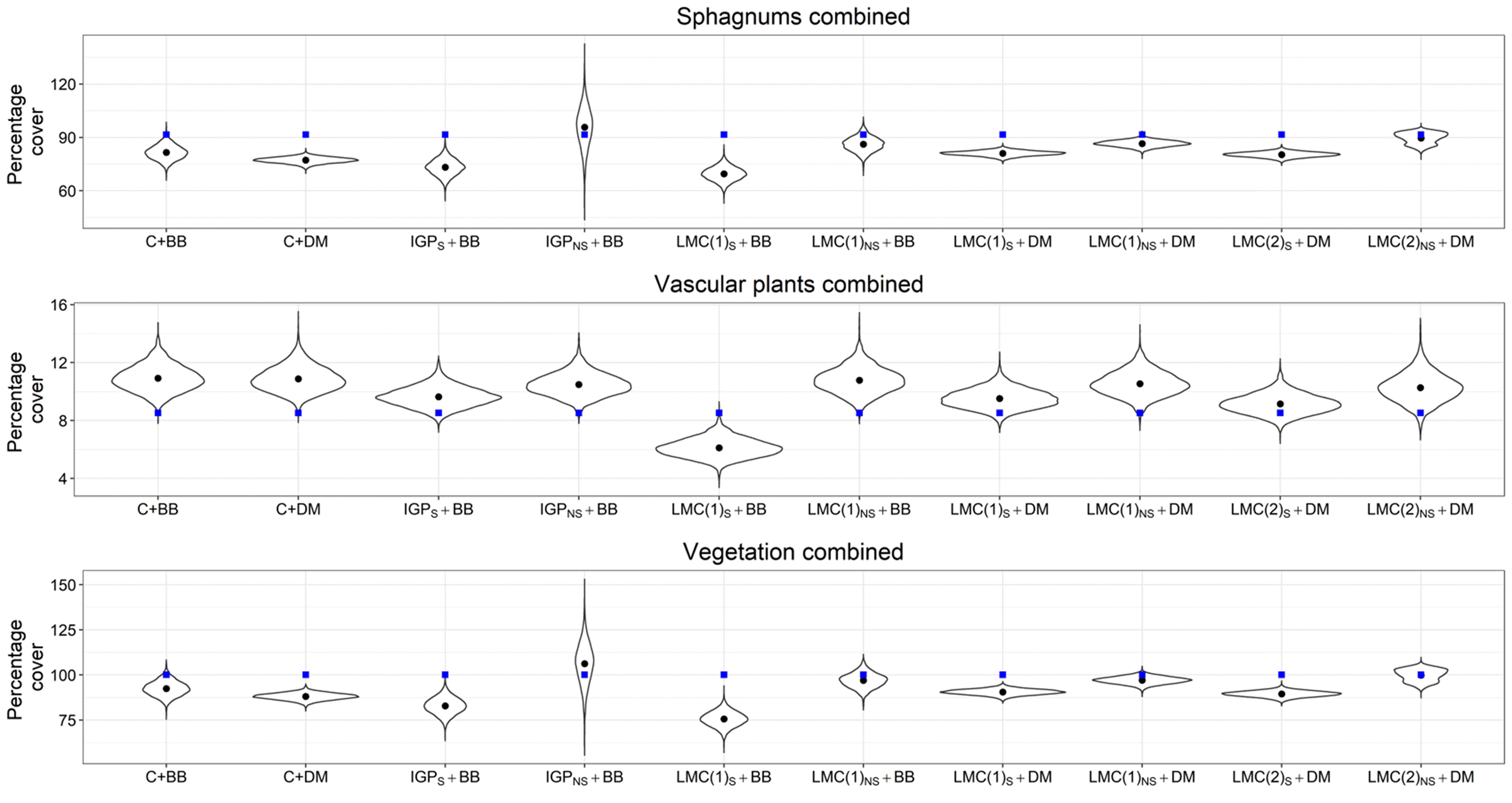}
\captionof{figure}{Posterior predictive distributions conditional on the full training data for the estimated combined sphagnum, vascular plants and total cover over the area for every model. Black dots show the posterior predictive means and blue squares represent the empircal average total percentage cover over the training data.  }
\label{fig:combined_totals}
\end{center}
\end{figure}

\begin{figure}  
\begin{center}  
\includegraphics{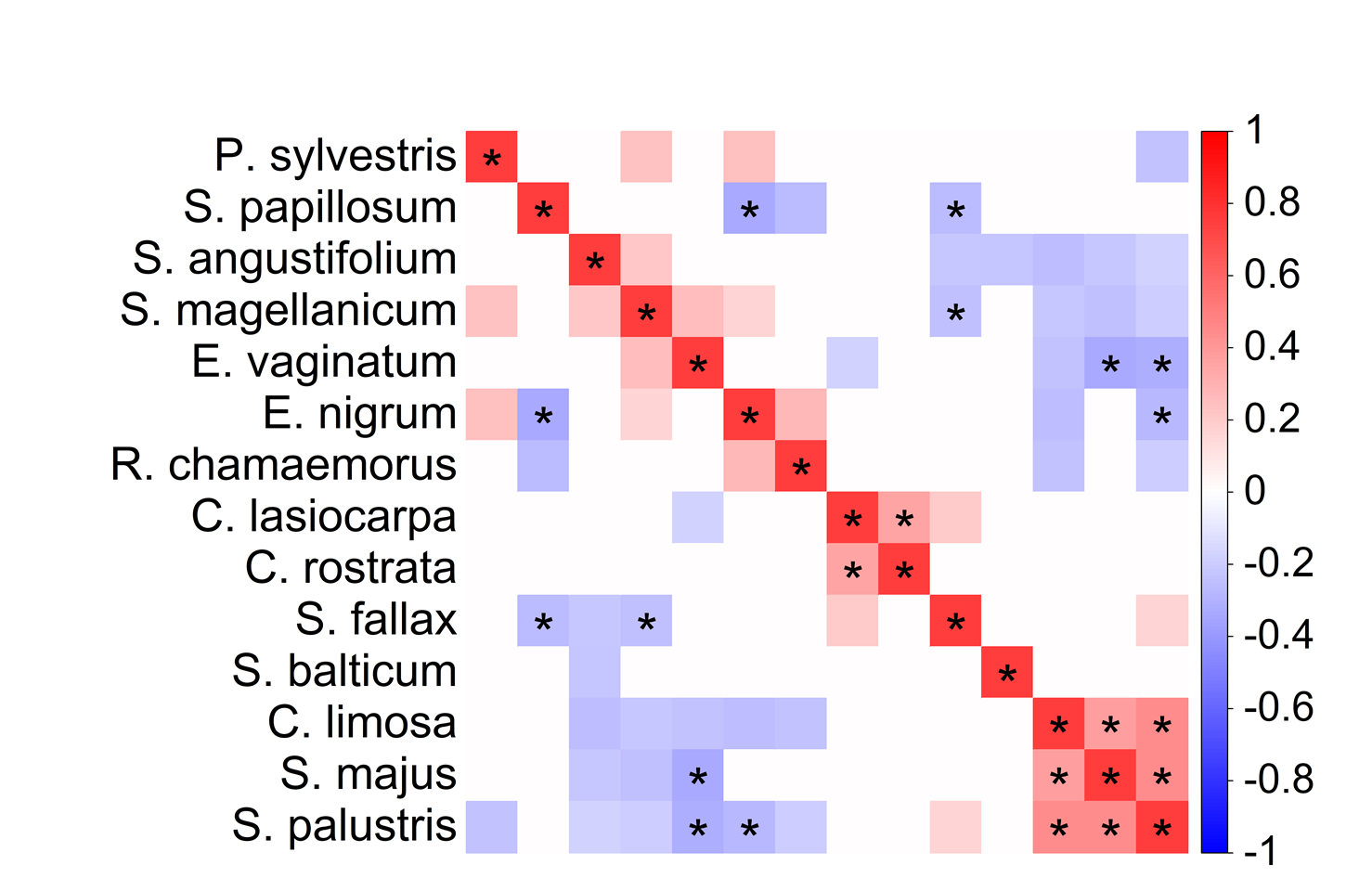}
\captionof{figure}{Posterior mean estimates of the interspecific correlations in the species niche preference in the LMC(1)$_\mathrm{NS}$+BB model. White cells indicate that the 80\% posterior credible interval of the correlation overlapped zero (i.e., weak support for interspecific correlation) and stars indicate that the 95\% posterio credible interval did not overlap zero (i.e. strong support for interspecific correlation).}
\label{fig:spatial-correlations_ICM_BB}
\end{center}
\end{figure}

\begin{table}[ht]
\renewcommand*{\arraystretch}{0.5}
\footnotesize
\centering
\caption{Species specific model comparison results with 10-fold cross-validation using log predictive density utility $CV_1$ together with the standard error estimates (se) and the Monte Carlo error estimate (me) for the CV estimates.}
\begin{tabular}[t]{lcccc}
\toprule

   & \textbf{1) C+BB} & \textbf{2) C+DM} & \textbf{3a) IGP$_\mathrm{S}$+BB} & \textbf{3b) IGP$_\mathrm{NS}$+BB}    \\

\textbf{Species}  & $CV_{1}$ (se/me) & $CV_{1}$ (se/me) & $CV_{1}$ (se/me) & $CV_{1}$ (se/me)  \\
\midrule
\textit{S. angustifolium} & -1.50 (2e-1/2e-4) & -1.50 (2e-1/1e-4) & -1.47 (2e-1/5e-4) & \textbf{-1.46}  (2e-1/5e-4)  \\
\textit{S. balticum} & -2.70 (1e-1/1e-4) & -2.72 (1e-1/1e-4) & \textbf{-2.51} (2e-1/6e-4) & \textbf{-2.51}  (2e-1/6e-4)  \\
\textit{S. magellanicum} & -3.22 (1e-1/2e-4) & -3.22 (1e-1/1e-4) & -3.05 (2e-1/5e-4) & -3.05 (2e-1/7e-4)  \\
\textit{S. majus} & -1.02 (2e-1/1e-4) & -1.01 (2e-1/5e-5) & -0.93 (1e-1/5e-4) & -0.94 (1e-1/4e-4)  \\
\textit{S. papillosum} & -4.11 (1e-1/2e-4) & -4.17 (1e-1/2e-4) & -4.18 (1e-1/1e-3) & -4.41 (1e-1/4e-3)  \\
\textit{S. fallax} & -2.47 (2e-1/1e-4) & -2.49 (2e-1/1e-4) & -2.33 (2e-1/7e-4) & \textbf{-2.31} (2e-1/7e-4)  \\
\textit{C. lasiocarpa} & -1.05 (1e-1/1e-4) & -1.05 (1e-1/1e-4) & -0.87 (1e-1/3e-4) & -0.88 (1e-1/3e-4)  \\
\textit{C. limosa} & -1.34 (1e-1/1e-4) & -1.34 (1e-1/1e-4) & -1.23 (1e-1/4e-4) & -1.24 (1e-1/3e-4)  \\
\textit{C. rostrata} & -0.90 (1e-1/1e-4) & -0.90 (1e-1/2e-4) & -0.82 (1e-1/3e-4) & -0.82 (1e-1/3e-4)  \\
\textit{e. nigrum} & -0.83 (1e-1/2e-4) & -0.83 (1e-1/2e-4) & -0.80 (1e-1/4e-4) & -0.80 (1e-1/4e-4)  \\
\textit{E. vaginatum} & -3.10 (1e-1/2e-4) & -3.10 (1e-1/2e-4) & -2.86 (1e-1/4e-4) & -2.87 (1e-1/5e-4)  \\
\textit{P. sylvestris} & -0.58 (9e-2/1e-4) & -0.58 (9e-2/1e-4) & -0.55 (9e-2/3e-4) & -0.55 (9e-2/3e-4)  \\
\textit{R. chamaemorus} & -0.65 (1e-1/1e-4) & -0.65 (1e-1/1e-4) & -0.60 (1e-1/5e-4) & -0.60 (1e-1/4e-4)  \\
\textit{S. palustris} & -1.69 (1e-1/1e-4) & -1.69 (1e-1/1e-4) & -1.42 (1e-1/4e-4) & -1.43 (1e-1/4e-4)  \\

\bottomrule
\\
\\
\midrule
& \textbf{5a) LMC(1)$_\mathrm{S}$+BB}  & \textbf{5b) LMC(1)$_\mathrm{NS}$+BB}    & \textbf{6a) LMC(1)$_\mathrm{S}$+DM}   & \textbf{6b) LMC(1)$_\mathrm{NS}$+DM}   \\

\textbf{Species}  & $CV_{1}$ (se/me) & $CV_{1}$ (se/me) & $CV_{1}$ (se/me) & $CV_{1}$ (se/me) \\
\midrule \\
\textit{S. angustifolium} & \textbf{-1.46} (2e-1/5e-4) & -1.47 (1.6e-1/5e-4) & -1.47 (2e-1/5e-4) & -1.53 (2e-1/1e-3)  \\
\textit{S. balticum} & \textbf{-2.51} (2e-1/6e-4) & \textbf{-2.51} (1.5e-1/1e-3) & \textbf{-2.51} (1e-1/4e-4) & -2.52 (2e-1/8e-4)  \\
\textit{S. magellanicum} & \textbf{-3.03} (2e-1/5e-4) & -3.09 (1.5e-1/9e-4) & \textbf{-3.03} (1e-1/5e-4) & -3.11 (2e-1/1e-3)  \\
\textit{S. majus} & \textbf{-0.91} (1e-1/5e-4) & -0.92 (1.4e-1/7e-4) & -0.92 (1e-1/5e-4) & -0.98 (2e-1/9e-4)  \\
\textit{S. papillosum} & \textbf{-4.10} (1e-1/8e-4) & -4.38 (1.7e-1/9e-3) & -4.18 (1e-1/9e-4) & -4.31 (2e-1/3e-3)  \\
\textit{S. fallax} & -2.34 (2e-1/7e-4) & -2.33 (1.8e-1/1e-3) & -2.34 (2e-1/6e-4) & -2.37 (2e-1/2e-3)  \\
\textit{C. lasiocarpa} & -0.87 (1e-1/3e-4) & \textbf{-0.86} (1.1e-1/4e-4) & -0.87 (1e-1/3e-4) & -0.87 (1e-1/4e-4)  \\
\textit{C. limosa} & \textbf{-1.22} (1e-1/3e-4) & \textbf{-1.22} (1.0e-1/4e-4) & \textbf{-1.22} (1e-1/3e-4) & \textbf{-1.22} (1e-1/5e-4)  \\
\textit{C. rostrata} & -0.82 (1e-1/3e-4) & \textbf{-0.81} (1.0e-1/3e-4) & \textbf{-0.81} (1e-1/3e-4) & \textbf{-0.81} (1e-1/3e-4)  \\
\textit{e. nigrum} & \textbf{-0.79} (1e-1/4e-4) & -0.80 (1.3e-1/6e-4) & \textbf{-0.79} (1e-1/4e-4) & -0.80 (1e-1/6e-4)  \\
\textit{E. vaginatum} & \textbf{-2.86} (1e-1/4e-4) & -2.88 (1.2e-1/4e-4) & \textbf{-2.86} (1e-1/5e-4) & -2.88 (1e-1/5e-4) \\
\textit{P. sylvestris} & \textbf{-0.54} (9e-2/3e-4) & -0.54 (8.8e-2/3e-4) & \textbf{-0.54} (9e-2/3e-4) & -0.55 (9e-2/3e-4)  \\
\textit{R. chamaemorus} & \textbf{-0.59} (1e-1/4e-4) & -0.60 (1.1e-1/4e-4) & \textbf{-0.59} (1e-1/4e-4) & \textbf{-0.59} (1e-1/5e-4)  \\
\textit{S. palustris} & \textbf{-1.40} (1e-1/4e-4) & -1.40 (1.1e-1/4e-4) & \textbf{-1.40} (1e-1/4e-4) & -1.42 (1e-1/5e-4)  \\

\bottomrule
\\
\\

\cmidrule(lr){1-4} 
& \textbf{7a) LMC(2)$_\mathrm{S}$+DM} & \textbf{7b) LMC(2)$_\mathrm{NS}$+DM} & \textbf{4a) IGP$_\mathrm{S}$+DM}   \\

\textbf{Species}  & $CV_{1}$ (se/me)   & $CV_{1}$ (se/me)& $CV_{1}$ (se/me)  \\
\midrule
\textit{S. angustifolium} & -1.47 (2e-1/4e-4)  & -1.72 (1.9e-1/1.9e-3) & -1.50 (2e-1/1e-3)  \\
\textit{S. balticum} & -2.52 (1e-1/4e-4)  & -3.05 (1.7e-1/1.9e-3) & -2.86 (2e-1/3e-3)  \\
\textit{S. magellanicum} & \textbf{-3.03} (1e-1/4e-4)  & -3.82 (1.9e-1/3.1e-3) & -3.36 (2e-1/3e-3)  \\
\textit{S. majus} & -0.92 (1e-1/5e-4)  & -1.43 (2.1e-1/2.4e-3) & -0.98 (1e-1/1e-3)  \\
\textit{S. papillosum} & -4.16 (1e-1/7e-4)  & -5.26 (1.8e-1/7.8e-3) & -4.30 (2e-1/3e-3)  \\
\textit{S. fallax} & -2.34 (2e-1/6e-4)  & -3.36 (2.5e-1/3.8e-3) & -2.40 (2e-1/2e-3)  \\
\textit{C. lasiocarpa} & -0.87 (1e-1/3e-4)  & -1.16 (1.4e-1/7.1e-4) & -0.87 (1e-1/3e-4)  \\
\textit{C. limosa} & -1.23 (1e-1/3e-4)  & -1.5 (1.2e-1/5.7e-4) & -1.23 (1e-1/4e-4)  \\
\textit{C. rostrata} & \textbf{-0.81} (1e-1/3e-4)  & -0.96 (1.2e-1/4.3e-4) & -0.82 (1e-1/3e-4)  \\
\textit{e. nigrum} & \textbf{-0.79} (1e-1/4e-4)  & -0.95 (1.5e-1/6.6e-4) & -0.80 (1e-1/4e-4)  \\
\textit{E. vaginatum} & \textbf{-2.86} (1e-1/4e-4)  & -3.31 (1.3e-1/8.1e-4) & \textbf{-2.86} (1e-1/4e-4)  \\
\textit{P. sylvestris} & \textbf{-0.54} (9e-2/3e-4)  & -0.59 (9e-2/3.4e-4) & -0.55 (9e-2/3e-4)  \\
\textit{R. chamaemorus} & \textbf{-0.59} (1e-1/4e-4)  & -0.71 (1.3e-1/4.9e-4) & -0.60 (1e-1/5e-4)  \\
\textit{S. palustris} & \textbf{-1.40} (1e-1/4e-4)  & -1.96 (1.5e-1/9.2e-4) & -1.42 (1e-1/4e-4) \\

\cmidrule(lr){1-4} 
\end{tabular}
\label{table:species-specific-CV1}

\end{table}

\begin{table}[ht]
\renewcommand*{\arraystretch}{0.5}
\centering
\footnotesize
\caption{Species specific model comparison results with 10-fold cross-validation using log predictive density utility $CV_3$ together with the standard error estimates (se) and the Monte Carlo error estimate (me) for the CV estimates.}
\begin{tabular}[t]{lccccccc}
\toprule

   & \textbf{1) C+BB} & \textbf{2) C+DM} & \textbf{3a) IGP$_\mathrm{S}$+BB} & \textbf{3b) IGP$_\mathrm{NS}$+BB}    \\

\textbf{Species}  & $CV_{3}$ (se/me) & $CV_{3}$ (se/me) & $CV_{3}$ (se/me) & $CV_{3}$ (se/me)  \\
\midrule

\textit{S. angustifolium} &  -30.0 (4/3e-3) & -30.0 (3/2e-3) & -29.4 (4/2e-2) & -29.2 (3/2e-2)  \\
\textit{S. balticum} &  -54.0 (3/2e-3) & -54.4 (3/2e-3) & -50.1 (4/2e-2) & \textbf{-49.9} (3/3e-2)  \\
\textit{S. magellanicum} &  -64.4 (4/3e-3) & -64.4 (4/2e-3) & -60.9 (4/3e-2) & -61.0 (4/4e-2)  \\
\textit{S. majus} & -20.3 (2/2e-3) & -20.2 (2/1e-3) & -18.7 (3/2e-2) & -18.8 (2/1e-2)  \\
\textit{S. papillosum} & -82.1 (3/3e-3) & -83.3 (2/3e-3) & -81.9 (3/5e-2) & -88.5 (3/3e-1)  \\
\textit{S. fallax} & -49.3 (4/2e-3) & -49.7 (4/2e-3) & -46.4 (4/3e-2) & -\textbf{46.2} (4/5e-2)  \\
\textit{C. lasiocarpa} & -21.1 (2/2e-3) & -21.1 (2/2e-3) & -17.4 (2/8e-3) & -17.5 (2/8e-3)  \\
\textit{C. limosa} & -26.7 (1/2e-3) & -26.7 (1/2e-3) & -24.5 (1/9e-3) & -24.7 (1/1e-2)  \\
\textit{C. rostrata} & -18.0 (3/2e-3) & -18.0 (2/2e-3) & -16.3 (2/8e-3) & -16.4 (2/7e-3)  \\
\textit{E. nigrum} & -16.6 (3/3e-3) & -16.6 (3/3e-3) & -15.9 (3/2e-2) & -15.9 (3/2e-2)  \\
\textit{E. vaginatum} & -62.0 (2/4e-3) & -62.0 (2/4e-3) & -57.1 (2/2e-2) & -57.1 (2/2e-2)  \\
\textit{P. sylvestris} & -11.5 (2/2e-3) & -11.5 (2/2e-3) & -10.9 (2/7e-3) & -10.9 (2/6e-3)  \\
\textit{R. chamaemorus} & -13.0 (3/2e-3) & -13.0 (3/2e-3) & -11.8 (3/2e-2) & -12.0 (3/1e-2)  \\
\textit{S. palustris} & -33.8 (2/2e-3) & -33.8 (2/2e-3) & -28.2 (2/1e-2) & -28.6 (2/1e-2)  \\

\bottomrule
\\
\\
   \cmidrule(lr){1-5}
& \textbf{5a) LMC(1)$_\mathrm{S}$+BB}  & \textbf{5b) LMC(1)$_\mathrm{NS}$+BB}    & \textbf{6a) LMC(1)$_\mathrm{S}$+DM}   & \textbf{6b) LMC(1)$_\mathrm{NS}$+DM}   \\
   
\textbf{Species}  & $CV_{3}$ (se/me) & $CV_{3}$ (se/me)  & $CV_{3}$ (se/me) & $CV_{3}$ (se/me)  \\
\cmidrule(lr){1-5} 

\textit{S. angustifolium} & -29.1 (3/1e-2) & \textbf{-28.8} (3/2e-2) & -29.2 (3/2e-2) & -30.5 (4/6e-2)  \\
\textit{S. balticum} & -50.1 (3/2e-2) & -50.3 (4/5e-2) & -50.0 (3/2e-2) & -50.5 (3/6e-2)  \\
\textit{S. magellanicum} & \textbf{-60.6} (4/2e-2) & -61.6 (4/9e-2) & \textbf{-60.6} (4/2e-2) & -62.3 (4/6e-2)  \\
\textit{S. majus} & \textbf{-18.3} (2/1e-2) & -18.8 (3/3e-2) & -18.5 (3/2e-2) & -19.7 (3/5e-2)  \\
\textit{S. papillosum} & \textbf{-81.3} (3/5e-2) & -94.1 (5/3e-2) & -83.1 (3/5e-2) & -86.3 (4/3e-1)  \\
\textit{S. fallax} & -46.5 (4/4e-2) & -47.3 (4/6e-2) & -46.7 (4/4e-2) & -48.1 (5/9e-2)  \\
\textit{C. lasiocarpa} & -17.4 (2/9e-3) & \textbf{-17.2} (2/1e-2) & -17.3 (2/9e-3) & \textbf{-17.2} (2/1e-2)  \\
\textit{C. limosa} & \textbf{-24.3} (1/1e-2) & -24.3 (1/1e-2) & -24.4 (1/9e-3) & -24.4 (1/2e-2)  \\
\textit{C. rostrata} & -16.3 (2/7e-3) & -16.2 (2/8e-3) & -16.3 (2/7e-3) & -16.3 (2/9e-3)  \\
\textit{E. nigrum} & -15.7 (3/1e-2) & -15.9 (3/4e-2) & -15.7 (3/1e-2) & -16.0 (3/2e-2)  \\
\textit{E. vaginatum} & \textbf{-56.9} (2/2e-2) & -57.5 (2/2e-2) & \textbf{-56.9} (2/2e-2) & -57.3 (2/2e-2)  \\
\textit{P. sylvestris} & -10.8 (2/6e-3) & -10.8 (2/7e-3) & -10.8 (2/6e-3) & -10.9 (2/7e-3)  \\
\textit{R. chamaemorus} & \textbf{-11.6} (3/1e-2) & -11.8 (3/3e-2) & \textbf{-11.6} (3/2e-2) & -11.7 (3/2e-2)  \\
\textit{S. palustris} & \textbf{-27.9} (2/1e-2) & -28.1 (2/2e-2) & \textbf{-27.9} (2/1e-2) & -28.3 (2/2e-2)  \\

\bottomrule
\\
\\
   \cmidrule(lr){1-4}
& \textbf{7a) LMC(2)$_\mathrm{S}$+DM} & \textbf{7b) LMC(2)$_\mathrm{NS}$+DM} & \textbf{4a) IGP$_\mathrm{S}$+DM}   \\
   
\textbf{Species}  & $CV_{3}$ (se/me) & $CV_{3}$ (se/me)  \\
\cmidrule(lr){1-4} 

\textit{S. angustifolium} &  -29.3 (3/1e-2)  & -32.1 (4/6e-2)& -30.5 (4/5e-2)  \\
\textit{S. balticum} & -50.3 (3/1e-2)  & -58.0 (4/7e-2)& -52.0 (3/6e-2)  \\
\textit{S. magellanicum} &  \textbf{-60.6} (4/1e-2)  & -72.0 (4/1e-1)& -62.1 (4/7e-2) \\
\textit{S. majus} & -18.5 (3/1e-2)  & -26.8 (3/2e-1)& -18.7 (2/3e-2)  \\
\textit{S. papillosum} &  -82.7 (3/5e-2)  & -97.8 (3/2e.1)& -83.2 (3/1e-1)  \\
\textit{S. fallax} &  -46.7 (4/3e-2)  & -61.5 (5/2e-1)& -47.9 (4/1e-1)  \\
\textit{C. lasiocarpa} &  \textbf{-17.2} (2/7e-3)  & -22.7 (3/4e-2) & -17.4 (2/8e-3)  \\
\textit{C. limosa} &  -24.5 (1/9e-3)  & -29.6 (8e-1/2e-2)& -24.5 (1/9e-3)  \\
\textit{C. rostrata} &  \textbf{-16.2} (2/6e-3)  & -19.2 (2/1e-2)& -16.3 (2/8e-3)  \\
\textit{E. nigrum} &  \textbf{-15.6} (3/2e-2)  & -18.7 (4/3e-2) & -15.9 (3/2e-2)  \\
\textit{E. vaginatum} & -57.1 (2/1e-2)  & -65.2 (2/8e-2)& -57.1 (2/1e-2)  \\
\textit{P. sylvestris} & \textbf{-10.7} (2/6e-3)  & -11.8 (2/9e-3) & -10.9 (2/7e-3)  \\
\textit{R. chamaemorus} & -11.7 (3/2e-2)  & -14.2 (3/3e-2)& -11.8 (3/2e-2) \\
\textit{S. palustris} & -28.0 (2/1e-2)  & -38.5 (2/1e-1) & -28.2 (2/1e-2) \\

\cmidrule(lr){1-4} 
\end{tabular}
\label{table:species-specific-CV3}
\end{table}

\end{document}